\documentclass[longauth]{aa}
\usepackage{graphicx}
\usepackage{color}
\usepackage[dvipsnames]{xcolor}
\usepackage{hyperref}
\usepackage{natbib}
\usepackage[title]{appendix}
\usepackage[colorinlistoftodos]{todonotes}
\usepackage{comment}
\usepackage{multirow}

\usepackage[symbol]{footmisc}

\usepackage{txfonts}
\usepackage{soul}

\newcommand{\wazp}{\textsc{W}a\textsc{ZP}}
\newcommand{\RM}{redMaPPer}
\newcommand{\spt}{\textsc{SPT}}
\newcommand{\act}{\textsc{ACT}}
\newcommand{\sz}{SZE}

\newcommand{\wazpYI}{\textsc{W}a\textsc{ZP}-Y1}

\newcommand{\wazpYVI}{\textsc{W}a\textsc{ZP}-Y6}

\newcommand{\phz}{\textsc{photo-$z$}}
\newcommand{\spz}{\textsc{spec-$z$}}

\newcommand{\zsp}{z_{\rm spec}}
\newcommand{\zph}{z_{\rm phot}}
\newcommand{\zdv}{$z_{\Delta v}$}
\newcommand{\zbcg}{$z_{\rm CG}$}

\newcommand{\zband}{$z$-band}

\newcommand{\zmax}{z_{max}}

\newcommand{\dt}{\Delta\theta}
\newcommand{\dz}{\Delta z}
\newcommand{\drp}{\Delta r_\perp}

\newcommand{\rich}{N_{gals}}
\newcommand{\zwazp}{z_{\wazp}}
\newcommand{\zsz}{z_{\rm\sz}}

\newcommand{\kspt}{$k_{\rm SPT}$}

\newcommand{\richI}{\rich^{Y1}}
\newcommand{\richVI}{\rich^{Y6}}

\newcommand{\actdeep}{\act$_{\rm deep}$}

\newcommand{\code}[1]{\texttt{#1}}
\newcommand{\reffigure}[1]{Fig.~\ref{fig:#1}}

\newcommand{\wazpweblink}{
    \footnote{
        \href{https://des.ncsa.illinois.edu/releases/y6a2/Y6cluster-wazp}{https://des.ncsa.illinois.edu/releases/y6a2/Y6cluster-wazp}
    }
}

\newcommand{\includegraphicsX}[2][]{
    \includegraphics[#1]{#2}
}

\newcommand{\Dv}{$\Delta v$}
\newcommand{\ns}{{\color{white}0}}

\begin{document}
%\linenumbers
%\modulolinenumbers[5]
%\nolinenumbers

\title{
    The final \wazp\ galaxy cluster catalog of the Dark Energy Survey and comparison with \sz\ data
}

\author{
    C. Benoist$^{1,2}$
    \and
    M. Aguena$^{2,3}$
    \and
    L. da Costa$^{2}$
    \and
    J. Gschwend$^{2}$
    \and
    S. Allam$^{4}$
    \and
    O. Alves$^{5}$
    \and
    F. Andrade-Oliveira$^{6}$
    \and
    D. Bacon$^{7}$
    \and
    L. Bleem$^{8}$
    \and
    D. Brooks$^{9}$
    \and
    A. Carnero Rosell$^{1,10,11}$
    \and
    J. Carretero$^{12}$
    \and
    F.J. Castander$^{13,14}$
    \and
    M. Costanzi$^{15,16,3}$
    \and
    J. De Vicente$^{17}$
    \and
    S. Desai$^{18}$
    \and
    S. Dodelson$^{19,20,4}$
    \and
    P. Doel$^{9}$
    \and
    S. Everett$^{21}$
    \and
    B. Flaugher$^{4}$
    \and
    J. Frieman$^{19,20,4}$
    \and
    J. Garcia-Bellido$^{22}$
    \and
    G. Giannini$^{12,20}$
    \and
    P. Giles$^{23}$
    \and
    R. Gruendl$^{24,25}$
    \and
    G. Gutierrez$^{4}$
    \and
    S. Hinton$^{26}$
    \and
    D.L. Hollowood$^{27}$
    \and
    K. Honscheid$^{28,29}$
    \and
    D. James$^{30}$
    \and
    K. Kuehn$^{31,32}$
    \and
    S. Lee$^{33}$
    \and
    J. Marshall$^{34}$
    \and
    J. Mena-Fernández$^{35}$
    \and
    R. Miquel$^{12,36}$
    \and
    A. Plazas Malagón$^{37,38}$
    \and
    K. Romer$^{23}$
    \and
    E. Sanchez$^{17}$
    \and
    B. Santiago$^{1,39}$
    \and
    I. Sevilla$^{17}$
    \and
    M. Smith$^{40}$
    \and
    E. Suchyta$^{41}$
    \and
    G. Tarle$^{5}$
    \and
    N. Weaverdyck$^{42,43}$
    \and
    J. Weller$^{44,45}$
    \and
    M.E. da Silva Pereira$^{46}$
}

\abstract{
    In this work, we present and characterize the galaxy cluster catalog detected by the \wazp\ cluster finder, which is not based on red-sequence identification, on the full six years of observations of the Dark Energy Survey (DES-Y6).
    The full catalog contains over 400k detected clusters with richnesses, $\rich$, above 5 and that reach redshifts up to $z=1.3$. We also provide a version of the catalog where the observation depth and richness computation are homogenized to be used for cosmology, containing 33k rich ($\rich$>25) clusters.
    We compare our results with the previous \wazp\ catalog obtained from the DES first-year data release (DES-Y1).
    We find that essentially all clusters within the common footprint and depth limit are recovered. The deeper observations on DES-Y6 and the more complete available spectroscopic redshift sample lead to improvements in the redshifts of the clusters, resulting in an average scatter of 1.4\% and offset of 0.2\%.
    The optical clusters are also cross-matched with Sunyaev Zel'dovich Effect (\sz) cluster samples detected by the South Pole Telescope (SPT) and the Atacama Cosmology Telescope (ACT).
    We find that essentially all \sz\ clusters with reasonable overlapping footprint have a corresponding \wazp\ cluster. Conversely, 90\% of the optical detections with richness greater than 150 have a counterpart in the deeper regions of the \sz\ surveys.
Based on cross-match with the \sz\ catalogs, we also find that 15-20\% of the \sz\ matched systems have more than one possible \wazp\ counterpart at the same redshift and within the \sz\ $R_{500c}$, indicating possible interacting or unrelaxed systems. Finally, given the optical and \sz\ beams, \wazp\ and \sz\ centerings are found to be consistent.  A more detailed study of the \sz-\wazp\ mass-richness relation will be presented in a separate paper.
}

\keywords{Galaxies: clusters: general -- Galaxies: distances and redshifts -- Methods: data analysis  --  Surveys}

\maketitle
%

%%%%%%%%%%%%%%%%%%%%%
\section{Introduction}
%%%%%%%%%%%%%%%%%%%%%

\nolinenumbers

Galaxy clusters are good tracers of the largest peaks in the matter density field, and key for studying the history of mass assembly in the universe. In particular, large cluster samples can be used to constrain the evolution of the halo mass function that characterizes the mass density of halos per volume unit \citep{Wei13}.  Detected as large concentrations of galaxies, hot gas, and dark matter, galaxy clusters also provide a unique environment for studying galaxy evolution, hydrodynamics of the intergalactic medium, and plasma physics. They also host the largest galaxies in the universe, where supermassive black holes can be found. All these elements make galaxy clusters fundamental systems for addressing cosmological and astrophysical questions.

As galaxy clusters are sensitive not only to the expansion history of the universe but also to the growth of structures, they provide cosmological constraints complementary to classical geometrical probes such as Supernova Type Ia (SNIa) and Baryonic Acoustic Oscillations (BAOs).  Many studies have shown the power of using the evolution of their abundance to constrain cosmology, including the dark energy equation of state, the neutrino mass, or modified gravity scenarios (e.g. \citealt{Allen11,Mantz15,Bohringer16,Bolliet20}).  Galaxy clusters present a particular case for cosmological application with promising results \citep{DES20,Cos22,To21,Lesci22,Fumagalli23,Sunayama23, 2024PhRvD.110h3510B,2024A&A...689A.298G}.  Tens of thousands of clusters have been detected in current optical surveys such as the Dark Energy Survey (DES) \citep{McC19,Aguena21}, and we expect to have an order-of-magnitude increase in the number of clusters from the optical surveys of the upcoming decade, such as the LSST \citep[Vera C. Rubin Observatory Legacy Survey of Space and Time]{LSST2009} and the optical/near-infrared Euclid Consortium
\citep{Euclid24}.

However, several limitations may hamper the use of galaxy cluster samples as a cosmological probe.  The first is the difficulty of deriving a precise selection function for a given cluster finder.  The characterization of the selection function remains a complex task, with different methodologies based either on realistic simulations of galaxy catalogs (e.g. Euclid Collaboration 2019 and references therein), or inferred from real data, or on a mix of the two approaches \citep{goto2002, kim2002, cos19}.  Simulation-driven approaches, within the confines of their assumptions, offer a valuable truth table for comparison, featuring clusters integrated within realistic large-scale structures. They establish a direct connection between galaxy clusters and dark matter haloes. Nevertheless, these methods depend on intricate modeling that has yet to accurately replicate all observed galaxy properties, particularly at higher redshifts \citep{derose19, To24}. Furthermore, mock catalogs often fail to capture the full spectrum of complexities and intricacies found in actual images, including defects arising during source extraction and classification stages. Addressing the cluster selection function using real data encounters inherent limitations, as there is no absolute reference available to directly compare the outcomes of any cluster finder. Nonetheless, valuable insights can be derived by cross-matching samples obtained from different optical cluster finders. These resulting samples may exhibit discrepancies not only due to varying physical assumptions but also as a result of divergent implementation details within the cluster finder algorithms \citep{Asc16,AguLim18}, or even discrepancies in how the algorithms handle specific features.

An alternative way to improve our understanding of cluster selection functions can also be achieved by cross-matching with samples detected at different wavelengths (e.g., \citealp {Sar15,Ble20,Grandis21,Upsdell23}), taking advantage that each wavelength domain reflects different tracers not affected by the same projection effects. Galaxy clusters can indeed be observed across the entire electromagnetic spectrum with specific signatures: Bremsstrahlung emission of the hot intra-cluster gas is visible in X-rays \citep{Felten66}; up-scattered Cosmic Microwave Background (CMB) photons by Compton interactions on the hot gas produce the Sunyaev-Zel'dovich Effect (\sz) at mm-wavelengths \citep {SZE72}; the cluster galaxies and the gravitationally-lensed background galaxies are detected in the visible and infrared wavelength domains \citep{kim2002,Ryk14,bellagamba2018,Adam2019,McC19,Aguena21}; and cosmic-rays and magnetic fields can be probed in the radio and gamma-ray bands \citep{Carillii02}.  Such an approach is also critical for addressing another important limitation of using clusters as a cosmological probe, the difficulty of inferring a mass from the available observables and associated calibrations. In addition to the intrinsic scatter between the halo masses and the available cluster mass proxies, clusters are not isolated systems. Although they are sensitive to structure growth, the drawback is precisely the many episodes of accretion and merging processes they undergo across cosmic time, causing deviations from the frequent assumptions of equilibrium and virialization (see, for instance \citealp{Seppi2021}).  With the latest (\citealt[eROSITA]{Merloni12}; \citealt[SPT-3G]{2014SPIE.9153E..1PB}) and upcoming (Euclid; LSST; \citealt[Simons Observatory]{2019JCAP...02..056A}; etc.) generation of very large and deep surveys, systematic multi-wavelength analysis of cluster samples will undoubtedly help characterize cluster detection processes, quantifying their dynamical state and deriving better-controlled mass proxies.

In this paper, we present the cluster catalog derived from the \wazp\ cluster finder applied to the final release of the Dark Energy Survey (DES-Y6, \citealt{Abb21}). The DES is currently one of the largest wide-area photometric surveys with a coverage of 1/8th of the sky with quality astrometry, photometry, and shape measurements. The resulting product is a unique catalog containing hundreds of thousands of optical clusters whose detection did not rely on the presence of a red-sequence, and with a thorough validation on all steps of the data handling, from the co-added objects to the final cluster detection. This work follows a similar study based on DES-Y1, the first-year release of the DES (\citealp{Aguena21}, hereafter Paper-1), where the \wazp\ cluster finder is described in detail and our results compared to the \RM\ catalog also extracted from the same dataset. Here, the \wazp\ catalog is cross-matched with \sz\ cluster detections by the \spt\ \citep[South Pole Telescope]{Car11} and the \act\  \citep[Atacama Cosmology Telescope]{Hil21}, two millimeter surveys that have a large overlap with DES-Y6.

This paper is organized as follows: section~\ref{sec:DESdata} describes the construction of the input DES galaxy catalog and the systematic maps; Section~\ref{sec:wazp} presents the \wazp\ catalog, evaluates its redshift performance with a sub-sample containing spectroscopic information, and performs the comparison with the DES-Y1 and external optical catalogs to assess the robustness of \wazp\ and enhancements of detections with the DES-Y6 data; Section~\ref{sec:compare} cross-matches \wazp\ and \sz\ clusters, evaluates their relative performances and studies the properties of the paired systems, in particular their relative centering and the mass-richness relation; and in Section 5 we present the conclusions of this work.

%%%%%%%%%%%%%%%%%%%%%
\section{Data}
\label{sec:DESdata}
%%%%%%%%%%%%%%%%%%%%%

%-------------------------------
\subsection{DES-Y6 Galaxy Catalog}
%-------------------------------

The DES is an optical/near-infrared imaging survey conducted \citep{Die16} using the Dark Energy Camera \citep{Fla15} mounted on the 4 m Blanco telescope at Cerro Tololo Inter-American Observatory in Chile that observed more than 500 million galaxies over a region of 5000 deg$^2$ in 5 bands ($g$, $r$, $i$, $z$, $Y$) \citep{DES05,Abb21,DESy6gold}. This study is based on the DES-Y6 COADD source catalog built by the DES science data release team for the six years of observations \citep{Abb21}, and on DES Y6 Gold source attributes \citep{DESy6gold}, including Single-Object Fitting (SOF) extinction corrected magnitudes. In addition to this catalog, a series of maps characterizing the survey performances designed to support cosmological analyses are also provided.

\begin{figure}
    \centering
    \includegraphicsX[scale=1]{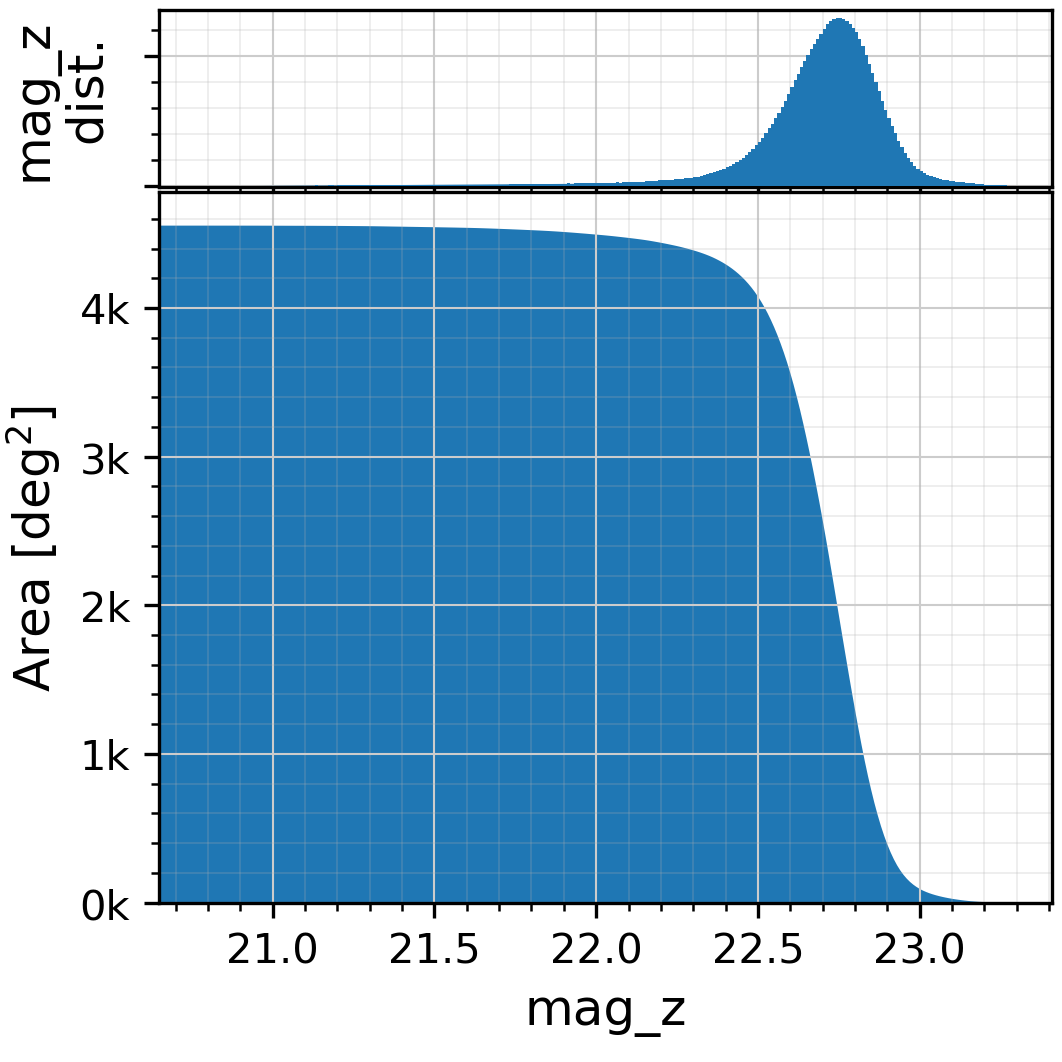}
    \caption{ Magnitude distribution (top panel) of the 10$\sigma$ DES-Y6 depth map on \zband, and the total area of the survey below said magnitude thresholds.}
    \label{fig:maglim}
\end{figure}

The galaxy catalog used in the present analysis was built using the DES Science Portal infrastructure described in \cite{Fau18}. Footprint maps are available in the HEALPix \citep{healpix} format (NSIDE=4096) where a coverage fraction of each pixel is provided. We considered only those with a coverage greater than 10\%. Then, regions contaminated by bright foreground sources (bright stars, globular clusters) or defects were discarded.  In addition, manual masking was required to enlarge masks around bright stars, stellar clusters, or around bright foreground galaxies. Finally, regions with high and irregular background levels were also manually removed by the authors of this work. These appeared mainly at the survey edges and in regions associated with galactic cirrus identified in the Planck All Sky Map at 857 GHz \citep{planck_HFI} (see Appendix~\ref{app:dust}). At these locations, concentrations of false positive galaxies or galaxies with poorer photometry could lead to false cluster detections. The resulting effective area of the final catalog is 4,545\,deg$^2$, corresponding to $\sim 90$\% of the total DES survey area. Fig.~\ref{fig:maglim} shows the final distribution of the survey area as a function of the 10$\sigma$ \zband\ limiting magnitude, which is our reference band in this work (see below). We point out that 98\% of the unmasked areas of the survey is very homogeneous, with a magnitude limit ranging between magnitudes 22.5 and 23.0 (see sections~\ref{sec:refband} and \ref{sec:homog}).

Considering the sources in the selected survey regions as described above, extreme color outliers ($g-r$, $r-i$, $i-z$ < -2 or >4) and high-confidence point sources were removed.  For the latter, we adopted the classification scheme provided by the DES data release team, based on the so-called EXTENDED\_FITVD parameter available within the internal release of DES-Y6 data \citep{DESy6gold,2022MNRAS.509.3547H}. This parameter is a small variation from the EXTENDED\_COADD described in the DES DR2 paper \citep{Abb21}. It combines the projected size of objects with information from the fit with models (bulge + disk) used for the photometry to estimate the object's "extendedness" better. Finally, a magnitude-limited sample was created from the selected region by removing objects with \zband\ mag > 23.4, corresponding to the peak of the galaxy counts. This leads to an average density of 15.84 galaxies/arcmin$^2$.

%-------------------------------
\subsection{Photometric redshifts}
\label{sec:zp_stats}
%-------------------------------

\begin{figure}
    \centering
    \includegraphicsX{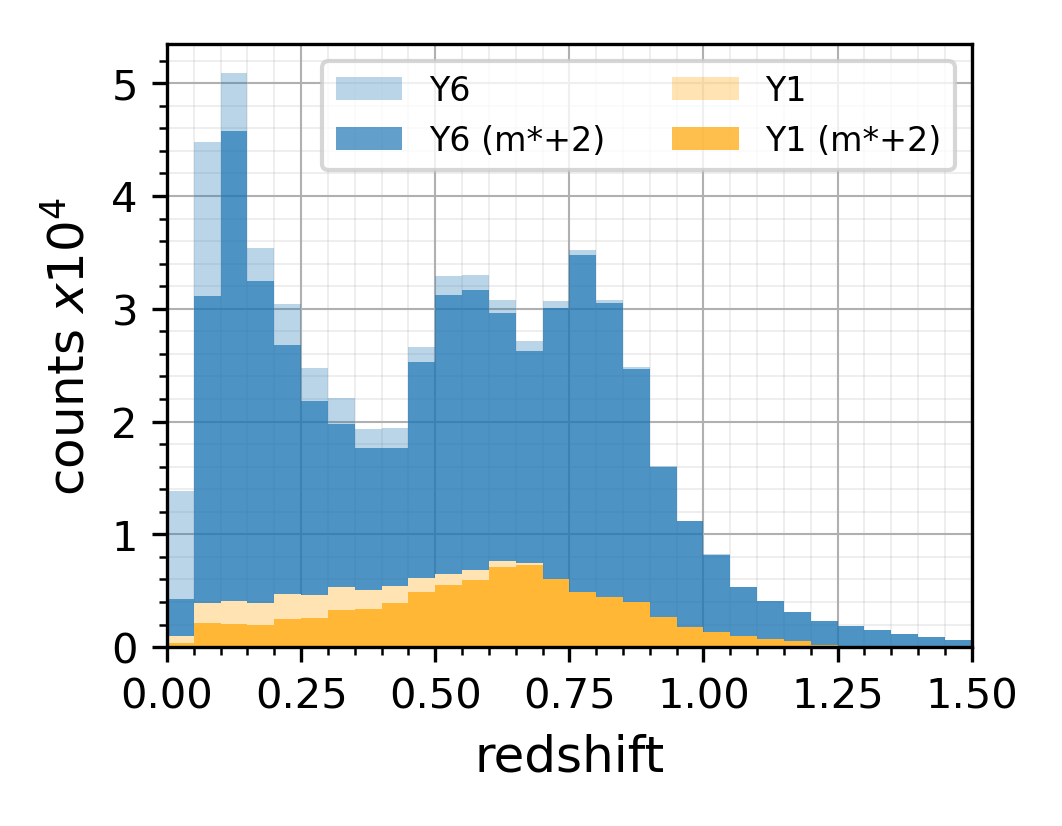}
    \caption{ Distribution of spectroscopic redshifts of galaxies used for training and validating DNF \phz s for this work (DES-Y6) and for Paper-1.  The complete sample used in the training and the $m^*(z)+2$ limited sample used by \wazp\ for cluster detection (solid areas) are shown as transparent and opaque histograms, respectively.}
    \label{fig:n_of_zspec}
\end{figure}

The photometric redshift estimates were based on the \textsc{DNF} algorithm (Directional Neighborhood Fitting, \citealt{deV16}), as it was done in our previous work based on DES-Y1 data (Paper-1).  DNF is a machine learning method that seeks the nearest neighbors in the hyperspace of observables using the combination of angular and Euclidean distances.  The DNF training was performed using an updated version of the spectroscopic redshift compilation described in \citet{GSC18}. The current spectro-photometric catalog contains 592,269 galaxies with good-quality redshift measurements successfully matched with DES-Y6 photometry (after quality cuts in DES data), ranging up to $z=2.0$ (Fig.~\ref{fig:n_of_zspec}). The recent addition of public \spz\ data from SDSS DR16 \citep{SDSSdr16} has made a major contribution, especially at lower redshifts. This sample is responsible for more than 50\% of \spz s successfully matched to DES objects in the DES-Y6 data.  In this work, Single-Object Fitting (SOF) magnitudes were used, as opposed to AUTO magnitudes used in Paper-1. The catalog was randomly divided into half for training and validation purposes.

\begin{figure}
    \centering
    \includegraphicsX{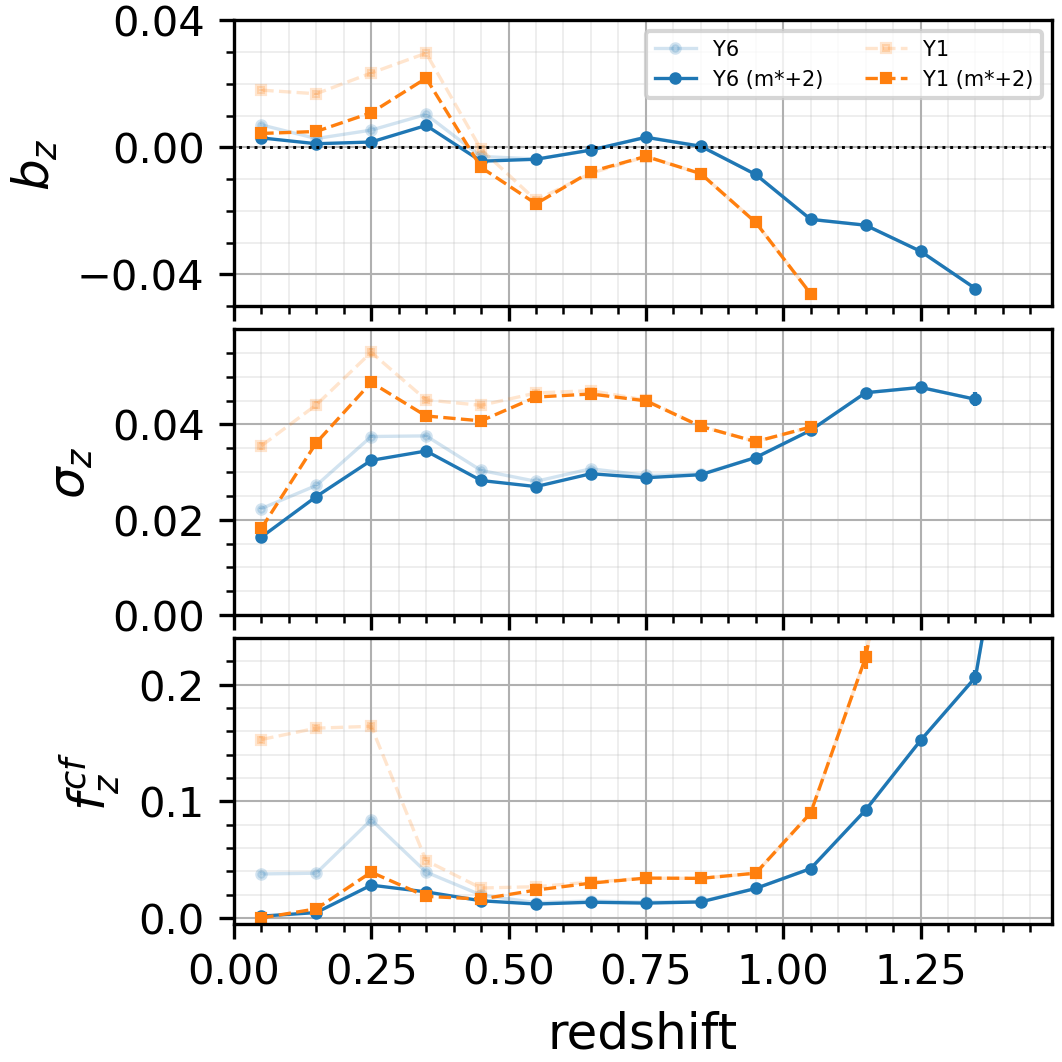}
    \caption{Comparison of \phz\ bias, scatter, and fraction of catastrophic failures as a function of \spz\ obtained with DES-Y6 (this work) and DES-Y1 (Paper-1) data. Metrics were computed using the complete samples (transparent lines) and samples brighter than $m^*(z)+2$  (full lines).}
    \label{fig:pz-metrics-per-z}
\end{figure}

The \phz\ quality is evaluated in terms of scatter $\sigma_z = std(\delta_z)$, bias $b_z =\left<\delta_z\right>$ and fraction of catastrophic failures $f^{cf}_z=\rm{frac}(|\delta_z|>0.15)$, where $\delta_z = (\zph-\zsp)/(1+\zsp)$.  The scatter and bias are computed, excluding objects with catastrophic failures.  These traditional metrics are compatible with the quality metrics threshold for dark energy studies defined as science requirements in the DES white paper \citep{DES05} and in the science verification paper \citep{San14}.  Compared to Paper-1, the global metrics in the range $0<z<1$ are significantly improved and the maximum available redshift increased, reaching $z\sim1.3$. In the common redshift range, the bias decreased from 0.7\% to 0.3\%, the scatter from 4.9\% to 3.2\%, and the fraction of catastrophic failures decreased from 6.8\% to 3.4\%.  The gain in \phz\ quality is substantial for particular redshift and magnitude ranges. The comparison of the metrics as a function of redshift in Fig.~\ref{fig:pz-metrics-per-z} shows that both metrics are significantly reduced at lower redshifts ($z \lesssim 0.4$), where a substantial increase of redshifts became available for training DNF (Fig.~\ref{fig:n_of_zspec}).  The bias is reduced by 60\% on average in the redshift range covered by the DES-Y1 sample.  The tail of the distribution in Fig.~\ref{fig:n_of_zspec} also shows a significant improvement in sampling the higher redshift regime ($z  \gtrsim 0.9$). If we now restrict the metrics to the brighter galaxies, the objects that are effectively used for detection in the various redshift slices (see the next section), the bias, scatter, and catastrophic failures decrease even more at low redshift. In that regime, the largest measured bias decreases to $\sim$ 1\% at the redshift of 0.35 in the case of DES-Y6. This improvement has a strong impact on the global shape of cluster counts, as discussed in section~\ref{sec:wazp}.

In conclusion, the new data set allowed us to extend the analysis to higher redshifts ($z<1.3$) than in Paper-1 ($z<1.1$) and significantly improve our results at low redshifts.  The higher DES-Y6 \phz\ quality relative to the DES-Y1 data set can be attributed to gains in both photometric and spectroscopic data. The former is due to a higher signal-to-noise in comparison to DES-Y1. The latter is mainly due to a much larger number of available \spz\ galaxies (approximately six times larger) and especially to the better coverage of magnitude-color-redshift space, leading to a training set representing much better the galaxy distribution being analyzed.

%-------------------------------
\subsection{Choice of the reference band}
\label{sec:refband}
%-------------------------------

In Paper-1, the reference band of our galaxy catalog for detecting clusters was the $i$-band. This choice was motivated by the more homogeneous coverage of that band compared to the $z$-band in the DES-Y1 data release. Here we reconsider this choice, in particular with regard to our interest in detecting clusters at higher redshifts.  To do this, we construct $z_{max}$ maps that correspond to the highest redshift where we can reliably measure the cluster richness, given the local magnitude limit of the survey. To estimate cluster richness, we essentially count galaxies down to a limiting magnitude $m^*(z_{cl}) + \delta mag$ where $m^*$ is the characteristic magnitude marking the knee of the luminosity function (see Paper-1) and $\delta mag$ is a fixed quantity, chosen here to be equal to 2. Therefore, $z_{max}$, the maximum redshift that can be reached,  satisfies the relation $m_{lim} = m^*(z_{max}) + \delta mag$, where $m_{lim}$ is the magnitude limit determined from the $10\sigma$ depth maps.

\begin{figure}
    \centering
    \includegraphicsX[scale=1]{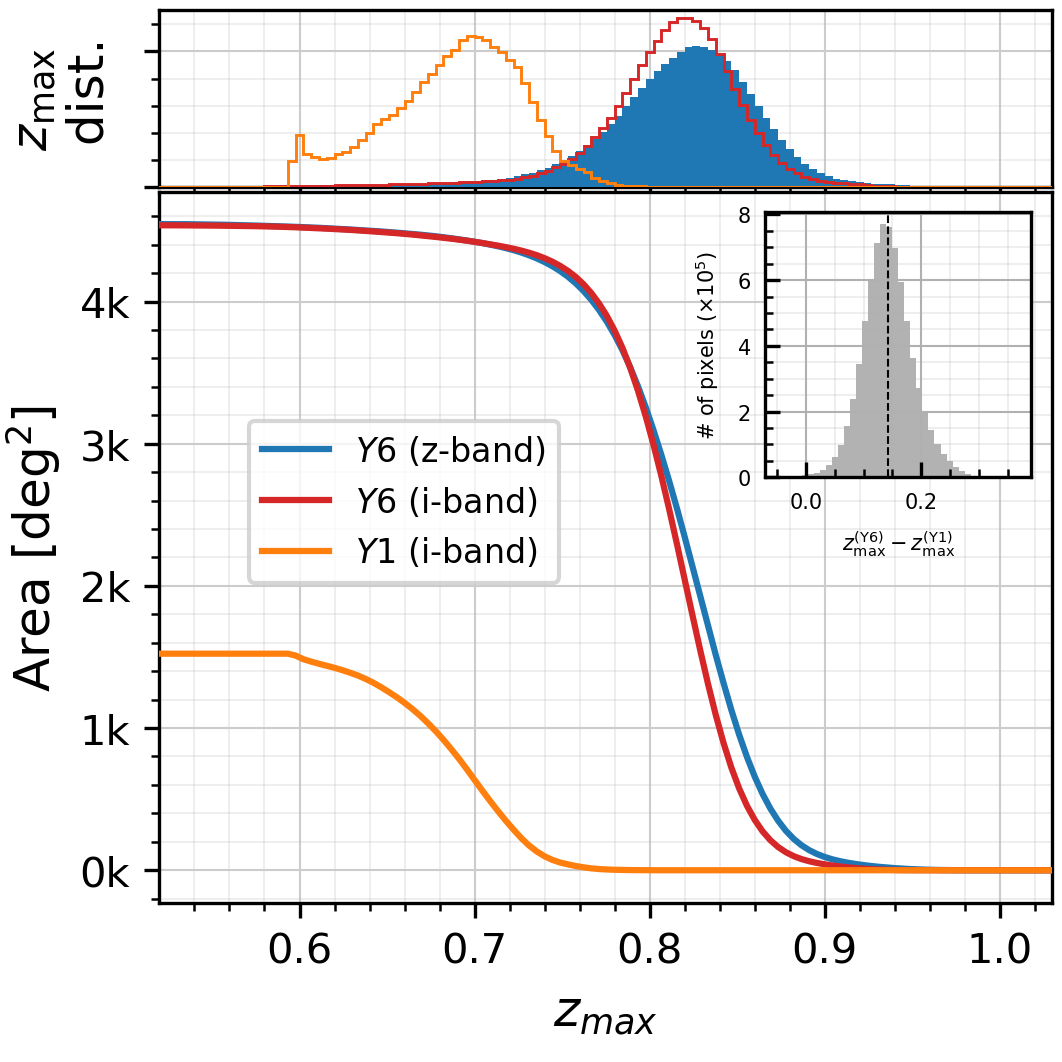}
    \caption{ Distribution of the values of the $z_{max}$ map of the DES-Y6 data release (top panel), that defines the maximum redshift where \wazp\ clusters are detected with complete richness, and the total area of the survey under said map (bottom panel).  The small panel displays the difference between the $z_{max}$ values of the pixels common to the DES-Y1 ($i$-band) and DES-Y6 ($z$-band) maps.}
    \label{fig:zmax}
\end{figure}

We compute these maps in the case of the $i-$ and $z-$ band magnitudes and compare them with Paper-1 results. Fig.~\ref{fig:zmax} shows the area of the survey that corresponds to a given redshift limit for cluster detection. The $z_{max}$ maps derived from the magnitudes of the DES-Y6 $i-$ and $z-$ bands are similar, covering approximately the same area at a given redshift, with a slight advantage in the case of the $z-$band, reaching redshifts larger by 0.1 in roughly half of the survey. Both cases show a clear gain when compared with the DES-Y1 data. Evaluating the common region of the two releases in a pixel-by-pixel comparison (smaller panel), there is an average increase of $z_{max} $ of $\sim 0.14$ and regions with an increase of up to 0.3. Therefore, considering the above discussion, we hereafter adopt the $z-$band as our reference band.

%-------------------------------
\subsection{Spatial homogeneity of photometric redshifts}
\label{sec:homog}
%-------------------------------

The bias and scatter of the photometric redshifts presented in section~\ref{sec:zp_stats} are evaluated by comparison with available spectroscopic redshift samples. However, these are concentrated in specific regions of the DES footprint. In order to evaluate the quality of photometric redshifts over the whole survey, one could use the knowledge of the depth maps in all bands entering the photometric redshift estimate. This rather complex approach may be limited again by the spectroscopic data that do not uniformly sample the wide diversity of depths in the survey. An alternative approach adopted here is to make use of $\zph^{\rm err}$, the estimated 1-$\sigma$ DNF errors.  To assess the significance of these photometric redshift errors, we compute the reduced $\chi^2$, $\chi_r^2 = \sum(\zph-\zsp)^2/(\zph^{\rm err})^2/N_{\rm dof}$, expected to be $\sim 1$. This is indeed the case with $\chi_r^2 = 1.02$ when we restrict to $\zph \geq 0.25$ and $\zph-\zsp$ within $\pm 3\sigma_{dz}$, with $\sigma_{dz}$ estimated in section~\ref{sec:zp_stats}.  Hence, $\zph^{\rm err}$ will be used to address the spatial homogeneity of the photometric redshift scatter as a function of location within the survey.

\begin{figure*}%[H]
    \centering
    \includegraphicsX[scale=1]{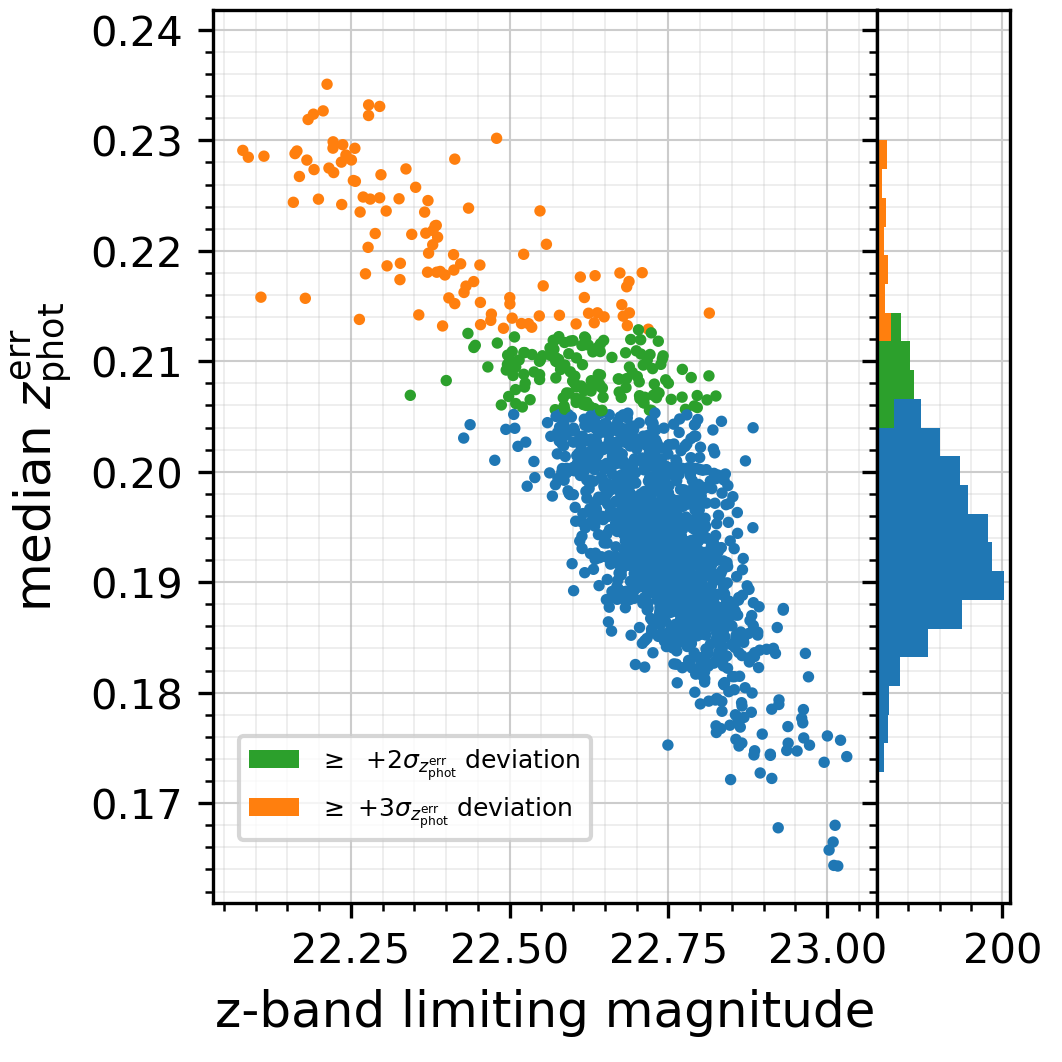}
    \includegraphicsX[scale=1]{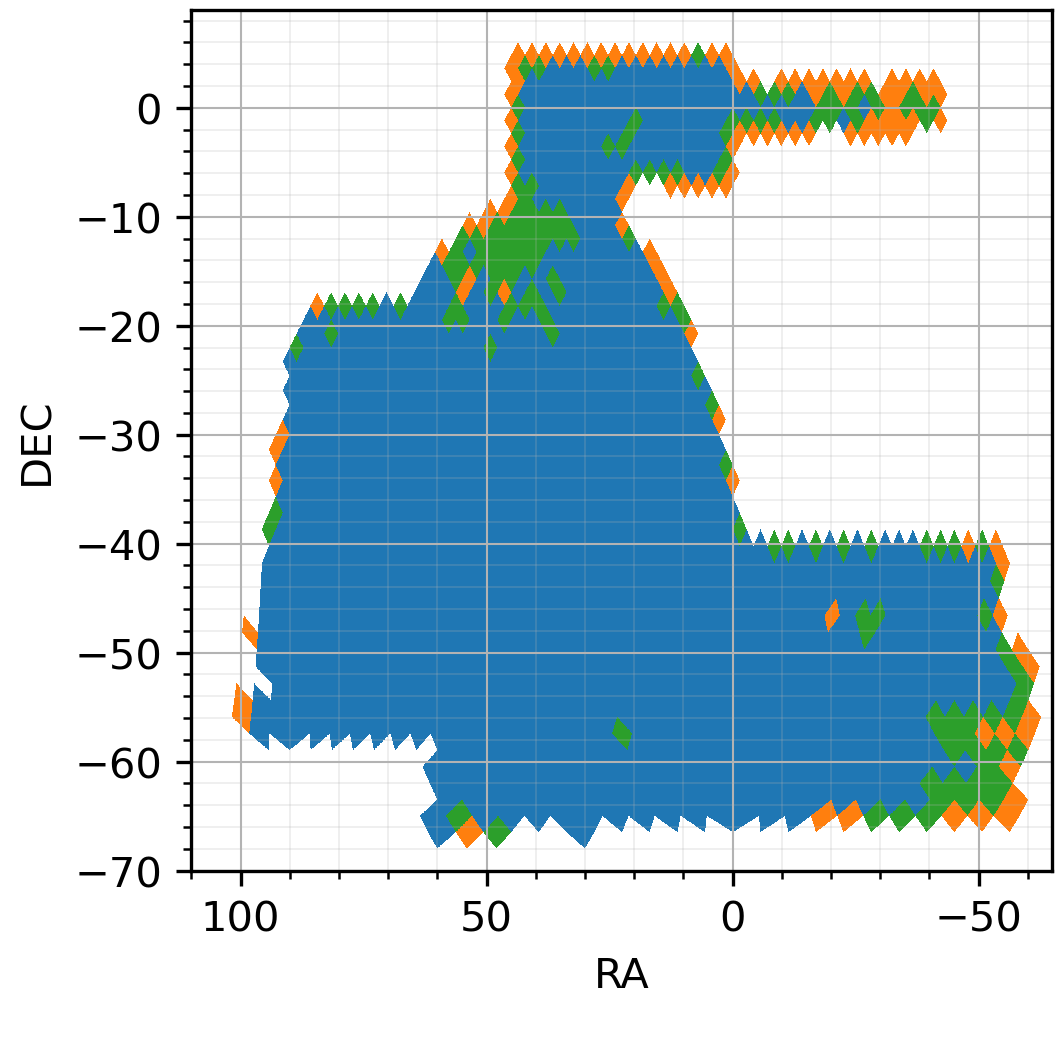}
    \caption{ Distribution of the $\zph$ error computed in cells of $\sim$ 3 deg$^2$ across the DES footprint and its relation to  the local z-band depth.  In green and orange are the cells deviating by more than 2- and 3-$\sigma$, respectively, from the median value of the survey.  The left panel shows that, in DES, the z-band depth captures well the global quality of the photometric redshifts.  The right panel shows that most of the variation of the $\zph$ errors occurs on the edges of the survey.}
    \label{fig:pdz_err_radec}
\end{figure*}

The left panel of Fig.~\ref{fig:pdz_err_radec} displays the median $\zph^{\rm err}$ computed in cells of $\sim 3$\,deg$^2$ covering the entire DES footprint.  As seen in the right histogram, it peaks at $<\zph^{\rm err}> = 0.19$, with a standard deviation $\sim 0.007$. This spatial variation is 7 times smaller than the average scatter per cell. This shows an excellent overall homogeneity of the photometric redshift properties in the survey. However, one can also notice a tail at high median $\zph^{\rm err}$. Cells deviating by more than 2- and 3-$\sigma$ are flagged and displayed in the R.A.-Dec. plane (Fig.~\ref{fig:pdz_err_radec}, right panel). Cells where photometric redshifts show the largest errors are clearly located at the edges of the survey, where 5-band coverage is not always guaranteed and where the depth is also lower. This corresponds to 0.5\% of the total survey area. Cells deviating by 2-$\sigma$ correspond to edges and also to regions affected by stronger extinction, as can be seen from the Planck All Sky Map at 857 GHz (Appendix \ref{app:dust}). These regions correspond to $\sim 3$\% of the survey area.

As can be seen in the left panel of Fig.~\ref{fig:pdz_err_radec}, the regions with noisier photometric redshifts also correspond to shallower z-band regions in the survey. As explained in the previous section, the z-band depth map is used to assign to each cluster detection a local maximum redshift for reliable characterization (e.g. a homogeneous richness). This quantity is provided in the \wazp\ catalog and can be used to identify potential problematic detections.

%%%%%%%%%%%%%%%%%%%%%
\section{The DES-Y6 \wazp\ Cluster Catalog}
\label{sec:wazp}
%%%%%%%%%%%%%%%%%%%%%

The Wavelet Z-Photometric (\wazp) cluster finder is designed for the optical detection of galaxy clusters from multi-wavelength photometric imaging galaxy surveys. It is based on the search for projected overdensities of galaxies in photometric redshift space and does not make any assumption on the existence of a red sequence. It relies only on weak assumptions on cluster radial profiles (maximum radius) and luminosity functions (redshift dependent faintest magnitude).  A full description of the \wazp\ detection method is presented in Paper-1, but we provide here a summary of the main steps.

Overdensity peaks are initially extracted from pixelized wavelet-based density maps constructed from the galaxy catalog in various photometric redshift slices. In each slice, galaxies are weighted based on their redshift PDF (here Gaussian).  Then, the peaks of successive slices are merged and the peaks with the largest SNR enter the final list of clusters. The accuracy of the cluster position on the sky is limited by the resolution of the wavelet maps chosen here to be 1/16\,Mpc.

Then, following the prescription of \cite{2016A&A...595A.111C}, cluster membership probabilities are computed. They depend on galaxy cluster-centric distances, magnitudes, and photometric redshifts.  Finally, the cluster members are used to compute the cluster richness ($\rich$), defined as the sum of membership probabilities of members brighter than ${\rm m}^{*}(z_{\rm cluster}) + \delta {\rm mag}$. This is a similar cut to the one evaluated for cluster detection (see Sec. \ref{sec:refband}), but with a more conservative value of 1.5 for $\delta {\rm mag}$.

\begin{figure}%[H]
    \includegraphicsX[scale=0.7]{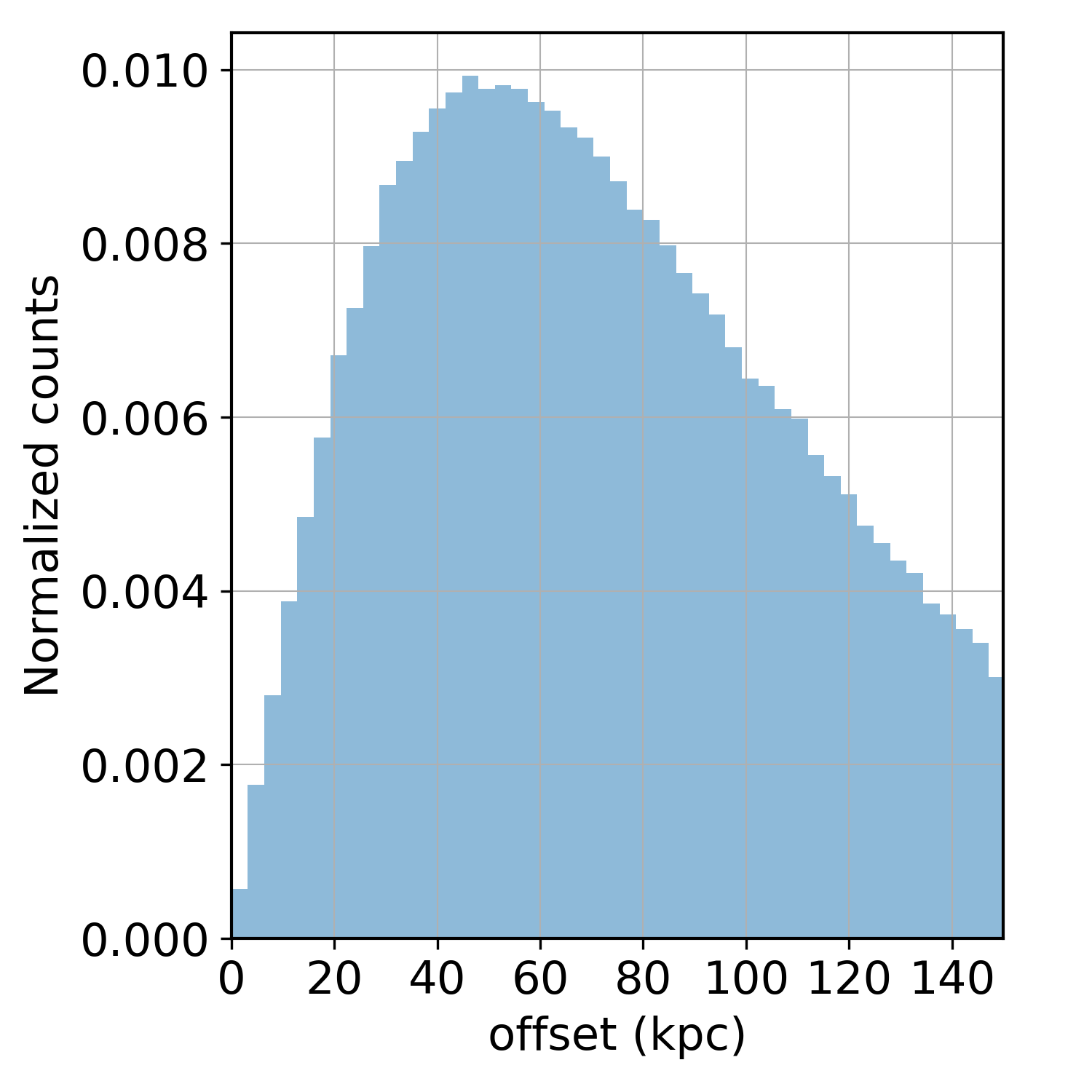}
    \caption{Distribution of the offset between the nominal \wazp\ center derived from smooth density maps and the associated brightest cluster member when it is closer than 150\,kpc from the \wazp\ center, i.e.\,$\sim 3$\,pixels in the \wazp\ density maps.}
    \label{fig:dist_centroid}
\end{figure}

As the centering of the \wazp\ algorithm is based on the identification of peaks in density maps, its precision is limited by the adopted pixelization scheme. By construction, in this work, the maps have a resolution of 1/16\,Mpc.  To estimate the induced detection beam or centering accuracy, we associate each detection with its brightest cluster member, which is expected to mark the center of the gravitational potential for relaxed clusters.  The distribution of the offsets relative to the associated density peaks is shown in Fig.~\ref{fig:dist_centroid}. It reaches a maximum for an offset of 45\,kpc, corresponding to one pixel diagonal length. This shows that most of the detected clusters have their brightest cluster member consistent with the peak of the local density map.

When the brightest cluster member lies within 150\,kpc ($\sim 3$ times the \wazp\ beam) from the nominal cluster center, we use it as a central galaxy.  Following this procedure, 67\% of \wazp\ clusters are assigned a central galaxy. These associated central galaxies are used throughout this work to associate a spectroscopic redshift to the clusters (section~\ref{sec:wazp_cat}). %and to evaluate the relative optical-SZE centering (section \ref{sec:sz_centering}).

%-------------------------------
\subsection{The "full" and "cosmology" cluster samples}
\label{sec:wazp_cat}
%-------------------------------

Following Paper-1, we analyze the DES-Y6 data using the \wazp\ cluster finder.  Two cluster samples are constructed: a "full sample" for which we include all detections regardless of the variations in depth over the survey (i.e. the cluster search over the entire survey extends to redshifts corresponding to the deepest region of the survey), and a "cosmology sample" where only clusters with richnesses not suffering from galaxy incompleteness are kept (as derived from the $z_{max}$ maps presented in
section~\ref{sec:refband}).

The \wazp\ detection based on the $z-$band leads to a {\it full} catalog of 416,947 clusters with richness $\rich \ge 5$ and a maximum redshift $z=1.32$, yielding a surface density of 91.74 clusters/deg$^2$.  The ten clusters with the highest signal-to-noise ratios are presented in Table~\ref{table:wazp_sample}.  If we restrict the catalog to higher richness ($\rich \ge 25$) we find 42,332 clusters and a density of 9.31 clusters/deg$^2$. We adopt this richness cut to define a cleaner sample to be used for the comparison of the properties of \wazp\ clusters with other catalogs. This value was adopted because it corresponded to \RM\ richness $\lambda = 20$ in Paper-1, and, as will be shown in section~\ref{sec:comp_y1}, the \wazp\ richness of the DES-Y6 and DES-Y1 \wazp\ clusters are compatible with each other.  Applying the cuts to obtain the cosmology catalog results in a catalog with 301k and 33k clusters for $\rich\geq5$ and $25$, respectively.

\setlength{\tabcolsep}{5pt}
\begin{table*}[h!]
\centering
\begin{tabular}{ccccccccc}
NAME & RA & DEC & REDSHIFT & REDSHIFT & SNR & $\rich$  & NMEM & RADIUS\\
(CLJHHMM.m$\pm$DDMM.m) & (degrees) & (degrees) &&(spec.)&&&&(arcmin) \\
\hline
\hline
CLJ023952.7-013418.9 &   \ns 39.9697 &  \ns-1.5719 & 0.391 $\pm$ 0.011 & 0.377 (\Dv)  & 31.6 &    271.8 $\pm$    13.4 & 621 & \ns 7.2 \\
CLJ225420.5-462048.7 &      343.5856 &    -46.3469 & 0.275 $\pm$ 0.013 &            - & 29.0 &    200.7 $\pm$    12.2 & 365 & \ns 8.9 \\
CLJ001420.5-302349.4 & \ns\ns 3.5856 &    -30.3970 & 0.291 $\pm$ 0.013 &            - & 26.5 &    264.2 $\pm$    14.4 & 442 & \ns 8.3 \\
CLJ000545.1-375116.8 & \ns\ns 1.4380 &    -37.8547 & 0.496 $\pm$ 0.012 &            - & 26.2 &    186.0 $\pm$    11.4 & 433 & \ns 6.4 \\
CLJ051258.1-384725.4 &   \ns 78.2421 &    -38.7904 & 0.325 $\pm$ 0.013 &            - & 25.9 &    203.3 $\pm$    11.8 & 425 & \ns 8.1 \\
CLJ032728.3-132622.6 &   \ns 51.8680 &    -13.4396 & 0.587 $\pm$ 0.015 &            - & 25.6 &    163.3 $\pm$    10.6 & 382 & \ns 5.9 \\
CLJ020309.4-201706.1 &   \ns 30.7892 &    -20.2850 & 0.449 $\pm$ 0.011 &            - & 25.2 &    201.2 $\pm$    11.7 & 483 & \ns 6.8 \\
CLJ023216.6-442057.3 &   \ns 38.0690 &    -44.3493 & 0.265 $\pm$ 0.014 &            - & 24.9 &    257.9 $\pm$    13.5 & 464 & \ns 8.2 \\
CLJ013152.5-133640.5 &   \ns 22.9689 &    -13.6113 & 0.214 $\pm$ 0.013 & 0.210 (\Dv)  & 24.7 &    232.5 $\pm$    13.3 & 358 &    10.9 \\
CLJ014151.0-141836.2 &   \ns 25.4624 &    -14.3101 & 0.253 $\pm$ 0.017 &            -  & 24.5 & \ns 92.5 $\pm$ \ns 8.0 & 200 & \ns 9.5 \\
\end{tabular}
\caption{
    \wazp\ detections with the highest signal-to-noise ratio. All clusters on this table are part of the "cosmology sample".
    The \wazp\ DES-Y6 cluster products with a full description of the catalogs and columns can be found are available in its online
    supplementary material (see Data Availability section).
}
\label{table:wazp_sample}
\end{table*}

\begin{figure}%[H]
    \includegraphicsX[scale=1]{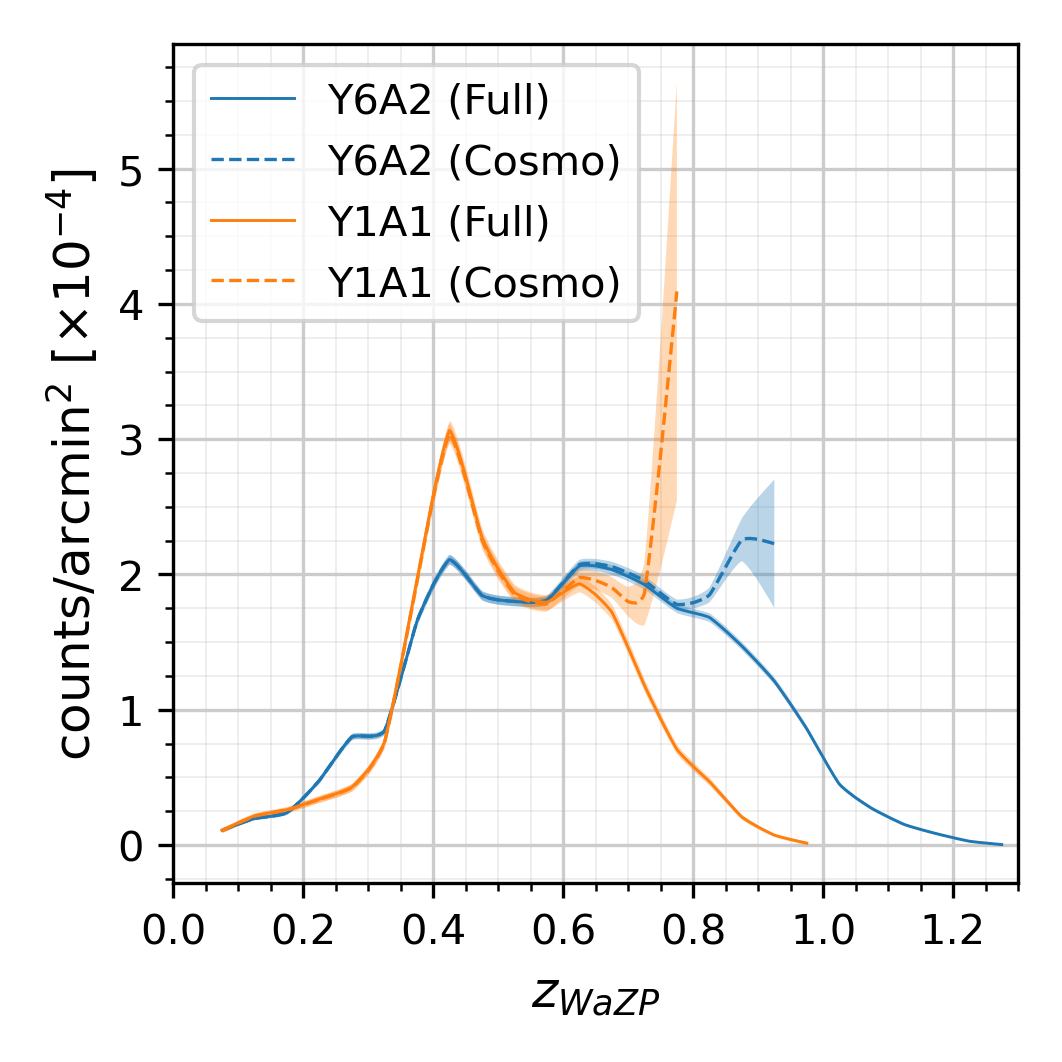}
    \caption{Cluster density per arcmin$^2$ of rich ($\rich \ge 25$) \wazp\ detections
    per area detected of \wazp\ detections on DES-Y1 and DES-Y6 data, considering the full and the cosmology (cluster $z$ < $z_{max}$ map) samples.}
    \label{fig:wazp_zr}
\end{figure}

In Fig.~\ref{fig:wazp_zr} we show the variation of the surface density of clusters as a function of redshift for the full and cosmology catalogs, for DES-Y1 and DES-Y6 data releases. As can be seen, the DES-Y6 full catalog is much deeper than DES-Y1. In addition, the strong density peak at $z\sim0.4$ for DES-Y1 appears to be artificial and a consequence of the much stronger photometric redshift bias relative to DES-Y6 data (section~\ref{sec:zp_stats}). The positive photometric redshift bias of Y1 at $z\sim 0.25-0.35$ indeed tends to concentrate galaxies (and clusters) located in the spectroscopic redshift range 0.3-0.4 at photometric redshift $\sim0.4$.  Starting at redshifts $z\approx0.8$ and above, an increasing fraction of the clusters have incomplete richnesses, and we see that the cluster density of the cosmology catalog becomes higher than the cluster density of the full catalog.

\begin{figure}%[H]
    \includegraphicsX[scale=1]{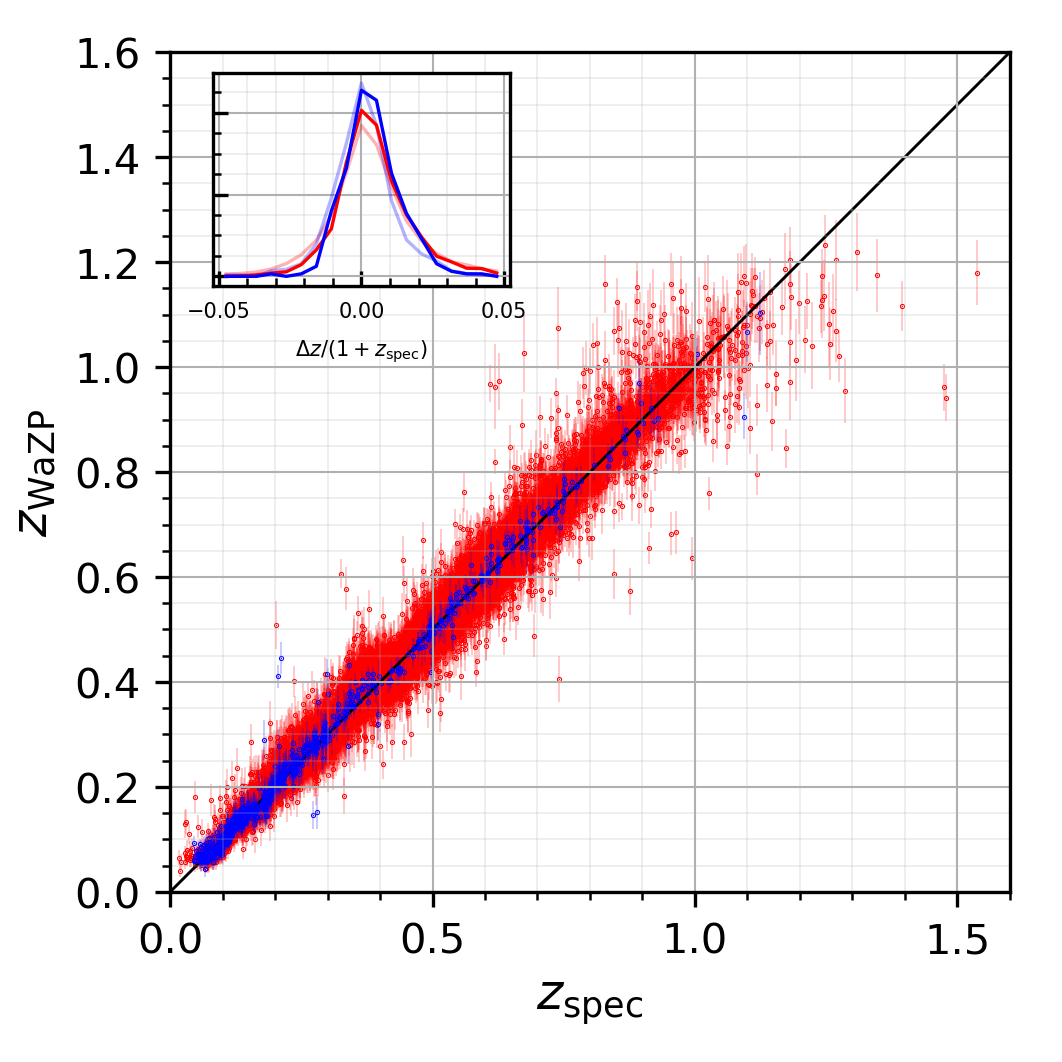}
    \caption{ Comparison of cluster redshifts derived by \wazp\ with spectroscopic redshift estimates (see section~\ref{sec:wazp_cat}). Clusters whose \spz s were estimated using \zdv\ (877 clusters) are represented by blue points and those estimated by \zbcg\ (23,850 clusters) are represented by red points. The smaller panel shows the total distribution of redshift offsets in each sample. Clusters with $\rich>25$ are represented by solid lines in the small panel.}
    \label{fig:wazp_zspec}
\end{figure}

To evaluate the quality of the cluster redshifts, we adopt the procedure of Paper-1, using cluster members that had a cross-match with galaxies with available spectroscopic data. Spectroscopic redshifts were first assigned to clusters by requiring at least 5 members within a 0.5\,Mpc projected radius and a line-of-sight velocity dispersion of $\Delta v$<2000 km/s from each other (these will be referred to as \zdv\ henceforth).  Then, to obtain a larger sample, the spectroscopic redshifts of central galaxies (see section~\ref{sec:wazp}) were assigned to the remaining clusters whenever available (referred to as \zbcg). This process resulted in 877 \zdv\ clusters and an additional 23,570 \zbcg\ clusters.  In Fig.~\ref{fig:wazp_zspec}, we compare \wazp\ redshifts, which are estimated from its members \phz s, with these spectroscopically assigned redshifts.  We see in the smaller panel that the scatter of both \spz\ samples is similar, with the \zbcg\ sample having a larger tail in the redshift offset distribution. One can also notice a systematic underestimation of \wazp\ redshifts at redshifts beyond $z\sim1$, which reflects the negative photometric redshift bias described in section~\ref{sec:zp_stats} (Fig.~\ref{fig:pz-metrics-per-z}).

In Table \ref{table:wazp_z_stat}, we detail the redshift scatter and offsets by \spz\ type and richness cut, limiting the sample to $\zsp<0.9$.  The overall scatter is 1.8\% with an offset of 0.4\%, which is improved by 30\% and 50\% considering only the \zdv\ subsample.  When restricted to richer clusters, the scatter of the \zbcg\ subsample has a much larger improvement (22\%) than the scatter of the \zdv\ subsample (8\%).  These results show that \wazp\ redshift estimates are very reliable and that better photometry leads to a significant improvement of the photometric redshift compared to the DES-Y1 cluster catalog characterized by $(b_z, \sigma_z)=(1.4\%, 2.6\%)$.

\setlength{\tabcolsep}{5pt}
\begin{table}[h!]
\centering
\begin{tabular}{c | c c c | c c c}
\multirow{2}{2.4em}{\spz} & \multicolumn{3}{c|}{$\rich>5$}& \multicolumn{3}{c}{$\rich>25$}\\
& N & $b_z$ & $\sigma_z$ & N & $b_z$ & $\sigma_z$ \\
\hline
\hline
all   & 24,447 & 0.004 & 0.018 & 2,933 & 0.005 & 0.014
\\
\zdv\ &  877 & 0.002 & 0.013 &   314 & 0.004 & 0.012   
\\
\zbcg\ & 23,570 & 0.004 & 0.018 & 2,619 & 0.005 & 0.014

\end{tabular}
\caption{{Offset ($b_z$) and scatter ($\sigma_z$) for the \wazp\ catalog estimated with spectroscopically assigned redshifts.}
}
\label{table:wazp_z_stat}
\end{table}

In our sample, there are 546 clusters with both \zdv\ and \zbcg.  Based on this subset, we can evaluate the possible misassignment of \wazp\ central galaxy caused by larger offsets on their photometric redshifts.  Assuming \zdv\ to be the more accurate measurement of a cluster redshift, we find a very small fraction of potentially central galaxy misidentification, with only 3 clusters having (\zbcg-\zdv)/(1+\zdv)$\geq3\times0.018$.

%-------------------------------
\subsection{Improvements from DES-Y1 to DES-Y6}
\label{sec:comp_y1}
%-------------------------------

In this section, we compare the \wazp\ DES-Y1 (Paper-1) and DES-Y6 (this paper) catalogs to evaluate how well our earlier results are reproduced and the contribution of the new data to improve our results in terms of depth and redshift estimate. In order to do that, we cross-match the two catalogs considering both angular and redshift proximity ($\dt$ and $\dz$ respectively), following the 4-step method defined in section~5.2.2 of Paper-1.  In this method, we do a preliminary match using clusters closer than 50\,kpc and compute the scatter ($\sigma_z$) and 99\% quantiles ($\sigma_{99}$) of $\dz$ from these pairs to establish a redshift window. Then, the clusters are matched in the following steps 1) Clusters with $\dt<1$\,arcmin; 2) Clusters with $\drp<300$\,kpc and $\dz<\sigma_z$, where $\drp$ is the projected distance between cluster centers at the mean redshift; 3) Clusters with $\drp<300$\,kpc and $\dz<3\sigma_z$; 4) Clusters with $\drp<$ cluster radius and $\dz<\max(\sigma_z, \sigma_{99})$;

We used the entire cluster catalogs with $\rich>5$, restricted to a common footprint, {\it i.e.} keeping all common pixels independently from their weight. This matching resulted in 54,036 DES-Y6 clusters and 50,301 DES-Y1 clusters with possible counterparts, which includes the possibility of multiple matches. When requiring unique pairs with a two-way matching, we found 48,436 pairs.

\begin{figure}%[H]
    \includegraphicsX[scale=1]{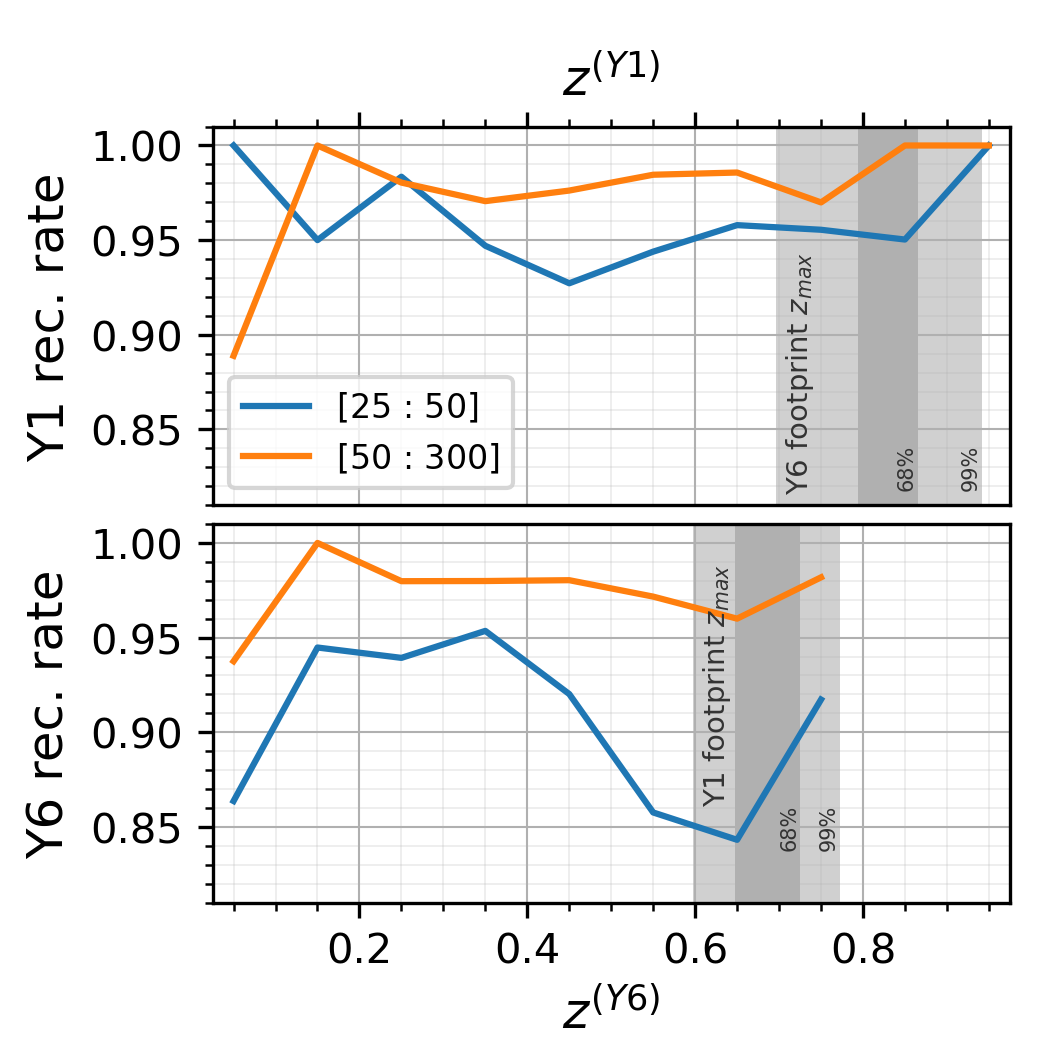} \caption{Recovery rate of \wazp\ catalogs among DES-Y1 and DES-Y6 data releases for different richness bins. Solid lines considers only clusters with high cover fraction and $z<z_{max}$ map, while dashed lines include all clusters in the shared footprint.}
\label{fig:wazpYImatch_cp} \end{figure}

To evaluate the relative recovery rate of the catalogs, we have only considered clusters within the common footprint and within a common $z_{max}$.  Initially, we evaluated only the clusters well covered by the other catalog footprint (solid lines in Fig.~\ref{fig:wazpYImatch_cp}), with a weighted coverage fraction above 80\% (see \citealt{Ryk12}).

With these considerations, we have found that 99\% of the DES-Y1 clusters are also detected on the DES-Y6 catalog, without dependence on redshift. Most of the unmatched systems are poor ($\rich<25$). However, we noticed that several rich $(\rich>50)$ DES-Y1 clusters were missing in the DES-Y6 catalog. A systematic visual inspection of these rich systems revealed that they correspond to bad regions not properly masked in the DES-Y1 catalog (mainly around bright stars). Most of them are at the periphery of bright star halos where large concentrations of false galaxy detections led to false positives of clusters.

Considering the other way around, rich clusters ($\rich>50$) of the DES-Y6 catalog were found to have a DES-Y1 counterpart 97\% of the time, while intermediate richness clusters ($25<\rich<50$) had counterparts 95\% of the time, with a progressive drop with redshift. This drop could be the result of a combined increased photometric redshift scatter and bias for DES-Y1 precisely in the redshift range of 0.45-0.65. The quantitative impact of the photometric redshift quality on the detection completeness is the subject of a separate paper based on numerical simulations.  These overall high recovery rates show the robustness of cluster detection across both data releases. They also show the impact of a lower photometric scatter that leads to a higher density of cluster detections, even at redshifts below $z_{max}$ (completeness redshift).

\begin{figure}%[H]
    \includegraphicsX[scale=1]{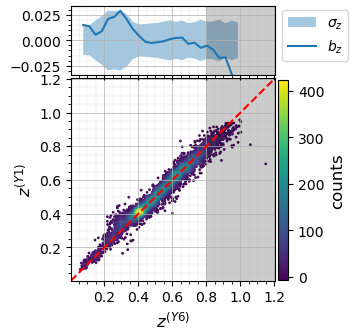}
    \caption{ Redshift comparison of \wazpYVI\ and \wazpYI\ matched clusters, and scatter and bias between them. The gray-shaded region corresponds to redshifts beyond the maximum value of the Y1 $\zmax$ map.}
    \label{fig:wazpYImatch_z}
\end{figure}

The redshift relation between matched clusters can be seen in Fig.~\ref{fig:wazpYImatch_z}, and is characterized by a bias and a scatter given as $(b_z, \sigma_z)=(0.006, 0.023)$. There is a peak in the redshift bias and scatter at $0.2<z<0.35$ $(b_z^{peak}, \sigma_z^{peak})=(0.027, 0.028)$, where DES-Y1 is known to be significantly more biased (as can be seen in Fig.~\ref{fig:pz-metrics-per-z}).

Common DES-Y1 and DES-Y6 \wazp\ clusters also showed very consistent centers, with 90\% of matched clusters having a center within 200\,kpc, and 40\% having the same center (within 1").  Essentially, all pairs with the same centering were composed of clusters with a central galaxy as their centers. The remaining pairs either had the peak of the wavelet map as the cluster center in both catalogs (23\% of cases), or a peak in one catalog and a central galaxy in the other (29\% of cases), or selected different galaxies as the central galaxy (8\%). We would like to remind the reader that the central galaxy assignment is merely a refinement of the \wazp\ centering, and the fact that the galaxy defined as central differs between Y1 and Y6 is affected more by different positions of the respective wavelet maps peaks than to a reclassification of the galaxy.

\begin{figure}%[H]
    \includegraphicsX[scale=1]{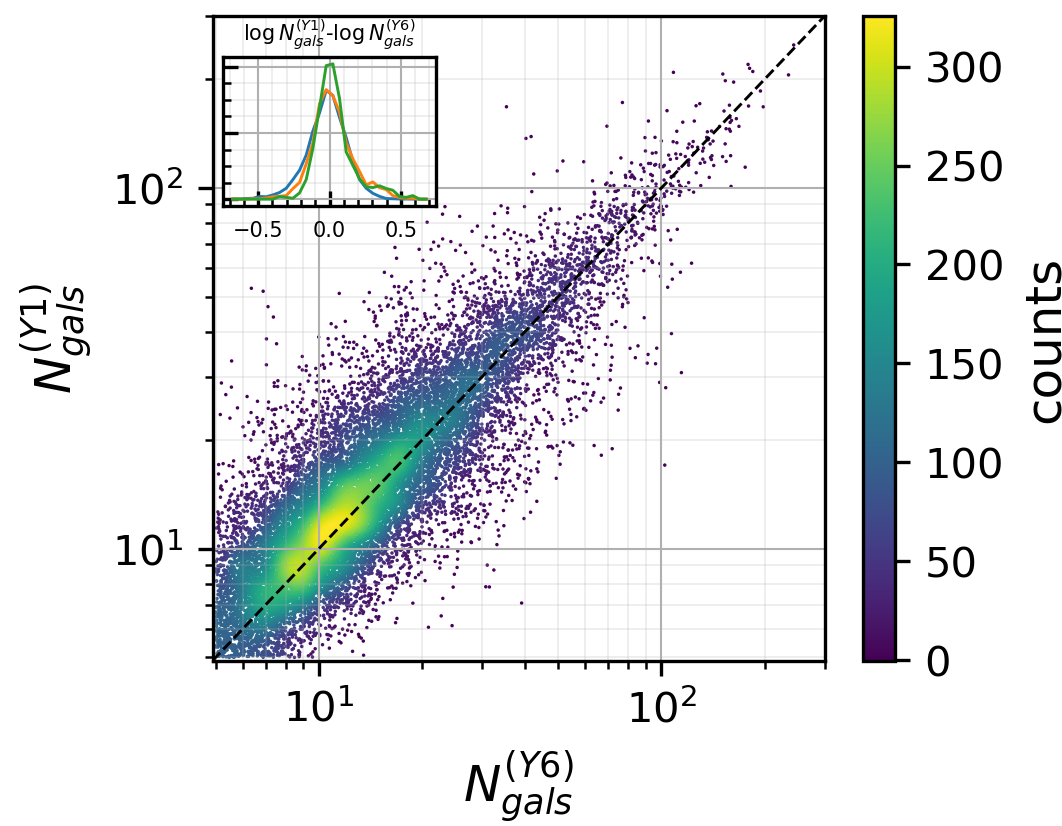}
    \caption{ Richness comparison of \wazpYVI\  and \wazpYI\ matched clusters with $\dt < 100\,$kpc and $\Delta z<0.023$.  The smaller panel shows the distribution of the difference between richnesses, with the colors (blue, orange, green) corresponding to the $\richVI$ bins: [10, 25, 50, 300].}
    \label{fig:wazpYImatch_m}
\end{figure}

To compare the richnesses assigned to the clusters by each DES data release, we restrict the sample to clusters with very similar centering (projected offset below 100\,kpc) and redshift ($\Delta z/(1+z)<0.023$).  We also restrict to clusters in the redshift range $0.1<z<0.6$, where all clusters are below the Y1 $z_{max}$ map. The resulting 18,130 common clusters are shown in Fig.~\ref{fig:wazpYImatch_m}. The smaller panel displays the distribution of $\Delta\log \rich=\log \richI-\log \richVI$ in three $\richVI$ bins: [10, 25, 50, 300] for illustration purposes. Using a thinner binning, we measure a constant scatter of $\sigma_{\Delta\log\rich}=0.13$ up to $\richVI=100$, after which it rapidly decreases to 0.02 at $\richVI=200$.  The offset in $\Delta\log\rich$ is $\sim 0.02$, with variation in different richnesses much smaller than $\sigma_{\Delta\log\rich}$.  No redshift dependence was found on the comparison of these richnesses.

Given that different types of magnitude (MAG\_AUTO in DES-Y1 and SOF in DES-Y6) and different bands (i-band for DES-Y1 and z-band for DES-Y6) were used, the richnesses of both catalogs show very good consistency. The careful inspection of very deviant outliers reveals that, in general, one of the runs identifies the galaxy overdensity as a large system while the other splits it into two parts.

\noindent

\newcommand{\clumpr}{\texttt{CLuMPR}} \newcommand{\wh}{\texttt{WH24}}
%-------------------------------
\subsection{Comparison to external optical cluster catalogs}
\label{sec:comp_optical}
%-------------------------------

Several optical cluster catalogs have considerable overlap with the DES footprint. In this section, we compare the counterparts between the largest catalogs and \wazp. We consider the recent \clumpr\ \citep{2024MNRAS.531.2285Y} and \wh\ \citep{2024ApJS..272...39W} catalogs, both of which have a full overlap with the \wazp\ footprint.

The matching between catalogs is done using an angular window of 5\,arcmin and a maximum redshift window of $0.05(1+z)$. Only clusters completely covered by the footprint of the counterpart catalog (within a $1$\,Mpc aperture) are considered in the recovery rates computations.

\begin{figure*}%[H]
    \includegraphicsX[scale=1]{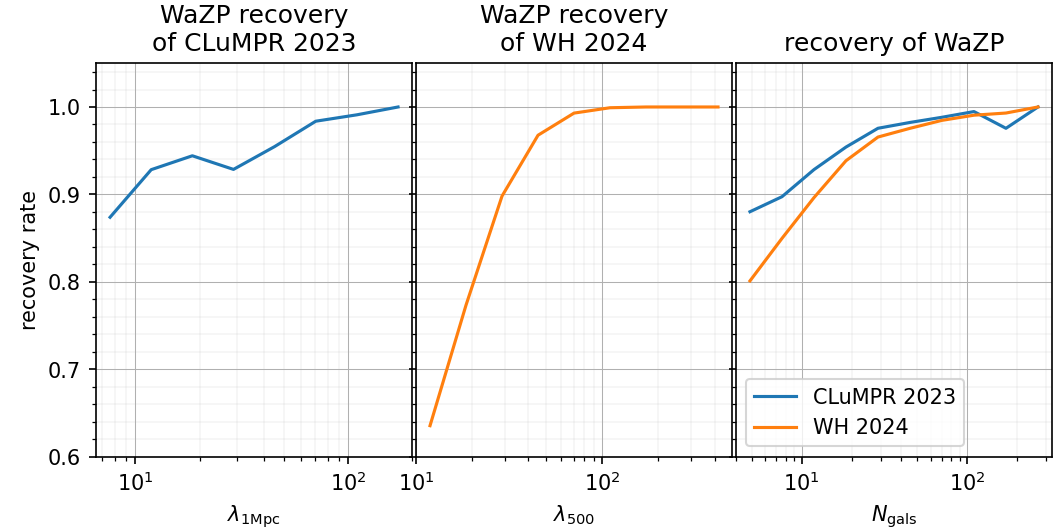}
    \caption{The recovery rates between \wazp\ clusters and other optical cluster catalogs as a function of \wazp\ richness.}
    \label{fig:wazpOPTmatch}
\end{figure*}

In Fig.~\ref{fig:wazpOPTmatch}, we display the recovery between \wazp\ and the other optical catalogs in the $0.2<z<1.0$ range as a function of richness. \wazp\ recovers over 90\% of \clumpr\ clusters with richness $\lambda_{\rm 1 Mpc}>9$ and \wh\ clusters with richness  $\lambda_{\rm 500}>30$. Alternatively, we also note that a high fraction of \wazp\ clusters are found in the other catalogs, with a recovery fraction above 95\% for $\rich>25$.

These comparisons show the robustness of \wazp\ detections, especially at the high richness end. However, we want to highlight that this evaluation excluded clusters with poor coverage in the common footprint between \wazp\ and each of the optical cluster catalogs. When we investigate regions not well covered by external catalogs, we find a large number of rich ($\rich>25$) clusters uniquely detected by \wazp. There are $\sim 13$k \wazp\ detections that are not present in \clumpr\ and $\sim 5$k not present in \wh.

%%%%%%%%%%%%%%%%%%%%%
\section{Optical vs. \sz\ cluster comparison}
\label{sec:compare}
%%%%%%%%%%%%%%%%%%%%%

In this section, we compare the \wazp\ DES-Y6 cluster catalog described above with the cluster catalogs derived from millimeter observations using the \sz.  Since the detection of \sz\ clusters is based on the presence of hot gas and is only weakly dependent on redshift, the selection function of \sz\ surveys differs significantly from optically selected cluster samples.  Comparison of optical and \sz\ cluster samples is used to investigate the correspondence between \sz\ detections with \wazp\ clusters and vice-versa.

To carry out these comparisons, we use the cluster catalogs derived from the South Pole Telescope (\spt) and the Atacama Cosmology Telescope (\act) surveys. As shown in Fig.~\ref{fig:footprints}, these two catalogs have significant overlaps with the DES-Y6 footprint.

%-------------------------------
\subsection{\spt\ and \act\ cluster samples}
\label{sec:matches}
%-------------------------------

\begin{figure}
    \includegraphics%[trim=.48cm 1.0cm 1.0cm.045cm, clip, scale=.8]
    {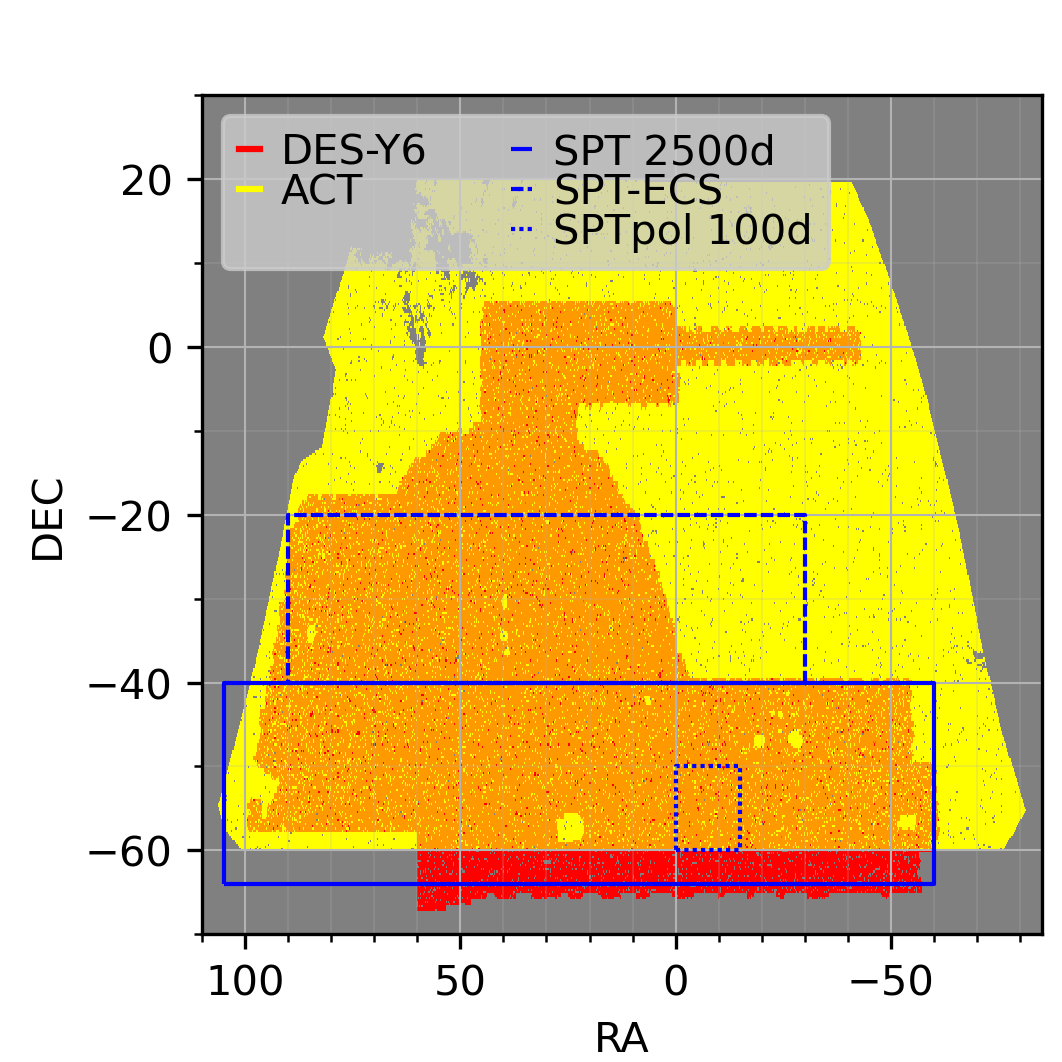}
    \caption{DES-Y6 footprint (red) compared to \act\ (yellow) and the different \spt\ (blue) footprints.}
    \label{fig:footprints}
\end{figure}

For comparison with \spt\, we constructed a sample that joins the SPT-SZ 2500d catalog \citep{Ble15} (with updated redshifts of \citealt{BOCQUET}), the SPT-ECS catalog \citep{Ble20} and the SPTpol 100d catalog \citep{Hua20}. The union of these three catalogs consists of 1,236 objects. However, there is overlap between these catalogs, and the same cluster could be found multiple times. We cross-match the clusters from different catalogs using a 2\,arcmin window to obtain a catalog with unique entries, resulting in 29 common pairs. These systems also have the same redshift across the different catalogs (27 pairs had exactly the same redshift value, and 2 were within 0.005 of each other, a difference much smaller than the redshift uncertainties of 0.04), confirming that they are the same cluster. We keep the cluster with the highest signal-to-noise ratio (SNR) as the best candidate in each case.  The final combined \spt\ catalog has a 3,343\,deg$^2$ overlap with the \wazpYVI\ footprint (Fig.~\ref{fig:footprints}), corresponding to a total of 816 unique \spt\ clusters with associated masses and redshifts.

\begin{figure*}
    \centering
    \includegraphics[]{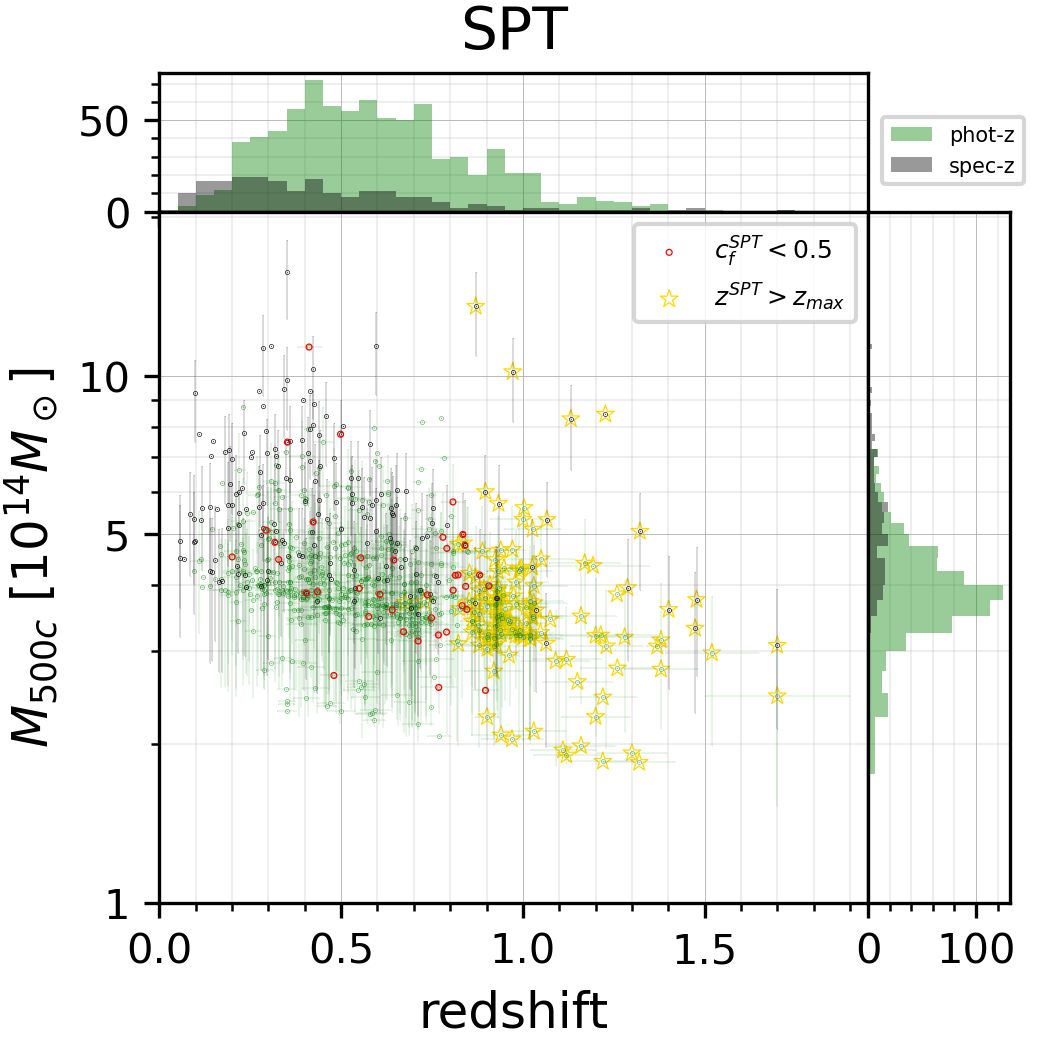}
    \includegraphics{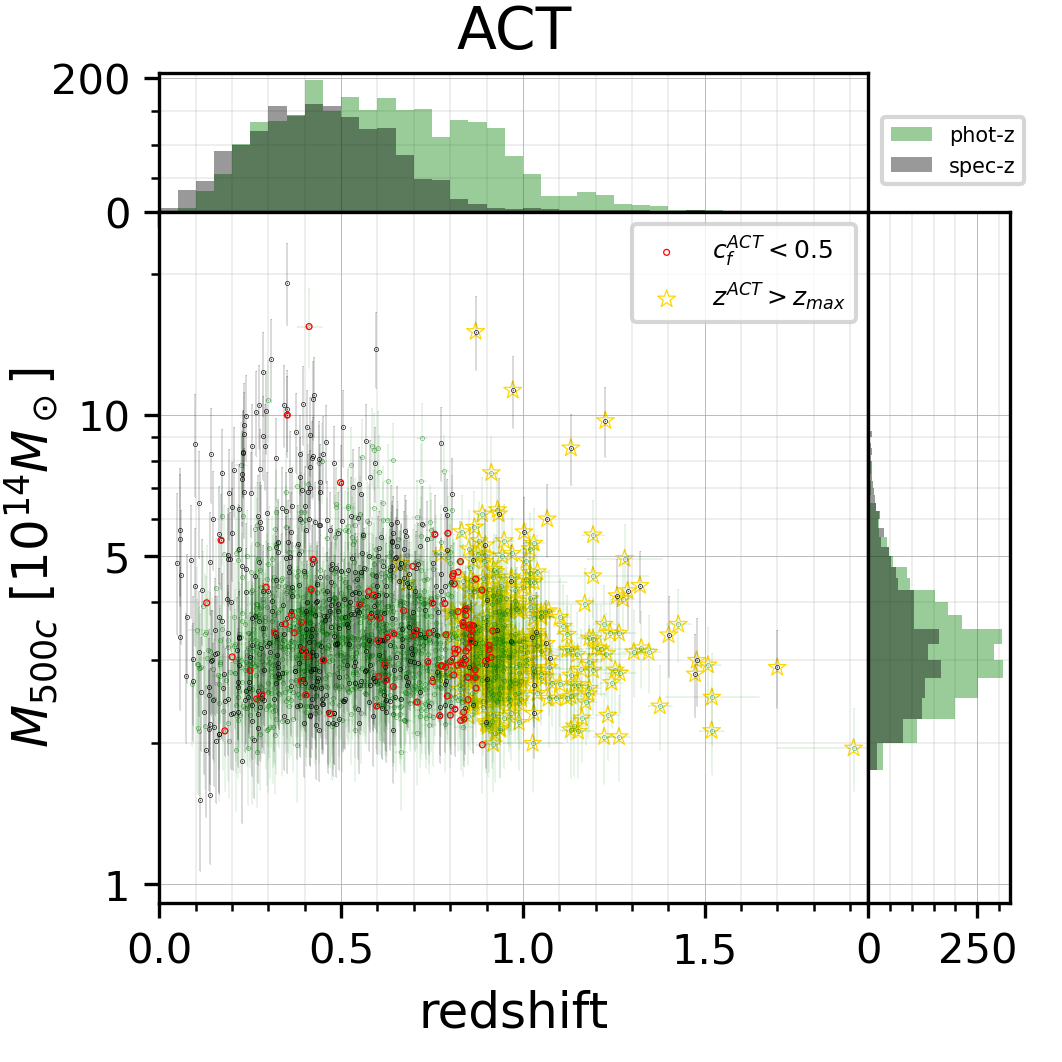}
    \caption{\sz\ clusters inside \wazp\ DES-Y6 footprint.  The colors of the points correspond the redshift type of the \sz\ counterparts: \spz\ shown in black and \phz\ in green. We also flagged clusters according to their coverage on the DES footprint: clusters with low coverage ($c_f<0.5$) are red, and clusters beyond the \wazp\ footprint $\zmax$ map are yellow.}
    \label{fig:SZ_mz}
\end{figure*}

In the case of the \act\ sample \citep{Hil21}, the common region with the \wazpYVI\ footprint consists of 4,243\,deg$^2$, and contains 1,873 clusters. In Fig.~\ref{fig:SZ_mz}, we show the masses and redshifts of each \sz\ cluster within the DES footprint color-coded by the type of redshift assigned to \sz.  Clusters poorly covered by the DES footprint (see definition of $c_f$ in the next section) or above the DES $z_{max}$ map are also flagged.  We see that a fraction of clusters are present in both \sz\ surveys, with \act\ covering a lower mass threshold and also having a larger fraction of high redshift clusters than \spt.

%%% Overlaps %%  wz.act     : 4,243 deg2 %%  wz.spt     : 3,343 deg2 %%  act.spt    : 4,060 deg2 %%  wz.act.spt : 3,035 deg2

%-------------------------------
\subsection{Cross-matching cluster catalogs}
\label{sec:sz_to_opt}
%-------------------------------

The cross-matching of DES-Y6 \wazp\ with both \sz\ cluster catalogs is based only on angular separation. First, an aperture of $2.6$\,arcmin (equivalent to 1\,Mpc at $z=0.5$) around each \sz\ cluster is considered.  The very few \sz\ systems without counterparts ($\lesssim$\,4\%) are cross-matched using a $5.3$\,arcmin aperture, more adequate for low redshift clusters (1\,Mpc at $z=0.2$). Multiple candidates are solved by selecting the richest \wazp\ cluster.  This choice is based on the assumption that the largest optical cluster detected on the same line of sight will be the source of the \sz\ emission and, in most cases, produce an optimal pairing. However, as discussed in Section \ref{sec:wazp_sz_counterparts}, in $\sim1\%$ of the cases, there is a secondary detection that is closer in redshift to the \sz\ cluster.

\begin{figure}
    \includegraphics {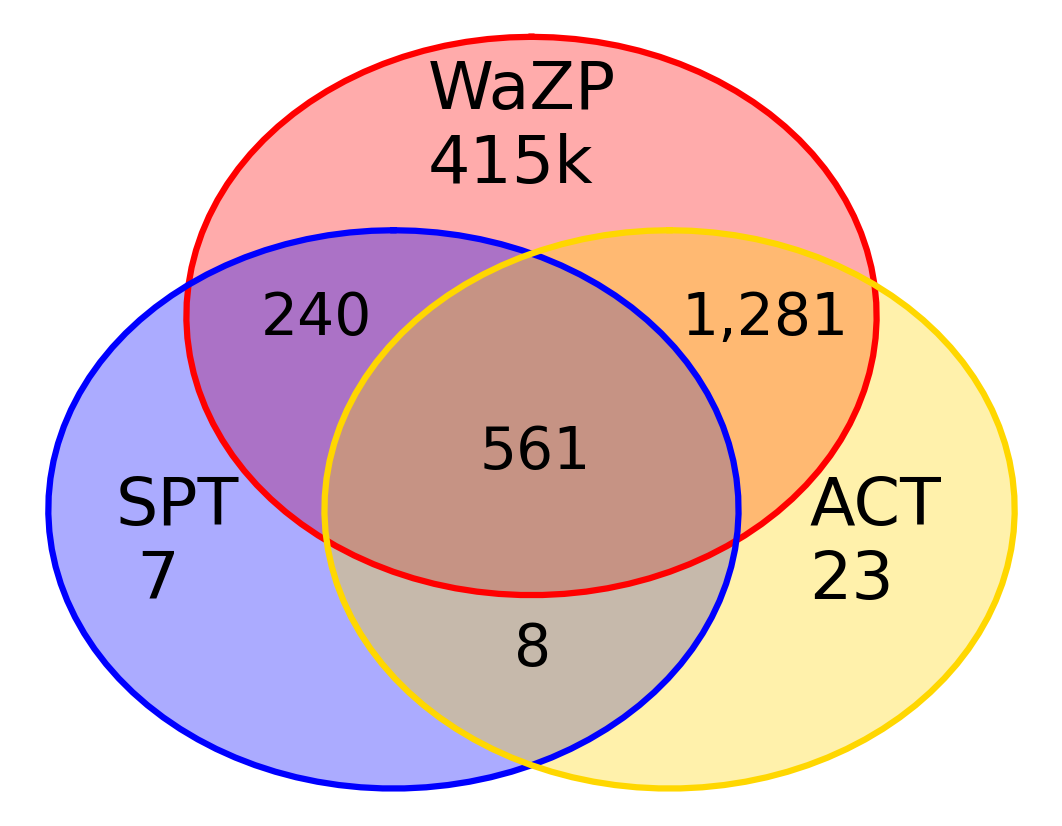}
    \caption{Diagram of the number of clusters matched considering the \wazp, \spt, and \act\ catalogs inside the DES footprint.}
    \label{fig:mt_sz_venn}
\end{figure}

This \sz\ - \wazp\ cross-matching resulted in 801 pairs out of the 816 \spt\ clusters and 1,842 pairs out of the 1,873 \act\ clusters. Some of these matches are common to the \spt\ and \act\ catalogs. Based on the same matching procedure, 570 \spt-\act\ pairs were found in the DES footprint, of which 561 also had a \wazp\ counterpart (see Fig.~\ref{fig:mt_sz_venn} for a summary of the intersections between the three samples). This means that $\sim 70$\% of \wazp-\spt\ pairs are also \wazp-\act\ pairs.
As an illustration, a mosaic of 9 \wazp\ clusters with their \act\ and \spt\ counterparts is shown in Appendix \ref{app:wazp_ex}.

\begin{figure*}
    \includegraphics{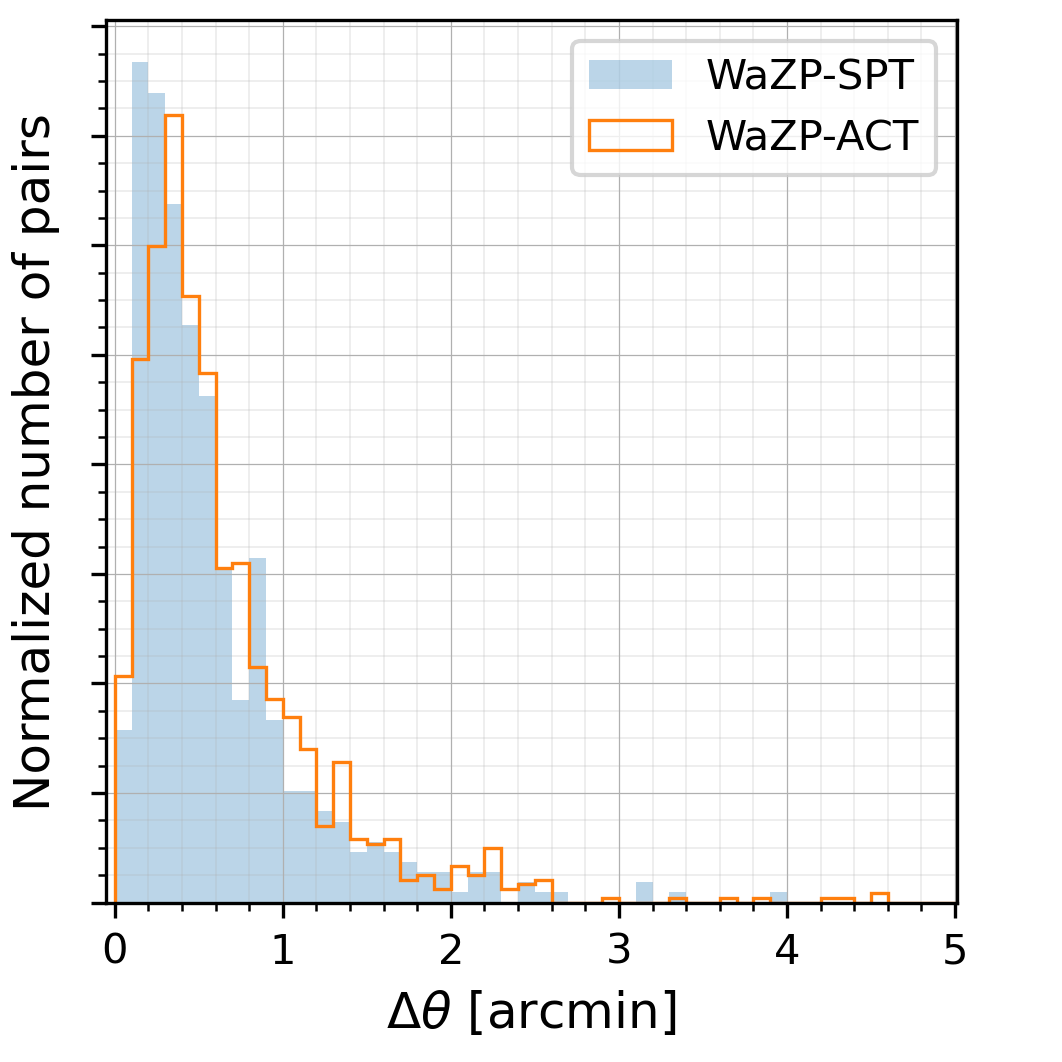}
    \includegraphics{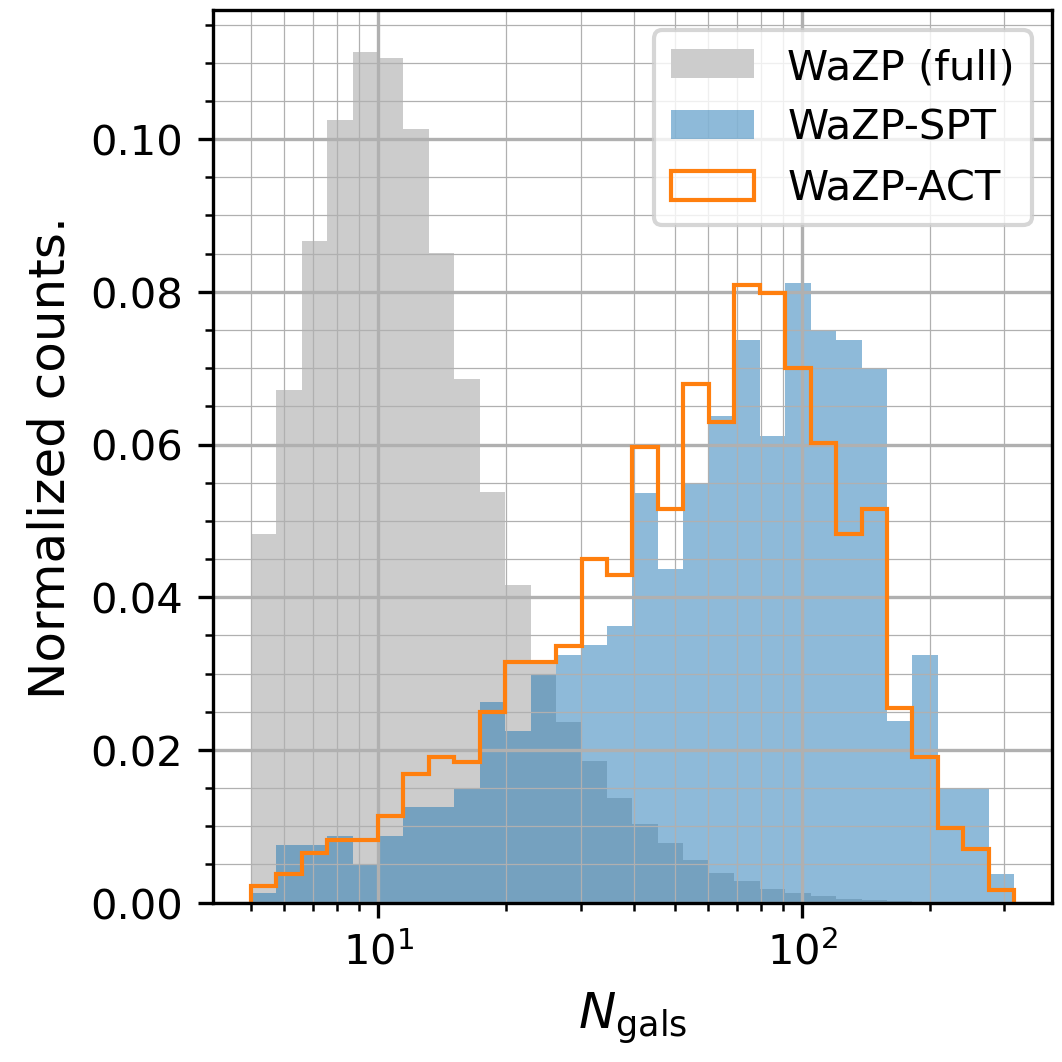}
    \caption{ Normalized distributions of \wazp-\sz\ matched clusters.  The angular separation between \wazp-\sz\ cluster centers of pairs with $c_f^{\rm SZ},c_f^{\rm WaZP}>0.8$ is shown on the left panel.  On the right panel, the distribution of the \wazp\ richnesses for the matched pairs is displayed, with the additional distribution of the full \wazp\ catalog for reference.  All distributions were normalized for their integral to be equal to unit.}
    \label{fig:SZ_da_arcmin_ngals}
\end{figure*}

Given the much larger surface density of the \wazp\ detections, we evaluate the possibility that a fraction of the matches corresponds to random associations.  Based on Fig.~\ref{fig:SZ_da_arcmin_ngals}, this is likely not the case. In the left panel, we verify that the associated optical counterparts have a projected median offset much smaller than the matching radii ($\sim 0.3$\,arcmin). In the right panel of Fig.~\ref{fig:SZ_da_arcmin_ngals}, we see that the median optical richness ($\sim 70$) of the matched clusters is much larger than the one of the entire sample ($\sim 10$).  We can estimate the probability of random associations as the expected number of \wazp\ clusters above a threshold richness inside a given aperture (if the number is larger than 1, we assume a 100\% chance). Considering the distribution in Fig.~\ref{fig:SZ_da_arcmin_ngals}, we assess this probability to be $\sim 6$\% for each \sz\ catalog.  Including the condition that the \wazp-\sz\ pairs also have consistent redshifts (which is shown in the next section), the chance of random associations becomes less than 0.6\%.

Considering the \sz\ systems that did not matched any \wazp\ detection, it was found that most of them are located beyond the local \wazpYVI\ $\zmax$ (3 for \spt\ and 8 for \act), or in regions only poorly covered by DES (8 for \spt\ and 16 for \act). These are identified by coverage fractions $c_f <50\%$, where $c_f$ is the Navarro-Frenk-White (NFW) weighted area (see \citealt{Ryk12}) fraction covered by the DES footprint in a $1$\,Mpc aperture around the \sz\ cluster.  The visual inspection of the four and seven remaining unmatched \spt\ and \act\ sources showed that two of them, common to both catalogs, do not have a clear galaxy concentration in the DES images. The others were all located near bright stars or foreground galaxies, significantly affecting cluster detection at optical wavelengths. These were not automatically discarded by our constraint on the local spatial coverage above either due to missing masks and/or because the $c_f$ calculation is based on the position of the \sz\ centers, which, in some cases may be far enough from the center of the associated optical system to miss the optical mask. This shows how complex the process of cross-matching clusters detected at different wavelengths can be, especially considering that they are affected by different observational effects.  In summary, for \sz\ clusters within the accessible redshift domain of DES-Y6, and as long as the local galaxy coverage is high enough, \wazp\ counterparts are found for all \sz\ systems.

\begin{figure}
    \includegraphics{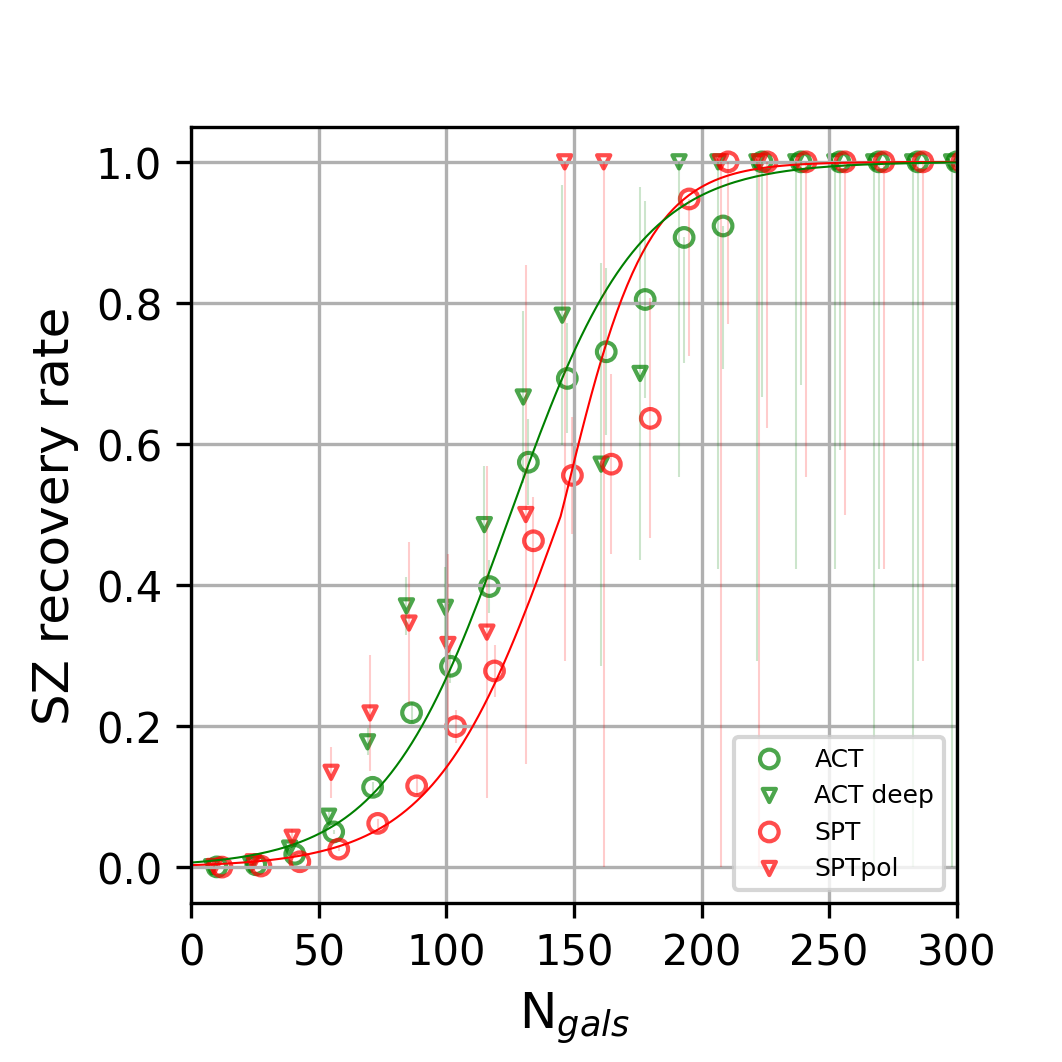}
    \caption{The recovery rates of \wazp\ clusters by \act\ or \spt\ as a function of \wazp\ richness, estimated as the fraction of \wazp\ clusters of a given richness for which there is an \act\ or \spt\ cluster within a radius of $5.3$\,arcmin. Only \wazp\ clusters with a 100\% coverage of the \sz\ footprints are considered. Errorbars are estimated assuming Poisson statistics.}
    \label{fig:missing_sz}
\end{figure}

Based on the same matching, we now analyze the fraction of \wazp\ clusters having an \sz\ counterpart as a function of the \wazp\ richness. This is shown in Fig.~\ref{fig:missing_sz}, where we only consider \wazp\ detections with a 100\% coverage within the match radius, with respect to each \sz\ footprint map.  The circles are the measured recovery rates for the \sz\ catalogs, and the lines are sigmoïd functions fits.  Also shown as triangles are the recovery rates for each \sz\ catalog considering only their deeper regions.  For the \spt\ survey, this is the SPTpol 100d region, while for \act\ (hereafter \actdeep), it corresponds to $\sim$900\,deg$^2$ with the lowest white noise, as shown in ~\citealt{Hil21} (white noise map, Figure 2).  For both \sz\ samples, we observe an increase of the recovery rate reaching 90\% completeness for richnesses greater than $\sim$175.  Despite their smaller size and larger statistical errors, SPTPol 100d and \actdeep\ show a better recovery rate than all combined \sz\ surveys.  On average, recovery rates exceed 90\% for richnesses above $\sim$175 (\spt\ 2500d, ECS, and \act), $\sim$140 (SPTpol 100d), and $\sim$160 (\actdeep).  The observed global trend is consistent with richer systems being more massive and likely detected by the \sz\ catalogs used here. The exact shape depends on both selection functions and the richness-mass scatter, which is investigated in a separate paper.

%-------------------------------
\subsection{The \wazp\ counterparts of \sz\ detections}
\label{sec:wazp_sz_counterparts}
%-------------------------------

The optical-\sz\ matching performed in the previous section is based only on the angular distance without any assumption on the redshift. The two \sz\ catalogs used to compare with our optical detection provide a spectroscopic redshift for one-third of the sample and a photometric redshift for the rest. The latter were assigned by cross-matching with the results of several optical cluster finders (e.g. \RM, Camira, AMICO, \clumpr, \wh) and/or from targeted follow-up observations \citep{Ble15,Ble20,Hua20,Hil21}.  Here, based on the redshift assignment of the ACT and SPT detections, we investigate whether our independent cross-match with \wazp\ clusters recovers the same pairs, in the limit of the DES-Y6 depth.

The assignment of a redshift to a \sz\ cluster is not always a trivial task, since one must first link the \sz\ signal to some group of galaxies. Sometimes, several possible counterparts can be identified, with the additional difficulty that the "true" redshift could also be beyond the reach of the optical survey under study.  Considering the much higher density of the optically selected sample, our angular \wazp - \sz\ cross-match may not necessarily lead to the identification of the same counterpart as was done by the \spt\ and \act\ collaborations. This can be tested by comparing \wazp\ redshifts with those assigned by \spt\ and \act.

\begin{figure*}
    \centering
    \includegraphics{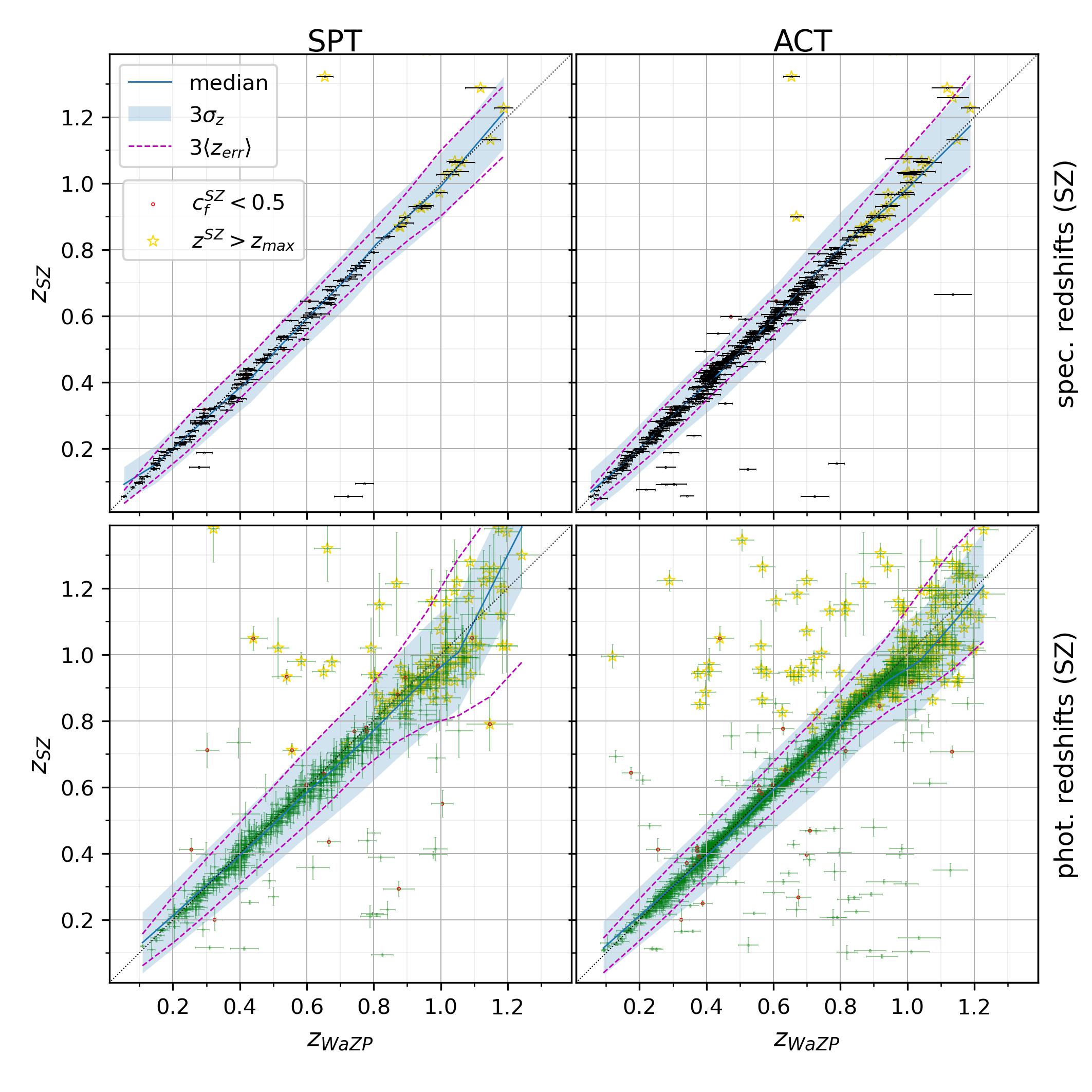}
    \caption{Redshift relation of \wazp-\spt\ (left) and \wazp-\act\ (right) matched clusters. The sample is split into clusters with spectroscopic \sz\ redshifts (top) and photometric redshifts (bottom). The blue-shaded regions correspond to 3 times the scatter of the sample with the outliers removed, and the dashed lines are 3 times the average error of the cluster redshifts.  \sz\ clusters with a lower coverage fraction (<0.5) are circled in red, and yellow stars highlight clusters beyond the $z_{max}$ map.}
    \label{fig:SZ_z}
\end{figure*}

In Fig.~\ref{fig:SZ_z}, based on our cross-matching, we compare the redshifts of the \wazp\ detections to those assigned by the \spt\ (left panels) and \act\ (right panels) collaborations.  We separate spectroscopic (top) and photometric (bottom) redshift assignments. In these four panels, we can clearly differentiate the systems for which we have likely identified and matched to the same optical counterpart (small redshift difference) and those for which we have selected another system (very discrepant redshifts). We differentiate these two cases statistically on the basis of the mean standard deviation of the redshift differences. We first clip obvious outliers defined by $|(\zsz-\zwazp)/(1+\zwazp)|>0.15$, which corresponds to more than 3 times the typical \wazp\ cluster redshift error (see section~\ref{sec:comp_y1}). We then compute the bias ($b_z$) and scatter ($\sigma_z$) of the resulting samples and define as consistent redshift pairs those with a difference within $3\sigma_z$, corresponding to the blue-shaded regions in the four panels.

We find that $\sim$93\% of the matches most likely correspond to the same optical counterpart as that assigned by the \spt\ and \act\ collaborations.
We find a different match in 7\% of cases. 
For these, if we exclude low redshift clusters ($z<0.2$) where the match radius is likely too small, and cases with the presence of nearby stars and defects, we find less than 2\% of of \sz\ clusters with disagreeing redshifts. They are discussed further down this section.

We also see that the common optical-\sz\ pairs have estimated redshifts compatible with redshifts uncertainties, with only $\sim $1-2\% of the pairs having an offset greater than 3 times the average error $\langle z_{\rm err} \rangle=\!\sqrt{\langle z_{\rm err} ^ {wazp} \rangle^2+\langle z_{\rm err} ^{sz} \rangle^2} $.  Furthermore, most of these outliers (55-60\%) were \sz\ clusters with either a lower cover fraction or beyond the $z_{max}$ map.

As seen in the upper panels of Fig.~\ref{fig:SZ_z}, there is no bias when \wazp\ redshifts are compared to the spectroscopic redshifts of \sz\ clusters in all ranges.  When evaluating photometric redshift \sz\ clusters (bottom panel), we find a bias at $z>1.0$. This may be due to the variety of sources and methods used to assign the photometric redshift to \sz\ clusters (see \citealt{BOCQUET,Ble15,Hua20,Hil21}).

In Table \ref{table:z_stat}, we give the bias and scatter values for each comparison. As can be seen, there is good agreement between the \wazp\ redshifts and the \sz\ spectroscopic values.  These metrics are comparable to those obtained by matching \wazp\ with all available spectroscopic sources, as described in section \ref{sec:wazp}.  Considering \sz\ clusters with photometric redshifts, we find, as expected, a larger scatter and bias.

\setlength{\tabcolsep}{5pt}
\begin{table}[h!]
\centering
\begin{tabular}{c | c c c | c c c}
\multirow{2}{2.4em}{z type} & \multicolumn{3}{c|}{\spt}& \multicolumn{3}{c}{\act}\\
%\hline
%& \multicolumn{}{|c|c|c|}{x & \spt\ & \act} \\
& N & $b_z$ & $\sigma_z$ & N & $b_z$ & $\sigma_z$ \\
%& \multicolumn{6}{c c c|c c c|}{\# of pairs & $b_z$ & $\sigma_z$ & \# of pairs & $b_z$ & $\sigma_z$}\\
\hline
\hline
%\multicolumn{7}{|c|c c c c c c}{
%spec. & 138 & 0.004 & 0.012 & 456 & 0.003 & 0.012
spec. & 160 & 0.004 & 0.011 & 456 & 0.003 & 0.012
%}
\\
%\multicolumn{7}{|c|c c c c c c}{
%phot. & 278 & 0.013 & 0.024 & 1,255 & 0.006 & 0.016
phot. & 583 & 0.009 & 0.021 & 1,255 & 0.006 & 0.016
%}
\end{tabular}
\caption{Offset ($b_z$) and scatter ($\sigma_z$) of the comparison between the redshifts of \wazp-\sz\ matched cluster pairs. The samples are split into \sz\ clusters with spectroscopic and photometric redshifts, with $N$ being the number of pairs of each sample.
}
\label{table:z_stat}
\end{table}

We now consider matched systems with discrepant redshifts ($\Delta z / (1+z) \geq 3\sigma_z$), 60 for SPT and 134 for ACT. The fact that these \sz\ detections were matched with a different optical concentration on the line of sight could be due to locally incomplete or too shallow optical data. It could also be due to the matching procedure, in particular at low redshift ($z\lesssim 0.2$), where mismatches may occur due to a too small angular matching radius.

\begin{figure*}
    \centering
    \includegraphics[scale=0.7]{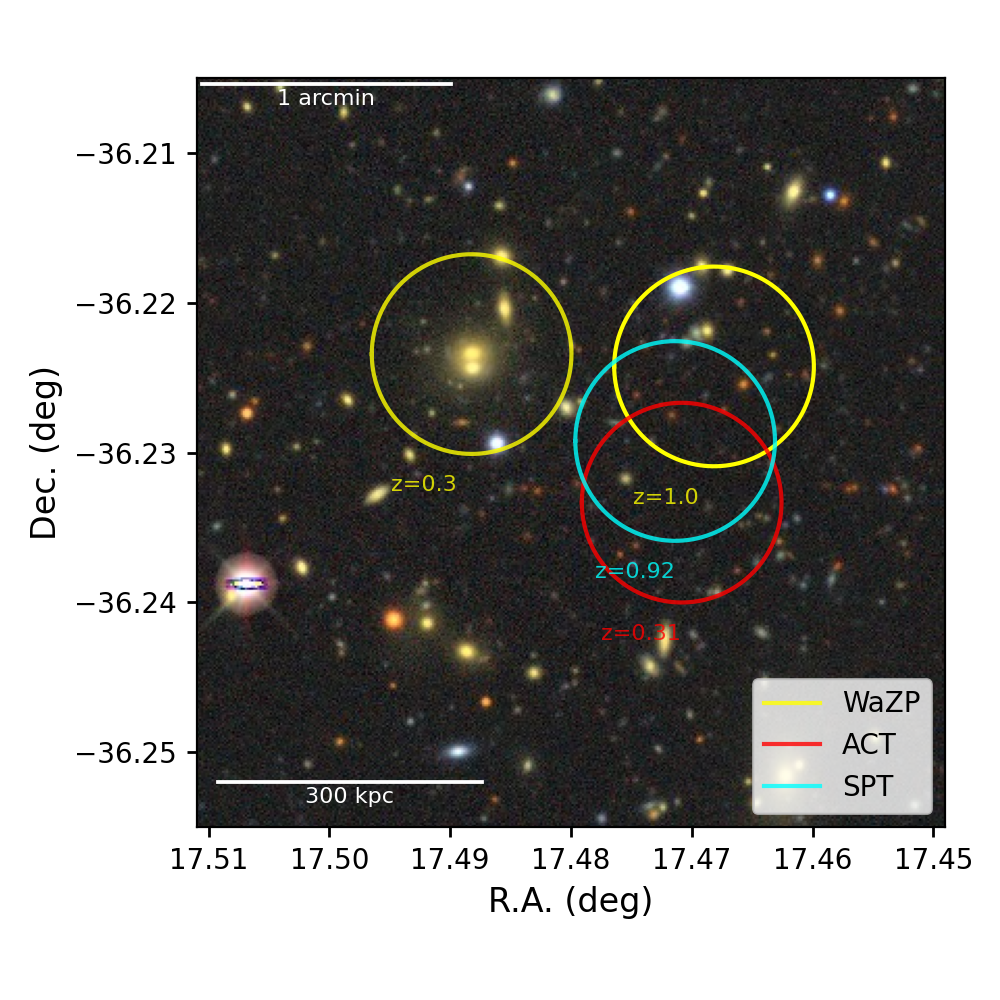}
    \includegraphics[scale=0.7]{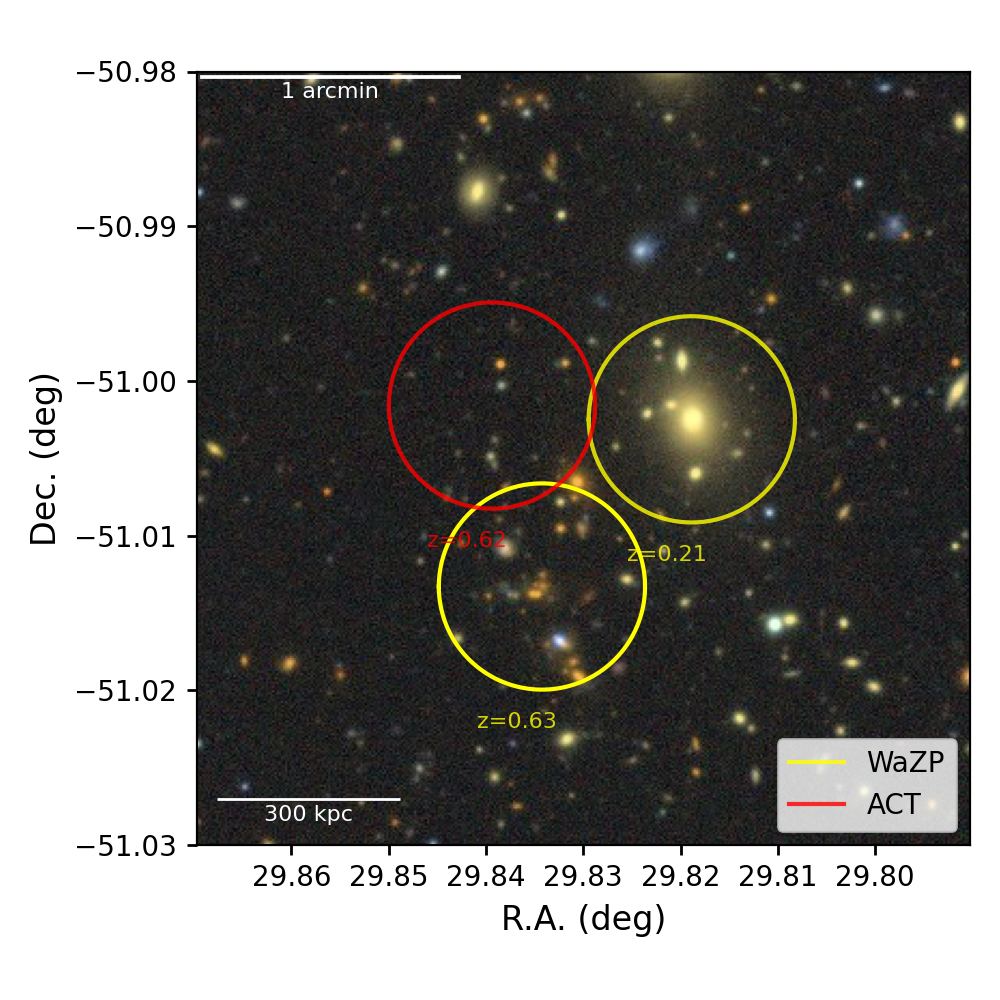}
    \includegraphics[scale=0.7]{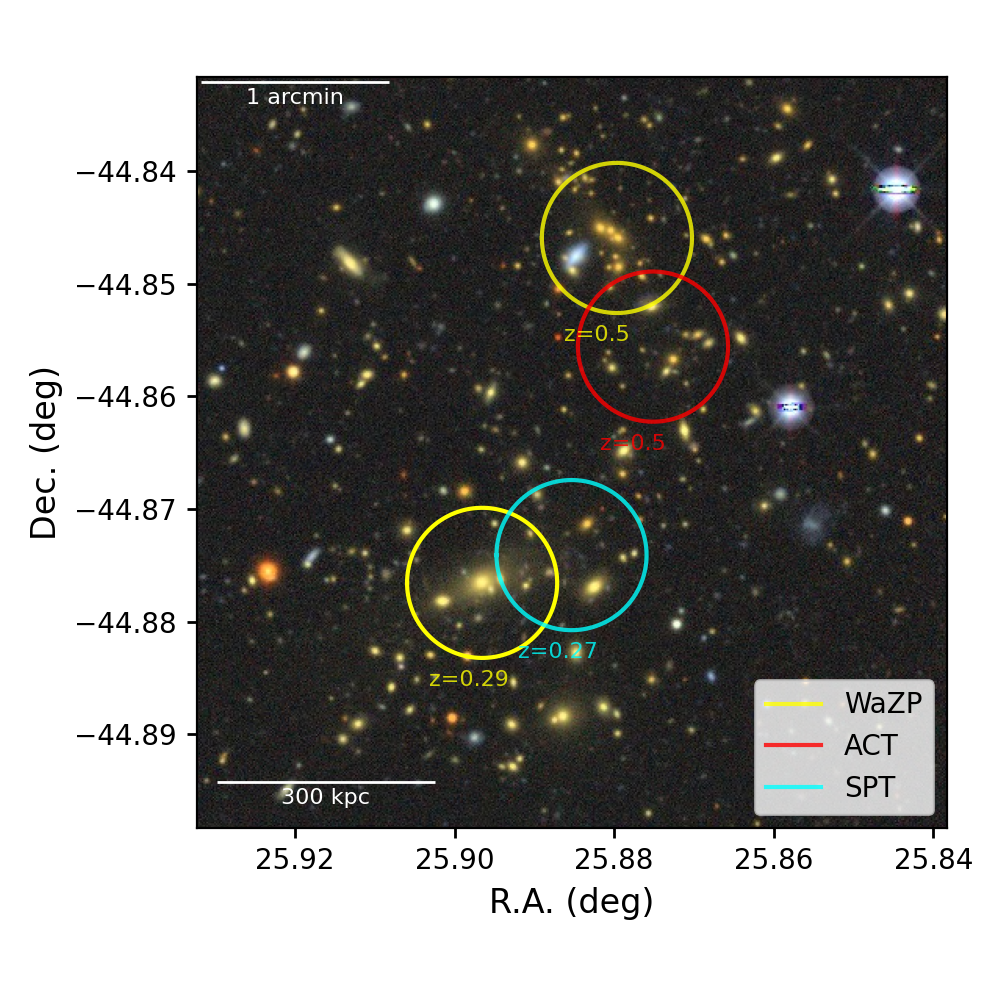}
    \includegraphics[scale=0.7]{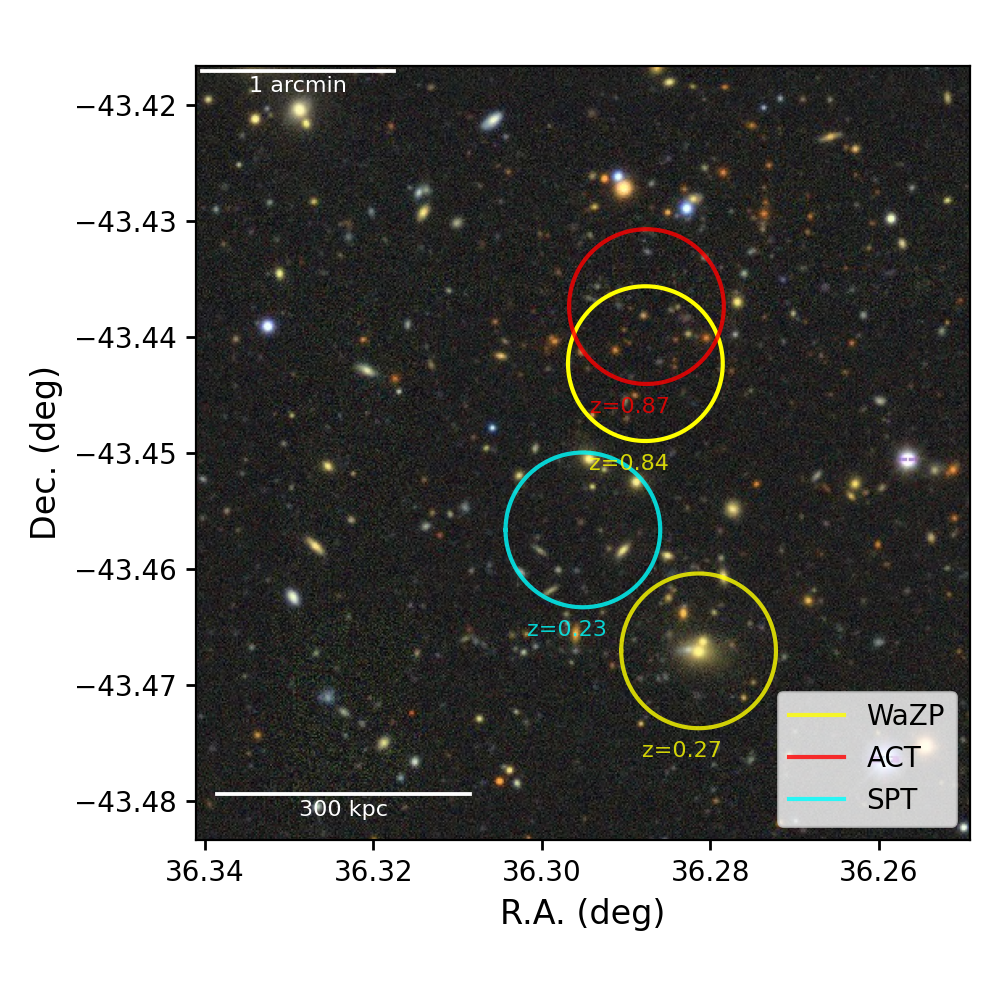}
    \caption{Four examples of \sz\ detections for which our detection and matching procedure assigned a primary redshift different from that of the \sz\ consortia. However, in these cases, a secondary counterpart on almost the same line of sight has a consistent redshift. These examples show the ambiguity of assigning an optical counterpart, with also different choices made by the \act\ and \spt\ groups.}
    \label{fig:superpo1}
\end{figure*}

In the end, we find that only 12 (SPT) and 31 (ACT) clusters with a good coverage fraction (>80\%), within \wazp's local $z_{max}$ map and with a redshift beyond $0.2$, were matched to a \wazp\ system at a redshift significantly different from the one assigned by the SZ collaborations.  It is interesting to note that among these, 10 SPT and 17 ACT clusters have a secondary, poorer, alternative \wazp\ counterpart with a consistent redshift. Based on visual inspection, half of these cases were wrongly matched due to the choice of the matching window or to the presence of nearby bright stars clipping members of clear galaxy concentrations likely to be the optical counterparts of the \sz\ detections.
 The other half shows the presence of several clear clusters at different redshifts and along the same line of sight, leading to ambiguities in the redshift assignment. In Fig.~\ref{fig:superpo1} we show four examples of such systems.

\begin{figure}
    \includegraphics[scale=0.7]{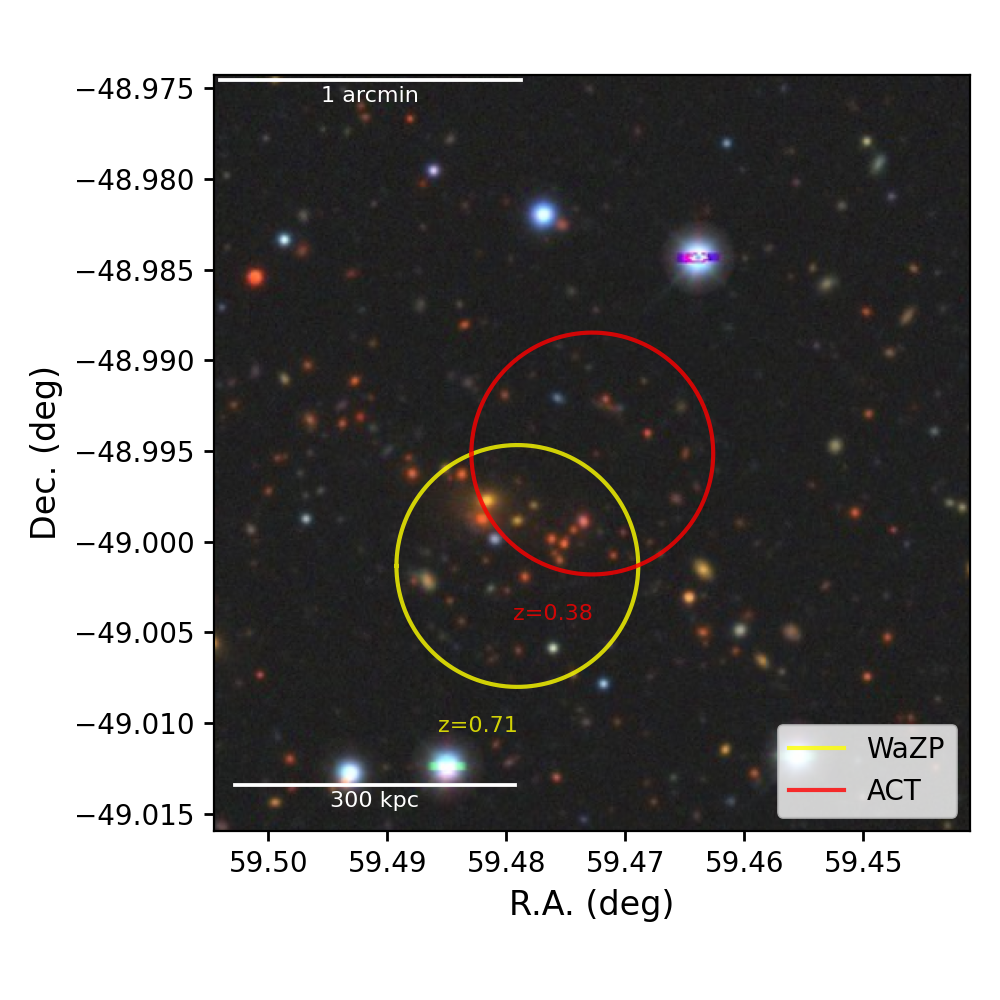}
    \caption{Case for which the \wazp\ counterpart to the \act\ detection is found at a very different redshift ($z=0.71$) from the \act\ redshift ($z=0.38$). This is an ambiguous case where two clear cD galaxies at redshifts $0.38$ and $0.71$ are located next to each other. However, only the $z=0.71$ system was rich enough to be detected by \wazp.
    }
    \label{fig:superpo2}
\end{figure}

Finally, we find that 2 SPT and 14 ACT clusters were matched to a \wazp\ detection at a different redshift without any other option along the line of sight. All these systems except one are suffering from the presence of nearby bright stars that affect optical detection and matching. Only one system can be considered as a robust alternative counterpart to the redshift assigned by the \act\ consortium. This case is shown in Fig.~\ref{fig:superpo2}.

In conclusion, when considering regions of DES-Y6 that are well covered and deep enough (i. e. $z_{\rm SPT}$ is below the local value of the DES $z_{max}$ map), we identify optical counterparts with consistent redshifts for the vast majority of \spt\ and \act\ clusters using the proposed matching procedure.  Only in less than 2\% of the cases are alternative counterparts found. Most of these are either low redshift systems ($z\le 0.2$) for which a larger matching radius would have been required or systems residing close to bright stars, which hampers the optical detection and/or our matching procedure.  Finally, among discrepant redshift associations, 15 \sz\ detections have two rich \wazp\ counterparts at significantly different redshifts along the line of sight, leading to ambiguous redshift assignments.

%-------------------------------
\subsection{Multiplicity of \wazp\ counterparts}
\label{sec:multiplicity}
%-------------------------------

\begin{figure}
    \includegraphics[scale=0.45]{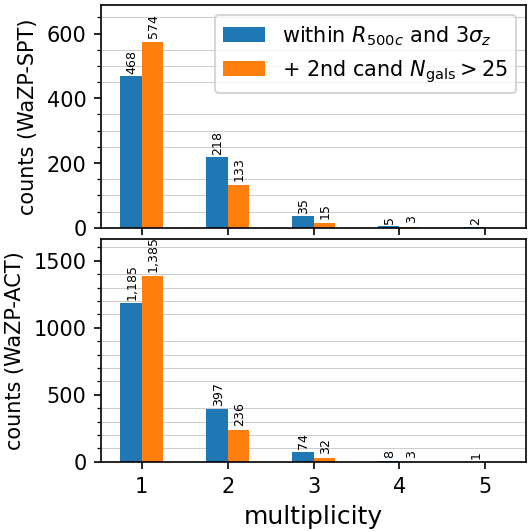}
    \caption{ Number of \wazp\ clusters within a cylinder composed of an aperture of $R_{500c}$ and height of 3$\sigma_z$ from Table~\ref{table:z_stat}.  In orange, only clusters whose richness was within 3 times the uncertainty of the most massive candidate, the criteria used to refine the matching.}
    \label{fig:multiplicity}
\end{figure}

In the previous section, we chose to assign to \sz\ detections the richest \wazp\ system along the line of sight. In the vast majority of cases this leads to redshift assignments consistent with those from the \act\ and \spt\ collaborations. However, we noticed that, within the redshift uncertainty and the \sz\ $R_{500c}$ radius, we could sometimes have additional, less rich, \wazp\ detections. This can be seen in Figure \ref{fig:multiplicity} where we display the number of \wazp\ candidates found within an aperture of $R_{500c}$ and in a redshift window derived from the redshift scatter shown in Table \ref{table:z_stat}. Without any restriction on \wazp\ richness, we find that $\sim $1/3 of the \sz\ clusters have more than one optical counterpart and, in very few cases, they have two or more additional optical counterparts. In order to avoid contamination by very low-richness systems, we can restrict the possible secondary optical detections to have richnesses above 25. This results in 85\% (ACT) and 80\% (SPT) of the \sz\ clusters having only one \wazp\ counterpart. The remaining 15-20\% \sz\ clusters display complex optical morphology, indicating possible on-going merging processes.

\begin{figure}
    \includegraphicsX[scale=0.7]{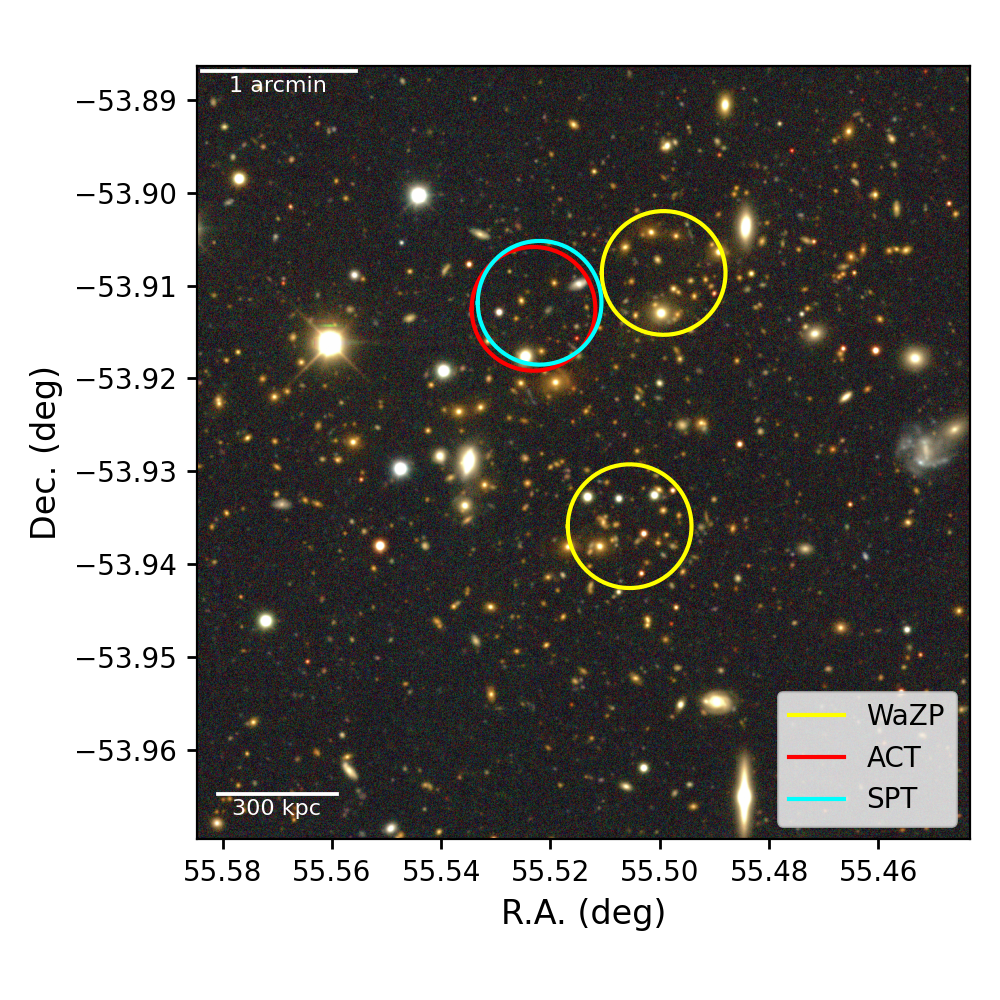}
    \caption{Example of a \sz\ cluster detected by both \act\ and \spt\ at $z=0.52$ with two \wazp\ counterparts also at the same redshift and both within an aperture of $R_{500c}$.}
    \label{fig:color_ex}
\end{figure}

An example of such a system is shown in Fig.~\ref{fig:color_ex}. For this system at redshift $z = 0.52$, the \sz\ detections match two \wazp\ clusters found at the same redshift, both within the ACT and SPT $R_{500c}$.

%\begin{itemize} \item Select the better candidate if $\Delta\rich<3\rich_{\rm err}$ \item  10-13\% of clusters have a secondary counterpart \item less than 5\% (38 SPT/84 ACT) of clusters are rematched \item about half of clusters with large offset in redshifts are fixed (9/14 for SPT and 16/27 for ACT) \end{itemize}

%-------------------------------
\subsection{\wazp\ {\it vs} \sz\ centering}
\label{sec:sz_centering}
%-------------------------------

In this section, we investigate the nature of the offsets between the centers of the \wazp\ and associated \sz\ clusters. For relaxed clusters, we expect the \sz\ and optical centers to be nearly identical, given their respective centering uncertainties.

As discussed in section~\ref{sec:multiplicity}, 15-20\% of \sz\ matched systems have more than one possible \wazp\ counterpart at the same redshift. This introduces some ambiguity to the estimation of the centering distribution and the model calibrated on it. Hence, to avoid confusion, the main analysis here will only consider systems with multiplicity 1. At the end of this section, we verify that this selection does not impact the main results and discuss the differences caused by including the entire matched sample.

\newcommand{\Saro}{Sar15} \newcommand{\Bleem}{Bl20} \newcommand{\Zent}{Zen20}

We follow the modeling proposed by \citet{Sar15} and \citet{Ble20} (referred as \Saro\ and \Bleem\ henceforth) based on the comparison of \RM\ and \spt\ centering.  The measured offset between the optical-\sz\ pairs is a convolution between the true offset and the intrinsic errors in the measurement of \sz\  and optical cluster centers.

The centering errors for the two \sz\ catalogs under study were modeled as Rayleigh distributions in their respective papers. In the case of \act, the associated standard deviation was empirically determined, and depends only on the SNR of the detections \citep{Hil21}:

\begin{equation}
    \sigma_{\rm{ACT}} = \frac{1.483}{\rm{SNR}_{2.4}}-0.012\;\; \rm{arcmin},
    \label{eq:dt_act}
\end{equation}

\noindent while the \spt\ error model is given by \citep{song12}:

\begin{equation}
    \sigma_{\rm{SPT}} = \frac{\sqrt{(k_{\rm SPT}\theta_c)^2+\theta_{beam}^2}}{\rm SNR} \;\;\rm{arcmin}.
    \label{eq:dt_spt}
\end{equation}

\noindent where $\theta_{beam}=1.2$\,arcmin is the \spt\ beam scale, $\theta_c$ is the cluster core (detection) scale, and \kspt\ is a parameter close to 1.

The equivalent beam for the optical center can be estimated either from the size of the pixels in the wavelet maps used for the \wazp\ detection or from the median separation between the central galaxy (when available, see section~\ref{sec:wazp}) and the centroid of the \wazp\ clusters. Both cases provide a similar value of $\sim$50\,kpc. The final model for the distributions of the \wazp-\sz\ offsets, is defined as the sum of Rayleigh distributions convolved with the \wazp\ centering uncertainty and with the \sz\ centering error models given above.  Here we found that two Rayleigh distributions can model the offset distribution well:

\begin{equation}
    P(x) = x \left( \frac{\rho_0}{\sigma_0^2}e^{-\frac{x^2}{2\sigma_0^2}} +\frac{1-\rho_0}{\sigma_1^2}e^{-\frac{x^2}{2\sigma_1^2}} \right),
\end{equation}

\noindent where $x=r/R_{500c}$, $\sigma_0$ and $\sigma_1$ are the standard deviations and $\rho_0$, ($1-\rho_0$) their relative amplitudes.  The resulting \wazp-\sz\ offset model is then compared with the observed offsets to calibrate its parameters (we also fixed \kspt$=1$, which produced similar results to applying the prior from \Bleem).

To ensure that the comparison between the alignment of clusters is not affected by holes in the DES/SPT/ACT footprints (see Fig. \ref{fig:footprints}) and depth variations in the DES footprint (see Section \ref{sec:refband}), only pairs of clusters where $c_f^{\rm \wazp}, c_f^{\rm\sz}>0.8$ are kept.  Additionally, pairs with discrepant redshifts (see previous section) were also excluded from this analysis.

\begin{figure*}
    \centering
    \includegraphics{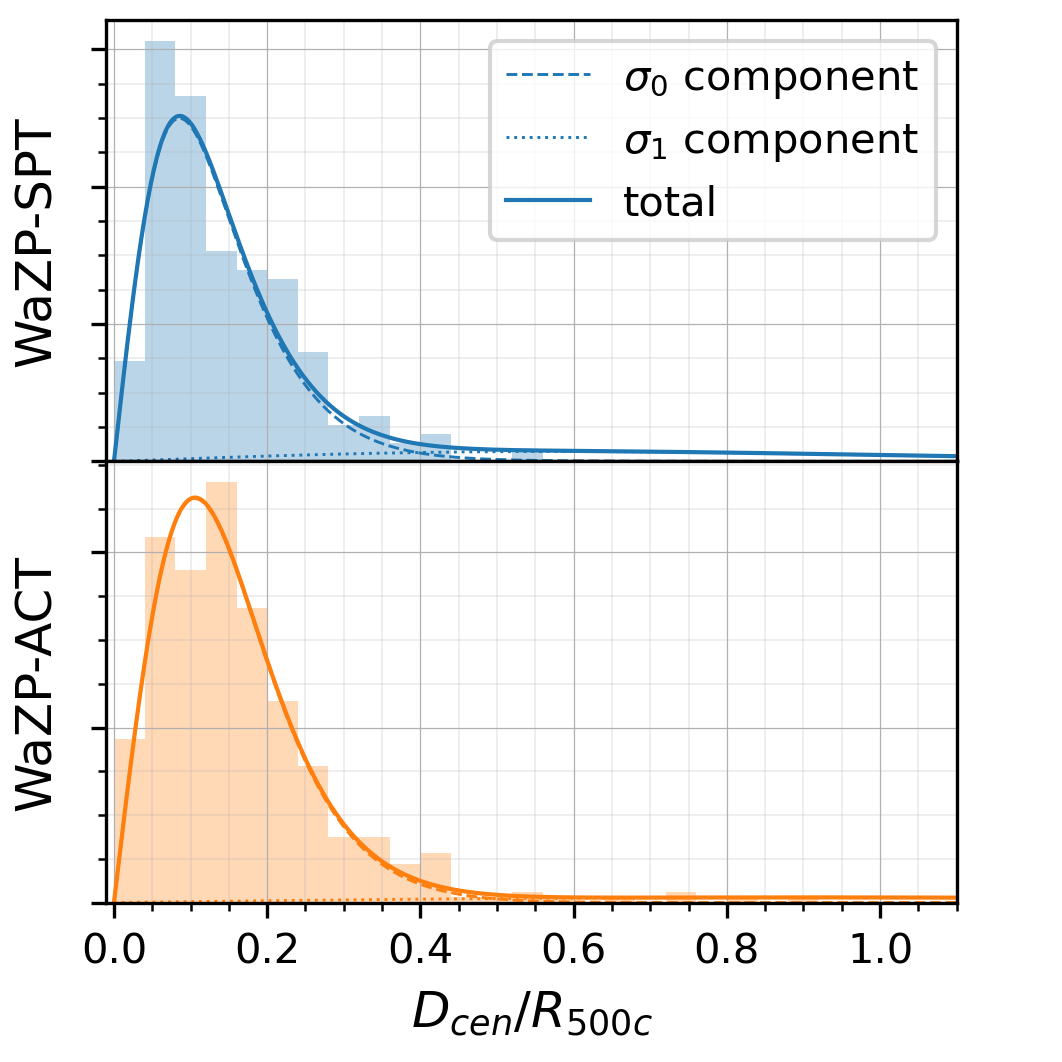}
    \includegraphics{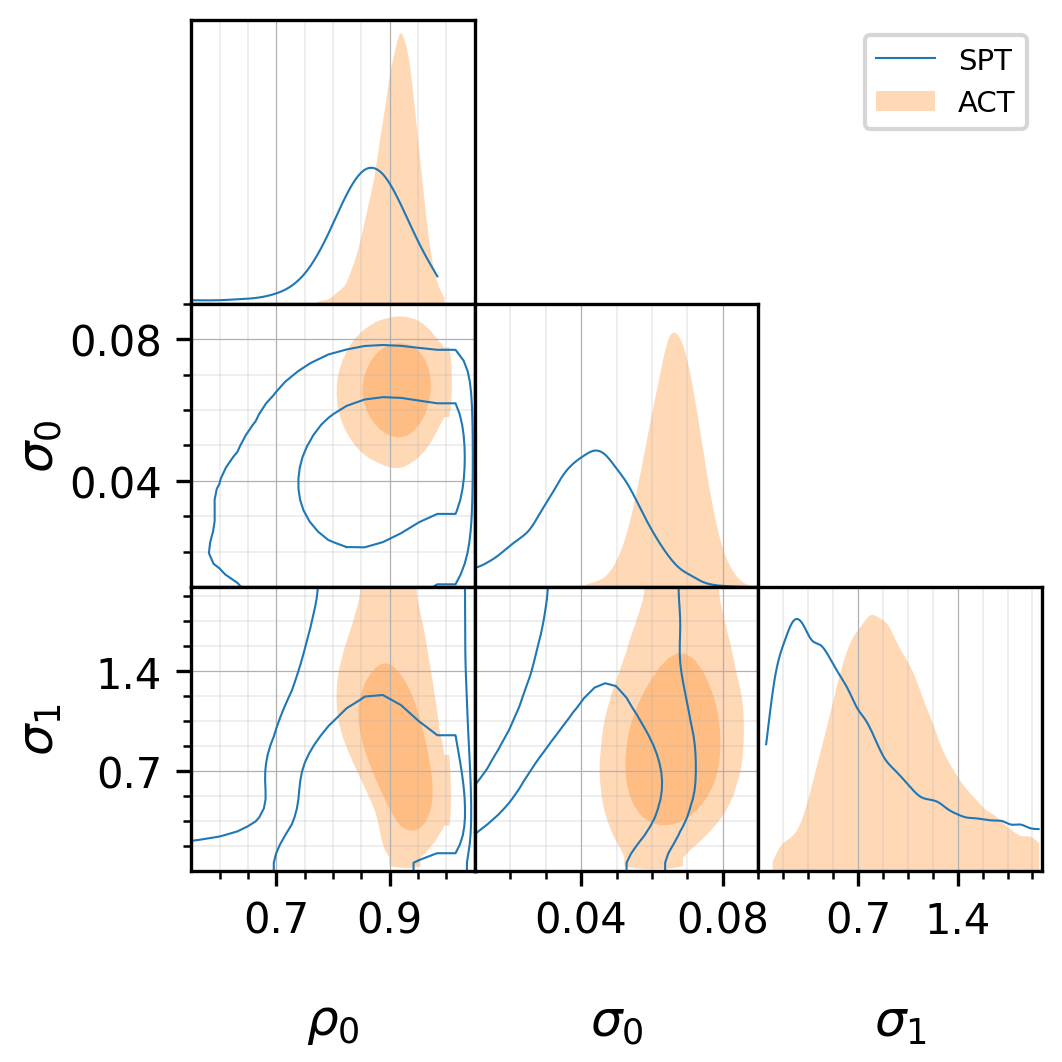}
    \caption{Left: Projected angular separation of matched clusters with SNR>5 and $c_f^{WaZP}, c_f^{\rm\sz}>0.8$. Histograms correspond to \wazp-\spt\ and \wazp-\act\ pairings, and the dashed lines represent the convolution of the offset models for each cluster catalog pairing. Right: Constraints of the \wazp\ offset model, best-fit values can be found in Table \ref{table:dtheta}.}
    \label{fig:SZ_da}
\end{figure*}

We constrain the parameters using Monte Carlo Markov Chains with the \code{emcee} package \citep{emcee}.  In the right panel of \reffigure{SZ_da}, we show the constraints obtained considering SNR$_{\rm SZ} \ge$ 5.  For both ACT and SPT, we find that a large fraction of the clusters ($\approx$90\%) have an offset well described by a Rayleigh distribution with $\sigma_0=0.04-0.07$, and a secondary population with $\sigma_1=0.6-0.9$.  In the left panel of \reffigure{SZ_da}, we show the measured offsets and the best-fit models decomposed into its two components.

We also calibrate the \wazp-\sz\ offset model for different SNR cuts, with results presented in Table \ref{table:dtheta}.  Comparison with \sz\ catalogs shows an increase in the well-centered population with SNR (although the variations are within 3 times the error-bars), while $\sigma_0$ and $\sigma_1$ remain constant within their uncertainties.  As a large fraction of \wazp-\act\ and \wazp-\spt\ are in common, we suspect that a significant part of the differences in the modeling are a consequence of the two different \sz\ error models.  This suspicion is reinforced by the fact that similar values in Table~\ref{table:dtheta} are found when we use only clusters found simultaneously in \wazp-\spt-\act.

To verify the robustness of the results obtained, we also constrain the centering model parameters including multiplicity>1 pairs in the \wazp-\sz\ samples.  We consider matched samples with different choices for solving multiple matches: most massive, highest SNR or closest candidate.  The values of $\rho_0$ and $\sigma_0$ remain consistent with Table \ref{table:dtheta}, while $\sigma_1$ is constrained around $0.3\pm0.1$,
with fluctuations of 0.05 from sample to sample.  This difference in $\sigma_1$ constraints (and the poor constraint power in the multiplicity = 1 sample) indicates that the population with larger offsets is composed of complex systems, with multiple possibilities of pairing. It also demonstrates how the constraints on $\sigma_1$ can be linked to the way the matching is performed.

The last rows of Table \ref{table:dtheta} contain values calibrated for \RM\ clusters \citep{Ryk12} as described in \Saro, \Bleem\ and \cite{Zen20}.  The fraction of the smaller offset population in the \wazp-\spt\ comparison is higher than all of these cases, although still within a 2-$\sigma$ level.  The values of $\sigma_0$ for \wazp-\spt\ are about half of those of \Saro\ and \Zent\ with significance of 1.5 and 4 $\sigma$, respectively.  Compared with \Bleem, we find that our $\sigma_0$ is larger. However, it is important to note that the offset model in \Bleem\ consists of three Rayleigh distributions.

    \setlength{\tabcolsep}{8pt}
    \renewcommand{\arraystretch}{1.5}
    \begin{table*}[h!]
    \centering
    \begin{tabular}{c c | c c c c}
    Catalog Pair & SNR$_{\rm SZ}>$  & $N_{\rm cl}$ & $\rho_0$ & $\sigma_0$ & $\sigma_1$ \\
    \hline
    \multirow{3}{2.4em}{\wazp-\act}
        & 4.0 &   805 & $0.80^{+0.03}_{-0.04}$ & $0.057^{+0.008}_{-0.010}$ & $0.32^{+0.03}_{-0.03}$ \\
        & 4.5 &   575 & $0.87^{+0.02}_{-0.03}$ & $0.067^{+0.008}_{-0.009}$ & $0.50^{+0.19}_{-0.13}$ \\
        & 5.0 &   409 & $0.91^{+0.03}_{-0.04}$ & $0.066^{+0.007}_{-0.007}$ & $0.90^{+0.44}_{-0.33}$ \\
    \hline
    \multirow{3}{2.4em}{\wazp-\spt}
        & 4.0 &   320 & $0.83^{+0.03}_{-0.04}$ & $0.031^{+0.014}_{-0.016}$ & $0.57^{+0.22}_{-0.16}$ \\
        & 4.5 &   279 & $0.84^{+0.04}_{-0.04}$ & $0.036^{+0.013}_{-0.015}$ & $0.68^{+0.28}_{-0.21}$ \\
        & 5.0 &   195 & $0.86^{+0.06}_{-0.07}$ & $0.042^{+0.013}_{-0.015}$ & $0.57^{+0.68}_{-0.35}$ \\
    \hline
    \RM-\spt\ 2015 & 4.5 & 25 & $0.63^{+0.15}_{-0.25}$ &  $0.07^{+0.03}_{-0.02}$ & $0.25^{+0.07}_{-0.06}$\\
    \RM-\spt\ 2020* & 5.0 & 249 & $0.68^{+0.07}_{-0.08}$ & $0.02^{+0.01}_{-0.01}$ & $0.15^{+0.03}_{-0.03}$\\
    \RM-\spt\ 2020b & 4.5 & 245 & $0.75\pm{0.05}$ & $0.08 \pm 0.01$ & $0.55 \pm 0.04$\\
    \end{tabular}
    \caption{Parameters of the offset model constrained with different SNR cuts. The \RM\ constraints correspond to \cite{Sar15}, \cite{Ble20}, and \cite{Zen20} respectively. * A model with three Gaussians was calibrated in \cite{Ble20}, only the parameters of the main two are shown here.
    }
    \label{table:dtheta}
    \end{table*}
    \renewcommand{\arraystretch}{1}

It has been suggested by \cite{Zen20} that the second Rayleigh distribution, with typical offsets 5-12 times larger than the primary one, consists of disturbed systems. In our case, as systems with multiple potential optical counterparts were removed for this modeling, most of the information is contained in the first Rayleigh distribution. Visual inspection of the few systems with offsets larger than $\sim 0.4 R_{500c}$ confirmed that they are not interacting systems, but instead reflect the centering accuracy of the combination of \sz\ and optical beams.

%%%%%%%%%%%%%%%%%%%%%
\section{Conclusions}
\label{sec:conc}
%%%%%%%%%%%%%%%%%%%%%

In this paper, we present a unique list of optically detected galaxy cluster candidates identified from the final galaxy catalog derived from DES-Y6 data using the \wazp\ cluster identifier introduced in Paper-1, whose detection does not rely on the presence of a red sequence of galaxies. Furthermore, using the recently published catalogs of clusters identified from millimeter surveys (SPT and ACT), we are able to carry out a compilation based on the cross-match of these surveys with unprecedented statistics.  We show that despite possible projection effects, rich systems, optically detected by \wazp\ in this work, indeed correspond to actual gravitationally bound systems identified by the SZ effect. We are also able to derive how the completeness of the \sz\ catalogs vary with optical richness.  The relation between the \wazp\ richness and the mass derived from the joint \wazp-\sz\ sample will be explored in future work.

Based on DES-Y6 data, the \wazp\ cluster finder produced a list of 417k clusters ($\rich\geq5$) over a total area of 4,545\,deg$^2$ and up to $z=1.32$. As the photometric depth varies across the survey, we also provide a subset of clusters for which the completeness of the cluster members is ensured to remain consistent across all redshifts. This cut results in a reduction of 25\% of the total number.  We are able to assign spectroscopic redshifts to over 25k \wazp\ clusters, and with this subsample, the estimated cluster redshift bias is 0.2\%, and the average scatter 1.4\%.

Comparison to DES-Y1 \wazp\ clusters (Paper-1) shows significant improvements due to deeper photometry and subsequent better quality of photometric redshifts, that also benefited from a much larger and fainter spectroscopic training set.  We also show the consistency of the detection method, with more than 95\% of the rich ($\rich$>50) clusters in common between both catalogs and having a compatible richness assignment.

The \wazp\ catalog was matched with detections of \spt\ (combination of three releases) and \act\ with an angular aperture corresponding to 1 and 2\,Mpc at $z=0.5$.  Essentially, all \sz\ sources well sampled by the DES-Y6 footprint and below its local redshift limit (801 for \spt\ and 1842 for \act) have a \wazp\ counterpart. The main exceptions are caused either by the presence of nearby bright objects ({\it i.e.} stars and foreground galaxies) or by a clear absence of galaxy concentrations in the DES images, possibly due to a too high redshift optical counterpart (two cases).

Although the matching did not use the cluster redshifts as a criterion, we find that 93\% of the pairs presented consistent redshifts. The mean offset between \wazp\ redshifts and those of the two \sz\ catalogs is similar, with an offset below the percent level and a scatter of 1-2\%. Comparing to spectroscopic redshift \sz\ clusters, the offset and scatter are 25-50\% smaller. Most pairs with discrepant redshifts either have a lower cover fraction, or the redshift of the \sz\ counterpart is above the local $\zmax$ map, or are affected by the presence of a nearby star that hampered optical detection and/or matching. However, a small subset of these pairs appears to have two rich optical counterparts along the same line of sight, leading to ambiguities in the redshift assignment.

When looking for \sz\ counterparts of optical detections, as expected, we find that the recovery rate is higher for richer clusters, being greater than 90\% for $\rich\gtrsim$175 clusters. The deeper regions of the \sz\ surveys are found to have a higher recovery of optical clusters, reaching 90\% for $\rich\gtrsim$150. The absence of counterparts is probably due to the SNR cut used in these two catalogs and to their respective selection functions.

We find that 15-20\% of the SZE matched systems have more than one possible \wazp\ counterpart at the same redshift and within the \sz\ $R_{500c}$, indicating possible interacting or unrelaxed systems.

The distribution of the offsets between the optical and \sz\ centerings is well modeled by a sum of two Rayleigh distributions.  The results using \spt\ and \act\ are consistent with each other: a larger population consisting of 80-90 \% of the clusters has a smaller scatter ( 3-6 \% of $R_{500} $) while the secondary population has 5 to 20 times larger offsets. When considering pairs with a multiplicity of 1, this population consists of very few systems with offsets larger than $\sim 0.4R_{500c} $. Their visual inspection does not show clear signs of unrelaxed systems but can instead be explained by the combination of optical and \sz\ beams.

This work provides a list of galaxy clusters optically detected from the final 5,000\,deg$^2$ of the DES-Y6 galaxy catalog, and the list of associated membership probabilities. Based on the comparison with external data (publicly available spectroscopy and SZE data), we also provide detailed properties of the centering of these detections, both in redshift and in position on the sky.  Cross-pairing with other optical cluster catalogs demonstrates the robustness of \wazp\ detections and highlights the new contribution of the presented catalog. Compared with the \sz\ data, it also indicates the good level of completeness and purity of the \wazp\ catalog, at least for massive/rich systems. In the following papers, we will address i) the mass-richness scaling relation based on SZE and X-ray clusters common to the DES-Y6 footprint and ii) the selection function of \wazp\ clusters based on numerical simulations.

%%%%%%%%%%%%%%%%%%%%%
\section*{Acknowledgments}
%%%%%%%%%%%%%%%%%%%%%

CB - developed the WaZP cluster finder, worked on the science analysis, data processing, and writing of the paper.
MA - developed additional software (ClEvaR) and methodologies used, worked on the science analysis, data processing, and writing of the paper.
LNdC - participated in the science analysis and writing of the paper. 
JG - combined all publicly available spectroscopic redshifts in the DES footprint and computed photometric redshifts for this study.

Funding for the DES Projects has been provided by the U.S. Department of Energy, the U.S. National Science Foundation, the Ministry of Science and Education of Spain, the Science and Technology Facilities Council of the United Kingdom, the Higher Education Funding Council for England, the National Center for Supercomputing Applications at the University of Illinois at Urbana-Champaign, the Kavli Institute of Cosmological Physics at the University of Chicago, the Center for Cosmology and Astro-Particle Physics at the Ohio State University, the Mitchell Institute for Fundamental Physics and Astronomy at Texas A\&M University, Financiadora de Estudos e Projetos, Funda{\c c}{\~a}o Carlos Chagas Filho de Amparo {\`a} Pesquisa do Estado do Rio de Janeiro, Conselho Nacional de Desenvolvimento Cient{\'i}fico e Tecnol{\'o}gico and the Minist{\'e}rio da Ci{\^e}ncia, Tecnologia e Inova{\c c}{\~a}o, the Deutsche Forschungsgemeinschaft and the Collaborating Institutions in the Dark Energy Survey. 

The Collaborating Institutions are Argonne National Laboratory, the University of California at Santa Cruz, the University of Cambridge, Centro de Investigaciones Energ{\'e}ticas, Medioambientales y Tecnol{\'o}gicas-Madrid, the University of Chicago, University College London, the DES-Brazil Consortium, the University of Edinburgh, the Eidgen{\"o}ssische Technische Hochschule (ETH) Z{\"u}rich, Fermi National Accelerator Laboratory, the University of Illinois at Urbana-Champaign, the Institut de Ci{\`e}ncies de l'Espai (IEEC/CSIC), the Institut de F{\'i}sica d'Altes Energies, Lawrence Berkeley National Laboratory, the Ludwig-Maximilians Universit{\"a}t M{\"u}nchen and the associated Excellence Cluster Universe, the University of Michigan, NSF NOIRLab, the University of Nottingham, The Ohio State University, the University of Pennsylvania, the University of Portsmouth, SLAC National Accelerator Laboratory, Stanford University, the University of Sussex, Texas A\&M University, and the OzDES Membership Consortium.

Based in part on observations at NSF Cerro Tololo Inter-American Observatory at NSF NOIRLab (NOIRLab Prop. ID 2012B-0001; PI: J. Frieman), which is managed by the Association of Universities for Research in Astronomy (AURA) under a cooperative agreement with the National Science Foundation.

The DES data management system is supported by the National Science Foundation under Grant Numbers AST-1138766 and AST-1536171. The DES participants from Spanish institutions are partially supported by MICINN under grants PID2021-123012, PID2021-128989 PID2022-141079, SEV-2016-0588, CEX2020-001058-M and CEX2020-001007-S, some of which include ERDF funds from the European Union. IFAE is partially funded by the CERCA program of the Generalitat de Catalunya.

We acknowledge support from the Brazilian Instituto Nacional de Ci\^encia
e Tecnologia (INCT) do e-Universo (CNPq grant 465376/2014-2).

This document was prepared by the DES Collaboration using the resources of the Fermi National Accelerator Laboratory (Fermilab), a U.S. Department of Energy, Office of Science, Office of High Energy Physics HEP User Facility. Fermilab is managed by Fermi Forward Discovery Group, LLC, acting under Contract No. 89243024CSC000002.

This scientific work uses data obtained from Inyarrimanha Ilgari Bundara / the Murchison Radio-astronomy Observatory. We acknowledge the Wajarri Yamaji People as the Traditional Owners and native title holders of the Observatory site. CSIRO’s ASKAP radio telescope is part of the Australia Telescope National Facility (https://ror.org/05qajvd42). Operation of ASKAP is funded by the Australian Government with support from the National Collaborative Research Infrastructure Strategy. ASKAP uses the resources of the Pawsey Supercomputing Research Centre. Establishment of ASKAP, Inyarrimanha Ilgari Bundara, the CSIRO Murchison Radio-astronomy Observatory and the Pawsey Supercomputing Research Centre are initiatives of the Australian Government, with support from the Government of Western Australia and the Science and Industry Endowment Fund. This paper includes archived data obtained through the CSIRO ASKAP Science Data Archive, CASDA (https://data.csiro.au).

The South Pole Telescope program is supported by the National Science Foundation (NSF) through award OPP- 1852617. Partial support is also provided by the Kavli Institute of Cosmological Physics at the University of Chicago.

MA is supported by the PRIN 2022 project EMC2 - Euclid Mission Cluster Cosmology: unlock the full cosmological utility of the Euclid photometric cluster catalog (code no. J53D23001620006).

%%%%%%%%%%%%%%%%%%%%%
\section*{Data Availability}
\label{sec:data_availability}
%%%%%%%%%%%%%%%%%%%%%
The \wazp\ DES-Y6 cluster products with a full description of the catalogs and columns can be found are available in its online supplementary material \wazpweblink.

%%%%%%%%%%%%%%%%%%%%%
\bibliographystyle{aa}
\bibliography{main}

\begin{thebibliography}{63}
\expandafter\ifx\csname natexlab\endcsname\relax\def\natexlab#1{#1}\fi

\bibitem[{{Abbott} {et~al.}(2021){Abbott}, {Adam{\'o}w}, {Aguena}, {Allam}, {Amon}, {Annis}, {Avila}, {Bacon}, {Banerji}, {Bechtol}, {Becker}, {Bernstein}, {Bertin}, {Bhargava}, {Bridle}, {Brooks}, {Burke}, {Carnero Rosell}, {Carrasco Kind}, {Carretero}, {Castander}, {Cawthon}, {Chang}, {Choi}, {Conselice}, {Costanzi}, {Crocce}, {da Costa}, {Davis}, {De Vicente}, {DeRose}, {Desai}, {Diehl}, {Dietrich}, {Drlica-Wagner}, {Eckert}, {Elvin-Poole}, {Everett}, {Evrard}, {Ferrero}, {Fert{\'e}}, {Flaugher}, {Fosalba}, {Friedel}, {Frieman}, {Garc{\'\i}a-Bellido}, {Gaztanaga}, {Gelman}, {Gerdes}, {Giannantonio}, {Gill}, {Gruen}, {Gruendl}, {Gschwend}, {Gutierrez}, {Hartley}, {Hinton}, {Hollowood}, {Honscheid}, {Huterer}, {James}, {Jeltema}, {Johnson}, {Kent}, {Kron}, {Kuehn}, {Kuropatkin}, {Lahav}, {Li}, {Lidman}, {Lin}, {MacCrann}, {Maia}, {Manning}, {Maloney}, {March}, {Marshall}, {Martini}, {Melchior}, {Menanteau}, {Miquel}, {Morgan}, {Myles}, {Neilsen}, {Ogando}, {Palmese}, {Paz-Chinch{\'o}n}, {Petravick},
  {Pieres}, {Plazas}, {Pond}, {Rodriguez-Monroy}, {Romer}, {Roodman}, {Rykoff}, {Sako}, {Sanchez}, {Santiago}, {Scarpine}, {Serrano}, {Sevilla-Noarbe}, {Smith}, {Smith}, {Soares-Santos}, {Suchyta}, {Swanson}, {Tarle}, {Thomas}, {To}, {Tremblay}, {Troxel}, {Tucker}, {Turner}, {Varga}, {Walker}, {Wechsler}, {Weller}, {Wester}, {Wilkinson}, {Yanny}, {Zhang}, {Nikutta}, {Fitzpatrick}, {Jacques}, {Scott}, {Olsen}, {Huang}, {Herrera}, {Juneau}, {Nidever}, {Weaver}, {Adean}, {Correia}, {de Freitas}, {Freitas}, {Singulani}, {Vila-Verde}, \& {Linea Science Server}}]{Abb21}
{Abbott}, T.~M.~C., {Adam{\'o}w}, M., {Aguena}, M., {et~al.} 2021, \apjs, 255, 20

\bibitem[{{Ade} {et~al.}(2019){Ade}, {Aguirre}, {Ahmed}, {Aiola}, {Ali}, {Alonso}, {Alvarez}, {Arnold}, {Ashton}, {Austermann}, {Awan}, {Baccigalupi}, {Baildon}, {Barron}, {Battaglia}, {Battye}, {Baxter}, {Bazarko}, {Beall}, {Bean}, {Beck}, {Beckman}, {Beringue}, {Bianchini}, {Boada}, {Boettger}, {Bond}, {Borrill}, {Brown}, {Bruno}, {Bryan}, {Calabrese}, {Calafut}, {Calisse}, {Carron}, {Challinor}, {Chesmore}, {Chinone}, {Chluba}, {Cho}, {Choi}, {Coppi}, {Cothard}, {Coughlin}, {Crichton}, {Crowley}, {Crowley}, {Cukierman}, {D'Ewart}, {D{\"u}nner}, {de Haan}, {Devlin}, {Dicker}, {Didier}, {Dobbs}, {Dober}, {Duell}, {Duff}, {Duivenvoorden}, {Dunkley}, {Dusatko}, {Errard}, {Fabbian}, {Feeney}, {Ferraro}, {Flux{\`a}}, {Freese}, {Frisch}, {Frolov}, {Fuller}, {Fuzia}, {Galitzki}, {Gallardo}, {Tomas Galvez Ghersi}, {Gao}, {Gawiser}, {Gerbino}, {Gluscevic}, {Goeckner-Wald}, {Golec}, {Gordon}, {Gralla}, {Green}, {Grigorian}, {Groh}, {Groppi}, {Guan}, {Gudmundsson}, {Han}, {Hargrave}, {Hasegawa}, {Hasselfield},
  {Hattori}, {Haynes}, {Hazumi}, {He}, {Healy}, {Henderson}, {Hervias-Caimapo}, {Hill}, {Hill}, {Hilton}, {Hilton}, {Hincks}, {Hinshaw}, {Hlo{\v{z}}ek}, {Ho}, {Ho}, {Howe}, {Huang}, {Hubmayr}, {Huffenberger}, {Hughes}, {Ijjas}, {Ikape}, {Irwin}, {Jaffe}, {Jain}, {Jeong}, {Kaneko}, {Karpel}, {Katayama}, {Keating}, {Kernasovskiy}, {Keskitalo}, {Kisner}, {Kiuchi}, {Klein}, {Knowles}, {Koopman}, {Kosowsky}, {Krachmalnicoff}, {Kuenstner}, {Kuo}, {Kusaka}, {Lashner}, {Lee}, {Lee}, {Leon}, {Leung}, {Lewis}, {Li}, {Li}, {Limon}, {Linder}, {Lopez-Caraballo}, {Louis}, {Lowry}, {Lungu}, {Madhavacheril}, {Mak}, {Maldonado}, {Mani}, {Mates}, {Matsuda}, {Maurin}, {Mauskopf}, {May}, {McCallum}, {McKenney}, {McMahon}, {Meerburg}, {Meyers}, {Miller}, {Mirmelstein}, {Moodley}, {Munchmeyer}, {Munson}, {Naess}, {Nati}, {Navaroli}, {Newburgh}, {Nguyen}, {Niemack}, {Nishino}, {Orlowski-Scherer}, {Page}, {Partridge}, {Peloton}, {Perrotta}, {Piccirillo}, {Pisano}, {Poletti}, {Puddu}, {Puglisi}, {Raum}, {Reichardt}, {Remazeilles},
  {Rephaeli}, {Riechers}, {Rojas}, {Roy}, {Sadeh}, {Sakurai}, {Salatino}, {Sathyanarayana Rao}, {Schaan}, {Schmittfull}, {Sehgal}, {Seibert}, {Seljak}, {Sherwin}, {Shimon}, {Sierra}, {Sievers}, {Sikhosana}, {Silva-Feaver}, {Simon}, {Sinclair}, {Siritanasak}, {Smith}, {Smith}, {Spergel}, {Staggs}, {Stein}, {Stevens}, {Stompor}, {Suzuki}, {Tajima}, {Takakura}, {Teply}, {Thomas}, {Thorne}, {Thornton}, {Trac}, {Tsai}, {Tucker}, {Ullom}, {Vagnozzi}, {van Engelen}, {Van Lanen}, {Van Winkle}, {Vavagiakis}, {Verg{\`e}s}, {Vissers}, {Wagoner}, {Walker}, {Ward}, {Westbrook}, {Whitehorn}, {Williams}, {Williams}, {Wollack}, {Xu}, {Yu}, {Yu}, {Zago}, {Zhang}, {Zhu}, \& {Simons Observatory Collaboration}}]{2019JCAP...02..056A}
{Ade}, P., {Aguirre}, J., {Ahmed}, Z., {et~al.} 2019, \jcap, 2019, 056

\bibitem[{{Aguena} {et~al.}(2021){Aguena}, {Benoist}, {da Costa}, {Ogando}, {Gschwend}, {Sampaio-Santos}, {Lima}, {Maia}, {Allam}, {Avila}, {Bacon}, {Bertin}, {Bhargava}, {Brooks}, {Carnero Rosell}, {Carrasco Kind}, {Carretero}, {Costanzi}, {De Vicente}, {Desai}, {Diehl}, {Doel}, {Everett}, {Evrard}, {Ferrero}, {Fert{\'e}}, {Flaugher}, {Fosalba}, {Frieman}, {Garc{\'\i}a-Bellido}, {Giles}, {Gruendl}, {Gutierrez}, {Hinton}, {Hollowood}, {Honscheid}, {James}, {Jeltema}, {Kuehn}, {Kuropatkin}, {Lahav}, {Melchior}, {Miquel}, {Morgan}, {Palmese}, {Paz-Chinch{\'o}n}, {Plazas}, {Romer}, {Sanchez}, {Santiago}, {Schubnell}, {Serrano}, {Sevilla-Noarbe}, {Smith}, {Soares-Santos}, {Suchyta}, {Tarle}, {To}, {Tucker}, \& {Wilkinson}}]{Aguena21}
{Aguena}, M., {Benoist}, C., {da Costa}, L.~N., {et~al.} 2021, \mnras, 502, 4435

\bibitem[{Aguena \& Lima(2018)}]{AguLim18}
Aguena, M. \& Lima, M. 2018, Phys. Rev. D, 98, 123529

\bibitem[{{Ahumada} {et~al.}(2020){Ahumada}, {Allende Prieto}, {Almeida}, {Anders}, {Anderson}, {Andrews}, {Anguiano}, {Arcodia}, {Armengaud}, {Aubert}, {Avila}, {Avila-Reese}, {Badenes}, {Balland}, {Barger}, {Barrera-Ballesteros}, {Basu}, {Bautista}, {Beaton}, {Beers}, {Benavides}, {Bender}, {Bernardi}, {Bershady}, {Beutler}, {Bidin}, {Bird}, {Bizyaev}, {Blanc}, {Blanton}, {Boquien}, {Borissova}, {Bovy}, {Brandt}, {Brinkmann}, {Brownstein}, {Bundy}, {Bureau}, {Burgasser}, {Burtin}, {Cano-D{\'\i}az}, {Capasso}, {Cappellari}, {Carrera}, {Chabanier}, {Chaplin}, {Chapman}, {Cherinka}, {Chiappini}, {Doohyun Choi}, {Chojnowski}, {Chung}, {Clerc}, {Coffey}, {Comerford}, {Comparat}, {da Costa}, {Cousinou}, {Covey}, {Crane}, {Cunha}, {Ilha}, {Dai}, {Damsted}, {Darling}, {Davidson}, {Davies}, {Dawson}, {De}, {de la Macorra}, {De Lee}, {Queiroz}, {Deconto Machado}, {de la Torre}, {Dell'Agli}, {du Mas des Bourboux}, {Diamond-Stanic}, {Dillon}, {Donor}, {Drory}, {Duckworth}, {Dwelly}, {Ebelke}, {Eftekharzadeh}, {Davis
  Eigenbrot}, {Elsworth}, {Eracleous}, {Erfanianfar}, {Escoffier}, {Fan}, {Farr}, {Fern{\'a}ndez-Trincado}, {Feuillet}, {Finoguenov}, {Fofie}, {Fraser-McKelvie}, {Frinchaboy}, {Fromenteau}, {Fu}, {Galbany}, {Garcia}, {Garc{\'\i}a-Hern{\'a}ndez}, {Garma Oehmichen}, {Ge}, {Geimba Maia}, {Geisler}, {Gelfand}, {Goddy}, {Gonzalez-Perez}, {Grabowski}, {Green}, {Grier}, {Guo}, {Guy}, {Harding}, {Hasselquist}, {Hawken}, {Hayes}, {Hearty}, {Hekker}, {Hogg}, {Holtzman}, {Horta}, {Hou}, {Hsieh}, {Huber}, {Hunt}, {Ider Chitham}, {Imig}, {Jaber}, {Jimenez Angel}, {Johnson}, {Jones}, {J{\"o}nsson}, {Jullo}, {Kim}, {Kinemuchi}, {Kirkpatrick}, {Kite}, {Klaene}, {Kneib}, {Kollmeier}, {Kong}, {Kounkel}, {Krishnarao}, {Lacerna}, {Lan}, {Lane}, {Law}, {Le Goff}, {Leung}, {Lewis}, {Li}, {Lian}, {Lin}, {Long}, {Longa-Pe{\~n}a}, {Lundgren}, {Lyke}, {Mackereth}, {MacLeod}, {Majewski}, {Manchado}, {Maraston}, {Martini}, {Masseron}, {Masters}, {Mathur}, {McDermid}, {Merloni}, {Merrifield}, {M{\'e}sz{\'a}ros}, {Miglio}, {Minniti},
  {Minsley}, {Miyaji}, {Mohammad}, {Mosser}, {Mueller}, {Muna}, {Mu{\~n}oz-Guti{\'e}rrez}, {Myers}, {Nadathur}, {Nair}, {Nandra}, {Correa do Nascimento}, {Nevin}, {Newman}, {Nidever}, {Nitschelm}, {Noterdaeme}, {O'Connell}, {Olmstead}, {Oravetz}, {Oravetz}, {Osorio}, {Pace}, {Padilla}, {Palanque-Delabrouille}, {Palicio}, {Pan}, {Pan}, {Parker}, {Paviot}, {Peirani}, {Ram{\'r}ez}, {Penny}, {Percival}, {Perez-Fournon}, {P{\'e}rez-R{\`a}fols}, {Petitjean}, {Pieri}, {Pinsonneault}, {Poovelil}, {Povick}, {Prakash}, {Price-Whelan}, {Raddick}, {Raichoor}, {Ray}, {Rembold}, {Rezaie}, {Riffel}, {Riffel}, {Rix}, {Robin}, {Roman-Lopes}, {Rom{\'a}n-Z{\'u}{\~n}iga}, {Rose}, {Ross}, {Rossi}, {Rowlands}, {Rubin}, {Salvato}, {S{\'a}nchez}, {S{\'a}nchez-Menguiano}, {S{\'a}nchez-Gallego}, {Sayres}, {Schaefer}, {Schiavon}, {Schimoia}, {Schlafly}, {Schlegel}, {Schneider}, {Schultheis}, {Schwope}, {Seo}, {Serenelli}, {Shafieloo}, {Shamsi}, {Shao}, {Shen}, {Shetrone}, {Shirley}, {Silva Aguirre}, {Simon}, {Skrutskie}, {Slosar},
  {Smethurst}, {Sobeck}, {Sodi}, {Souto}, {Stark}, {Stassun}, {Steinmetz}, {Stello}, {Stermer}, {Storchi-Bergmann}, {Streblyanska}, {Stringfellow}, {Stutz}, {Su{\'a}rez}, {Sun}, {Taghizadeh-Popp}, {Talbot}, {Tayar}, {Thakar}, {Theriault}, {Thomas}, {Thomas}, {Tinker}, {Tojeiro}, {Toledo}, {Tremonti}, {Troup}, {Tuttle}, {Unda-Sanzana}, {Valentini}, {Vargas-Gonz{\'a}lez}, {Vargas-Maga{\~n}a}, {V{\'a}zquez-Mata}, {Vivek}, {Wake}, {Wang}, {Weaver}, {Weijmans}, {Wild}, {Wilson}, {Wilson}, {Wolthuis}, {Wood-Vasey}, {Yan}, {Yang}, {Y{\`e}che}, {Zamora}, {Zarrouk}, {Zasowski}, {Zhang}, {Zhao}, {Zhao}, {Zheng}, {Zheng}, {Zhu}, \& {Zou}}]{SDSSdr16}
{Ahumada}, R., {Allende Prieto}, C., {Almeida}, A., {et~al.} 2020, \apjs, 249, 3

\bibitem[{{Allen} {et~al.}(2011){Allen}, {Evrard}, \& {Mantz}}]{Allen11}
{Allen}, S.~W., {Evrard}, A.~E., \& {Mantz}, A.~B. 2011, \araa, 49, 409

\bibitem[{{Ascaso} {et~al.}(2017){Ascaso}, {Mei}, {Bartlett}, \& {Ben{\'\i}tez}}]{Asc16}
{Ascaso}, B., {Mei}, S., {Bartlett}, J.~G., \& {Ben{\'\i}tez}, N. 2017, \mnras, 464, 2270

\bibitem[{{Bechtol} {et~al.}(2025){Bechtol}, {Sevilla-Noarbe}, {Drlica-Wagner}, {Yanny}, {Gruendl}, {Sheldon}, {Rykoff}, {De Vicente}, {Adamow}, {Anbajagane}, {Becker}, {Bernstein}, {Carnero Rosell}, {Gschwend}, {Gorsuch}, {Hartley}, {Jarvis}, {Jeltema}, {Kron}, {Manning}, {O'Donnell}, {Pieres}, {Rodr{\'\i}guez-Monroy}, {Sanchez Cid}, {Tabbutt}, {Toribio San Cipriano}, {Tucker}, {Weaverdyck}, {Yamamoto}, {Abbott}, {Aguena}, {Alarc{\'o}n}, {Allam}, {Amon}, {Andrade-Oliveira}, {Avila}, {Bernardinelli}, {Bertin}, {Blazek}, {Brooks}, {Burke}, {Carretero}, {Castander}, {Cawthon}, {Chang}, {Choi}, {Conselice}, {Costanzi}, {Crocce}, {da Costa}, {Davis}, {Desai}, {Diehl}, {Dodelson}, {Doel}, {Doux}, {Fert{\'e}}, {Flaugher}, {Fosalba}, {Frieman}, {Garc{\'\i}a-Bellido}, {Gatti}, {Gaztanaga}, {Giannini}, {Gruen}, {Gutierrez}, {Herner}, {Hinton}, {Hollowood}, {Honscheid}, {Huterer}, {Jeffrey}, {Krause}, {Kuehn}, {Lahav}, {Lee}, {Lidman}, {Lima}, {Lin}, {Marshall}, {Mena-Fern{\'a}ndez}, {Miquel}, {Mohr}, {Muir}, {Myles},
  {Ogando}, {Palmese}, {Plazas Malag{\'o}n}, {Porredon}, {Prat}, {Raveri}, {Romer}, {Roodman}, {Samuroff}, {Sanchez}, {Scarpine}, {Smith}, {Soares-Santos}, {Suchyta}, {Tarle}, {Troxel}, {Vikram}, {Walker}, {Weller}, {Wiseman}, \& {Zhang}}]{DESy6gold}
{Bechtol}, K., {Sevilla-Noarbe}, I., {Drlica-Wagner}, A., {et~al.} 2025, arXiv e-prints, arXiv:2501.05739

\bibitem[{{Bellagamba} {et~al.}(2018){Bellagamba}, {Roncarelli}, {Maturi}, \& {Moscardini}}]{bellagamba2018}
{Bellagamba}, F., {Roncarelli}, M., {Maturi}, M., \& {Moscardini}, L. 2018, \mnras, 473, 5221

\bibitem[{{Benson} {et~al.}(2014){Benson}, {Ade}, {Ahmed}, {Allen}, {Arnold}, {Austermann}, {Bender}, {Bleem}, {Carlstrom}, {Chang}, {Cho}, {Cliche}, {Crawford}, {Cukierman}, {de Haan}, {Dobbs}, {Dutcher}, {Everett}, {Gilbert}, {Halverson}, {Hanson}, {Harrington}, {Hattori}, {Henning}, {Hilton}, {Holder}, {Holzapfel}, {Irwin}, {Keisler}, {Knox}, {Kubik}, {Kuo}, {Lee}, {Leitch}, {Li}, {McDonald}, {Meyer}, {Montgomery}, {Myers}, {Natoli}, {Nguyen}, {Novosad}, {Padin}, {Pan}, {Pearson}, {Reichardt}, {Ruhl}, {Saliwanchik}, {Simard}, {Smecher}, {Sayre}, {Shirokoff}, {Stark}, {Story}, {Suzuki}, {Thompson}, {Tucker}, {Vanderlinde}, {Vieira}, {Vikhlinin}, {Wang}, {Yefremenko}, \& {Yoon}}]{2014SPIE.9153E..1PB}
{Benson}, B.~A., {Ade}, P.~A.~R., {Ahmed}, Z., {et~al.} 2014, 9153, 91531P

\bibitem[{{Bleem} {et~al.}(2020){Bleem}, {Bocquet}, {Stalder}, {Gladders}, {Ade}, {Allen}, {Anderson}, {Annis}, {Ashby}, {Austermann}, {Avila}, {Avva}, {Bayliss}, {Beall}, {Bechtol}, {Bender}, {Benson}, {Bertin}, {Bianchini}, {Blake}, {Brodwin}, {Brooks}, {Buckley-Geer}, {Burke}, {Carlstrom}, {Rosell}, {Carrasco Kind}, {Carretero}, {Chang}, {Chiang}, {Citron}, {Moran}, {Costanzi}, {Crawford}, {Crites}, {da Costa}, {de Haan}, {De Vicente}, {Desai}, {Diehl}, {Dietrich}, {Dobbs}, {Eifler}, {Everett}, {Flaugher}, {Floyd}, {Frieman}, {Gallicchio}, {Garc{\'\i}a-Bellido}, {George}, {Gerdes}, {Gilbert}, {Gruen}, {Gruendl}, {Gschwend}, {Gupta}, {Gutierrez}, {Halverson}, {Harrington}, {Henning}, {Heymans}, {Holder}, {Hollowood}, {Holzapfel}, {Honscheid}, {Hrubes}, {Huang}, {Hubmayr}, {Irwin}, {James}, {Jeltema}, {Joudaki}, {Khullar}, {Klein}, {Knox}, {Kuropatkin}, {Lee}, {Li}, {Lidman}, {Lowitz}, {MacCrann}, {Mahler}, {Maia}, {Marshall}, {McDonald}, {McMahon}, {Melchior}, {Menanteau}, {Meyer}, {Miquel}, {Mocanu},
  {Mohr}, {Montgomery}, {Nadolski}, {Natoli}, {Nibarger}, {Noble}, {Novosad}, {Padin}, {Palmese}, {Parkinson}, {Patil}, {Paz-Chinch{\'o}n}, {Plazas}, {Pryke}, {Ramachandra}, {Reichardt}, {Remolina Gonz{\'a}lez}, {Romer}, {Roodman}, {Ruhl}, {Rykoff}, {Saliwanchik}, {Sanchez}, {Saro}, {Sayre}, {Schaffer}, {Schrabback}, {Serrano}, {Sharon}, {Sievers}, {Smecher}, {Smith}, {Soares-Santos}, {Stark}, {Story}, {Suchyta}, {Tarle}, {Tucker}, {Vanderlinde}, {Veach}, {Vieira}, {Wang}, {Weller}, {Whitehorn}, {Wu}, {Yefremenko}, \& {Zhang}}]{Ble20}
{Bleem}, L.~E., {Bocquet}, S., {Stalder}, B., {et~al.} 2020, \apjs, 247, 25

\bibitem[{{Bleem} {et~al.}(2015){Bleem}, {Stalder}, {de Haan}, {Aird}, {Allen}, {Applegate}, {Ashby}, {Bautz}, {Bayliss}, {Benson}, {Bocquet}, {Brodwin}, {Carlstrom}, {Chang}, {Chiu}, {Cho}, {Clocchiatti}, {Crawford}, {Crites}, {Desai}, {Dietrich}, {Dobbs}, {Foley}, {Forman}, {George}, {Gladders}, {Gonzalez}, {Halverson}, {Hennig}, {Hoekstra}, {Holder}, {Holzapfel}, {Hrubes}, {Jones}, {Keisler}, {Knox}, {Lee}, {Leitch}, {Liu}, {Lueker}, {Luong-Van}, {Mantz}, {Marrone}, {McDonald}, {McMahon}, {Meyer}, {Mocanu}, {Mohr}, {Murray}, {Padin}, {Pryke}, {Reichardt}, {Rest}, {Ruel}, {Ruhl}, {Saliwanchik}, {Saro}, {Sayre}, {Schaffer}, {Schrabback}, {Shirokoff}, {Song}, {Spieler}, {Stanford}, {Staniszewski}, {Stark}, {Story}, {Stubbs}, {Vanderlinde}, {Vieira}, {Vikhlinin}, {Williamson}, {Zahn}, \& {Zenteno}}]{Ble15}
{Bleem}, L.~E., {Stalder}, B., {de Haan}, T., {et~al.} 2015, \apjs, 216, 27

\bibitem[{{Bocquet} {et~al.}(2019){Bocquet}, {Dietrich}, {Schrabback}, {Bleem}, {Klein}, {Allen}, {Applegate}, {Ashby}, {Bautz}, {Bayliss}, {Benson}, {Brodwin}, {Bulbul}, {Canning}, {Capasso}, {Carlstrom}, {Chang}, {Chiu}, {Cho}, {Clocchiatti}, {Crawford}, {Crites}, {de Haan}, {Desai}, {Dobbs}, {Foley}, {Forman}, {Garmire}, {George}, {Gladders}, {Gonzalez}, {Grandis}, {Gupta}, {Halverson}, {Hlavacek-Larrondo}, {Hoekstra}, {Holder}, {Holzapfel}, {Hou}, {Hrubes}, {Huang}, {Jones}, {Khullar}, {Knox}, {Kraft}, {Lee}, {von der Linden}, {Luong-Van}, {Mantz}, {Marrone}, {McDonald}, {McMahon}, {Meyer}, {Mocanu}, {Mohr}, {Morris}, {Padin}, {Patil}, {Pryke}, {Rapetti}, {Reichardt}, {Rest}, {Ruhl}, {Saliwanchik}, {Saro}, {Sayre}, {Schaffer}, {Shirokoff}, {Stalder}, {Stanford}, {Staniszewski}, {Stark}, {Story}, {Strazzullo}, {Stubbs}, {Vanderlinde}, {Vieira}, {Vikhlinin}, {Williamson}, \& {Zenteno}}]{BOCQUET}
{Bocquet}, S., {Dietrich}, J.~P., {Schrabback}, T., {et~al.} 2019, \apj, 878, 55

\bibitem[{{Bocquet} {et~al.}(2024){Bocquet}, {Grandis}, {Bleem}, {Klein}, {Mohr}, {Schrabback}, {Abbott}, {Ade}, {Aguena}, {Alarcon}, {Allam}, {Allen}, {Alves}, {Amon}, {Anderson}, {Annis}, {Ansarinejad}, {Austermann}, {Avila}, {Bacon}, {Bayliss}, {Beall}, {Bechtol}, {Becker}, {Bender}, {Benson}, {Bernstein}, {Bhargava}, {Bianchini}, {Brodwin}, {Brooks}, {Bryant}, {Campos}, {Canning}, {Carlstrom}, {Carnero Rosell}, {Carrasco Kind}, {Carretero}, {Castander}, {Cawthon}, {Chang}, {Chang}, {Chaubal}, {Chen}, {Chiang}, {Choi}, {Chou}, {Citron}, {Corbett Moran}, {Cordero}, {Costanzi}, {Crawford}, {Crites}, {da Costa}, {Pereira}, {Davis}, {Davis}, {DeRose}, {Desai}, {de Haan}, {Diehl}, {Dobbs}, {Dodelson}, {Doux}, {Drlica-Wagner}, {Eckert}, {Elvin-Poole}, {Everett}, {Everett}, {Ferrero}, {Fert{\'e}}, {Flores}, {Frieman}, {Gallicchio}, {Garc{\'\i}a-Bellido}, {Gatti}, {George}, {Giannini}, {Gladders}, {Gruen}, {Gruendl}, {Gupta}, {Gutierrez}, {Halverson}, {Harrison}, {Hartley}, {Herner}, {Hinton}, {Holder},
  {Hollowood}, {Holzapfel}, {Honscheid}, {Hrubes}, {Huang}, {Hubmayr}, {Huff}, {Huterer}, {Irwin}, {James}, {Jarvis}, {Khullar}, {Kim}, {Knox}, {Kraft}, {Krause}, {Kuehn}, {Kuropatkin}, {K{\'e}ruzor{\'e}}, {Lahav}, {Lee}, {Leget}, {Li}, {Lin}, {Lowitz}, {MacCrann}, {Mahler}, {Mantz}, {Marshall}, {McCullough}, {McDonald}, {McMahon}, {Mena-Fern{\'a}ndez}, {Menanteau}, {Meyer}, {Miquel}, {Montgomery}, {Myles}, {Natoli}, {Navarro-Alsina}, {Nibarger}, {Noble}, {Novosad}, {Ogando}, {Omori}, {Padin}, {Pandey}, {Paschos}, {Patil}, {Pieres}, {Plazas Malag{\'o}n}, {Porredon}, {Prat}, {Pryke}, {Raveri}, {Reichardt}, {Roberson}, {Rollins}, {Romero}, {Roodman}, {Ruhl}, {Rykoff}, {Saliwanchik}, {Salvati}, {S{\'a}nchez}, {Sanchez}, {Sanchez Cid}, {Saro}, {Schaffer}, {Secco}, {Sevilla-Noarbe}, {Sharon}, {Sheldon}, {Shin}, {Sievers}, {Smecher}, {Smith}, {Somboonpanyakul}, {Sommer}, {Stalder}, {Stark}, {Stephen}, {Strazzullo}, {Suchyta}, {Tarle}, {To}, {Troxel}, {Tucker}, {Tutusaus}, {Varga}, {Veach}, {Vieira}, {Vikhlinin},
  {von der Linden}, {Wang}, {Weaverdyck}, {Weller}, {Whitehorn}, {Wu}, {Yanny}, {Yefremenko}, {Yin}, {Young}, {Zebrowski}, {Zhang}, {Zohren}, {Zuntz}, {(SPT}, \& {DES Collaborations)}}]{2024PhRvD.110h3510B}
{Bocquet}, S., {Grandis}, S., {Bleem}, L.~E., {et~al.} 2024, \prd, 110, 083510

\bibitem[{{B{\"o}hringer} \& {Chon}(2016)}]{Bohringer16}
{B{\"o}hringer}, H. \& {Chon}, G. 2016, Modern Physics Letters A, 31, 1640008

\bibitem[{{Bolliet} {et~al.}(2020){Bolliet}, {Brinckmann}, {Chluba}, \& {Lesgourgues}}]{Bolliet20}
{Bolliet}, B., {Brinckmann}, T., {Chluba}, J., \& {Lesgourgues}, J. 2020, \mnras, 497, 1332

\bibitem[{{Carilli} \& {Taylor}(2002)}]{Carillii02}
{Carilli}, C.~L. \& {Taylor}, G.~B. 2002, \araa, 40, 319

\bibitem[{{Carlstrom} {et~al.}(2011){Carlstrom}, {Ade}, {Aird}, {et~al.}}]{Car11}
{Carlstrom}, J.~E., {Ade}, P.~A.~R., {Aird}, K.~A., {et~al.} 2011, \pasp, 123, 568

\bibitem[{{Castignani} \& {Benoist}(2016)}]{2016A&A...595A.111C}
{Castignani}, G. \& {Benoist}, C. 2016, \aap, 595, A111

\bibitem[{{Costanzi} {et~al.}(2021){Costanzi}, {Saro}, {Bocquet}, {Abbott}, {Aguena}, {Allam}, {Amara}, {Annis}, {Avila}, {Bacon}, {Benson}, {Bhargava}, {Brooks}, {Buckley-Geer}, {Burke}, {Carnero Rosell}, {Carrasco Kind}, {Carretero}, {Choi}, {da Costa}, {Pereira}, {De Vicente}, {Desai}, {Diehl}, {Dietrich}, {Doel}, {Eifler}, {Everett}, {Ferrero}, {Fert{\'e}}, {Flaugher}, {Fosalba}, {Frieman}, {Garc{\'\i}a-Bellido}, {Gaztanaga}, {Gerdes}, {Giannantonio}, {Giles}, {Grandis}, {Gruen}, {Gruendl}, {Gupta}, {Gutierrez}, {Hartley}, {Hinton}, {Hollowood}, {Honscheid}, {James}, {Jeltema}, {Krause}, {Kuehn}, {Kuropatkin}, {Lahav}, {Lima}, {MacCrann}, {Maia}, {Marshall}, {Menanteau}, {Miquel}, {Mohr}, {Morgan}, {Myles}, {Ogando}, {Palmese}, {Paz-Chinch{\'o}n}, {Plazas}, {Rapetti}, {Reichardt}, {Romer}, {Roodman}, {Ruppin}, {Salvati}, {Samuroff}, {Sanchez}, {Scarpine}, {Serrano}, {Sevilla-Noarbe}, {Singh}, {Smith}, {Soares-Santos}, {Stark}, {Suchyta}, {Swanson}, {Tarle}, {Thomas}, {To}, {Tucker}, {Varga}, {Wechsler},
  {Zhang}, {DES}, \& {SPT Collaborations}}]{Cos22}
{Costanzi}, M., {Saro}, A., {Bocquet}, S., {et~al.} 2021, \prd, 103, 043522

\bibitem[{{{Costanzi}, M. \& {Rozo}, E.} {et~al.}(2019){{Costanzi}, M. \& {Rozo}, E.}, {Rykoff}, {Farahi}, {Jeltema}, {Evrard}, {Mantz}, {Gruen}, {Mandelbaum}, {DeRose}, {McClintock}, {Varga}, {Zhang}, {Weller}, {Wechsler}, \& {Aguena}}]{cos19}
{{Costanzi}, M. \& {Rozo}, E.}, {Rykoff}, E.~S., {Farahi}, A., {et~al.} 2019, \mnras, 482, 490

\bibitem[{{De Vicente} {et~al.}(2016){De Vicente}, {S{\'a}nchez}, \& {Sevilla-Noarbe}}]{deV16}
{De Vicente}, J., {S{\'a}nchez}, E., \& {Sevilla-Noarbe}, I. 2016, \mnras, 459, 3078

\bibitem[{{DeRose} {et~al.}(2019){DeRose}, {Wechsler}, {Becker}, {Busha}, {Rykoff}, {MacCrann}, {Erickson}, {Evrard}, {Kravtsov}, {Gruen}, {Allam}, {Avila}, {Bridle}, {Brooks}, {Buckley-Geer}, {Carnero Rosell}, {Carrasco Kind}, {Carretero}, {Castander}, {Cawthon}, {Crocce}, {da Costa}, {Davis}, {De Vicente}, {Dietrich}, {Doel}, {Drlica-Wagner}, {Fosalba}, {Frieman}, {Garcia-Bellido}, {Gutierrez}, {Hartley}, {Hollowood}, {Hoyle}, {James}, {Krause}, {Kuehn}, {Kuropatkin}, {Lima}, {Maia}, {Menanteau}, {Miller}, {Miquel}, {Ogand o}, {Plazas Malag{\'o}n}, {Romer}, {Sanchez}, {Schindler}, {Serrano}, {Sevilla-Noarbe}, {Smith}, {Suchyta}, {Swanson}, {Tarle}, \& {Vikram}}]{derose19}
{DeRose}, J., {Wechsler}, R.~H., {Becker}, M.~R., {et~al.} 2019, arXiv e-prints, arXiv:1901.02401

\bibitem[{{DES Collaboration} {et~al.}(2020){DES Collaboration}, {Abbott}, {Aguena}, {Alarcon}, {Allam}, {Allen}, {Annis}, {Avila}, {Bacon}, {Bechtol}, {Bermeo}, {Bernstein}, {Bertin}, {Bhargava}, {Bocquet}, {Brooks}, {Brout}, {Buckley-Geer}, {Burke}, {Carnero Rosell}, {Carrasco Kind}, {Carretero}, {Castander}, {Cawthon}, {Chang}, {Chen}, {Choi}, {Costanzi}, {Crocce}, {da Costa}, {Davis}, {De Vicente}, {DeRose}, {Desai}, {Diehl}, {Dietrich}, {Dodelson}, {Doel}, {Drlica-Wagner}, {Eckert}, {Eifler}, {Elvin-Poole}, {Estrada}, {Everett}, {Evrard}, {Farahi}, {Ferrero}, {Flaugher}, {Fosalba}, {Frieman}, {Garc{\'\i}a-Bellido}, {Gatti}, {Gaztanaga}, {Gerdes}, {Giannantonio}, {Giles}, {Grandis}, {Gruen}, {Gruendl}, {Gschwend}, {Gutierrez}, {Hartley}, {Hinton}, {Hollowood}, {Honscheid}, {Hoyle}, {Huterer}, {James}, {Jarvis}, {Jeltema}, {Johnson}, {Johnson}, {Kent}, {Krause}, {Kron}, {Kuehn}, {Kuropatkin}, {Lahav}, {Li}, {Lidman}, {Lima}, {Lin}, {MacCrann}, {Maia}, {Mantz}, {Marshall}, {Martini}, {Mayers}, {Melchior},
  {Mena-Fern{\'a}ndez}, {Menanteau}, {Miquel}, {Mohr}, {Nichol}, {Nord}, {Ogando}, {Palmese}, {Paz-Chinch{\'o}n}, {Plazas}, {Prat}, {Rau}, {Romer}, {Roodman}, {Rooney}, {Rozo}, {Rykoff}, {Sako}, {Samuroff}, {S{\'a}nchez}, {Sanchez}, {Saro}, {Scarpine}, {Schubnell}, {Scolnic}, {Serrano}, {Sevilla-Noarbe}, {Sheldon}, {Smith}, {Smith}, {Suchyta}, {Swanson}, {Tarle}, {Thomas}, {To}, {Troxel}, {Tucker}, {Varga}, {von der Linden}, {Walker}, {Wechsler}, {Weller}, {Wilkinson}, {Wu}, {Yanny}, {Zhang}, {Zhang}, {Zuntz}, \& {DES Collaboration}}]{DES20}
{DES Collaboration}, {Abbott}, T.~M.~C., {Aguena}, M., {et~al.} 2020, \prd, 102, 023509

\bibitem[{{Diehl} {et~al.}(2016){Diehl}, {Neilsen}, {Gruendl}, {Yanny}, {Abbott}, {Aleksi{\'c}}, {Allam}, {Annis}, {Balbinot}, {Baumer}, {Beaufore}, {Bechtol}, {Bernstein}, {Birrer}, {Bonnett}, {Brout}, {Bruderer}, {Buckley-Geer}, {Capozzi}, {Carnero Rosell}, {Castander}, {Cawthon}, {Chang}, {Clerkin}, {Covarrubias}, {Cuhna}, {D'Andrea}, {da Costa}, {Das}, {Davis}, {Dietrich}, {Drlica-Wagner}, {Elliott}, {Eifler}, {Etherington}, {Flaugher}, {Frieman}, {Fausti Neto}, {Fern{\'a}ndez}, {Furlanetto}, {Gangkofner}, {Gerdes}, {Goldstein}, {Grabowski}, {Gupta}, {Hamilton}, {Head}, {Helsby}, {Hollowood}, {Honscheid}, {James}, {Johnson}, {Johnson}, {Jouvel}, {Kacprzac}, {Kent}, {Kessler}, {Kim}, {Krause}, {Krawiec}, {Kremin}, {Kron}, {Kuhlmann}, {Kuropatkin}, {Lahav}, {Lasker}, {Li}, {Luque}, {Maccrann}, {March}, {Marshall}, {Mondrik}, {Morganson}, {Mudd}, {Nadolski}, {Nugent}, {Melchior}, {Menanteau}, {Nagasawa}, {Nord}, {Ogando}, {Old}, {Palmese}, {Petravick}, {Plazas}, {Pujol}, {Queiroz}, {Reil}, {Romer},
  {Rosenfeld}, {Roodman}, {Rooney}, {Sako}, {Salvador}, {S{\'a}nchez}, {S{\'a}nchez {\'A}lvaro}, {Santiago}, {Schooneveld}, {Schubnell}, {Sheldon}, {Smith}, {Smith}, {Soares-Santos}, {Sobreira}, {Soumagnac}, {Spinka}, {Tie}, {Tucker}, {Vikram}, {Vivas}, {Walker}, {Wester}, {Wiesner}, {Wilcox}, {Williams}, {Zenteno}, {Zhang}, \& {Zhang}}]{Die16}
{Diehl}, H.~T., {Neilsen}, E., {Gruendl}, R., {et~al.} 2016, in Observatory Operations: Strategies, Processes, and Systems VI, Vol. 9910, 99101D

\bibitem[{{Euclid Collaboration} {et~al.}(2019){Euclid Collaboration}, {Adam}, {Vannier}, {Maurogordato}, {Biviano}, {Adami}, {Ascaso}, {Bellagamba}, {Benoist}, {Cappi}, {D{\'\i}az-S{\'a}nchez}, {Durret}, {Farrens}, {Gonzalez}, {Iovino}, {Licitra}, {Maturi}, {Mei}, {Merson}, {Munari}, {Pell{\'o}}, {Ricci}, {Rocci}, {Roncarelli}, {Sarron}, {Amoura}, {Andreon}, {Apostolakos}, {Arnaud}, {Bardelli}, {Bartlett}, {Baugh}, {Borgani}, {Brodwin}, {Castander}, {Castignani}, {Cucciati}, {De Lucia}, {Dubath}, {Fosalba}, {Giocoli}, {Hoekstra}, {Mamon}, {Melin}, {Moscardini}, {Paltani}, {Radovich}, {Sartoris}, {Schultheis}, {Sereno}, {Weller}, {Burigana}, {Carvalho}, {Corcione}, {Kurki-Suonio}, {Lilje}, {Sirri}, {Toledo-Moreo}, \& {Zamorani}}]{Adam2019}
{Euclid Collaboration}, {Adam}, R., {Vannier}, M., {et~al.} 2019, \aap, 627, A23

\bibitem[{{Euclid Collaboration} {et~al.}(2024){Euclid Collaboration}, {Mellier}, {Abdurro'uf}, {Acevedo Barroso}, {Ach{\'u}carro}, {Adamek}, {Adam}, {Addison}, {Aghanim}, {Aguena}, {Ajani}, {Akrami}, {Al-Bahlawan}, {Alavi}, {Albuquerque}, {Alestas}, {Alguero}, {Allaoui}, {Allen}, {Allevato}, {Alonso-Tetilla}, {Altieri}, {Alvarez-Candal}, {Alvi}, {Amara}, {Amendola}, {Amiaux}, {Andika}, {Andreon}, {Andrews}, {Angora}, {Angulo}, {Annibali}, {Anselmi}, {Anselmi}, {Arcari}, {Archidiacono}, {Aric{\`o}}, {Arnaud}, {Arnouts}, {Asgari}, {Asorey}, {Atayde}, {Atek}, {Atrio-Barandela}, {Aubert}, {Aubourg}, {Auphan}, {Auricchio}, {Aussel}, {Aussel}, {Avelino}, {Avgoustidis}, {Avila}, {Awan}, {Azzollini}, {Baccigalupi}, {Bachelet}, {Bacon}, {Baes}, {Bagley}, {Bahr-Kalus}, {Balaguera-Antolinez}, {Balbinot}, {Balcells}, {Baldi}, {Baldry}, {Balestra}, {Ballardini}, {Ballester}, {Balogh}, {Ba{\~n}ados}, {Barbier}, {Bardelli}, {Baron}, {Barreiro}, {Barrena}, {Barriere}, {Barros}, {Barthelemy}, {Bartolo}, {Basset},
  {Battaglia}, {Battisti}, {Baugh}, {Baumont}, {Bazzanini}, {Beaulieu}, {Beckmann}, {Belikov}, {Bel}, {Bellagamba}, {Bella}, {Bellini}, {Benabed}, {Bender}, {Benevento}, {Bennett}, {Benson}, {Bergamini}, {Bermejo-Climent}, {Bernardeau}, {Bertacca}, {Berthe}, {Berthier}, {Bethermin}, {Beutler}, {Bevillon}, {Bhargava}, {Bhatawdekar}, {Bianchi}, {Bisigello}, {Biviano}, {Blake}, {Blanchard}, {Blazek}, {Blot}, {Bosco}, {Bodendorf}, {Boenke}, {B{\"o}hringer}, {Boldrini}, {Bolzonella}, {Bonchi}, {Bonici}, {Bonino}, {Bonino}, {Bonvin}, {Bon}, {Booth}, {Borgani}, {Borlaff}, {Borsato}, {Bosco}, {Bose}, {Botticella}, {Boucaud}, {Bouche}, {Boucher}, {Boutigny}, {Bouvard}, {Bouwens}, {Bouy}, {Bowler}, {Bozza}, {Bozzo}, {Branchini}, {Brando}, {Brau-Nogue}, {Brekke}, {Bremer}, {Brescia}, {Breton}, {Brinchmann}, {Brinckmann}, {Brockley-Blatt}, {Brodwin}, {Brouard}, {Brown}, {Bruton}, {Bucko}, {Buddelmeijer}, {Buenadicha}, {Buitrago}, {Burger}, {Burigana}, {Busillo}, {Busonero}, {Cabanac}, {Cabayol-Garcia}, {Cagliari},
  {Caillat}, {Caillat}, {Calabrese}, {Calabro}, {Calderone}, {Calura}, {Camacho Quevedo}, {Camera}, {Campos}, {Canas-Herrera}, {Candini}, {Cantiello}, {Capobianco}, {Cappellaro}, {Cappelluti}, {Cappi}, {Caputi}, {Cara}, {Carbone}, {Cardone}, {Carella}, {Carlberg}, {Carle}, {Carminati}, {Caro}, {Carrasco}, {Carretero}, {Carrilho}, \& {Carron Duque}}]{Euclid24}
{Euclid Collaboration}, {Mellier}, Y., {Abdurro'uf}, {et~al.} 2024, arXiv e-prints, arXiv:2405.13491

\bibitem[{{Fausti Neto} {et~al.}(2018){Fausti Neto}, {da Costa}, {Carnero}, {Gschwend}, {Ogando}, {Sobreira}, {Maia}, {Santiago}, {Rosenfeld}, {Singulani}, {Adean}, {Nunes}, {Campisano}, {Brito}, {Soares}, {Vila-Verde}, {Abbott}, {Abdalla}, {Allam}, {Benoit-L{\'e}vy}, {Brooks}, {Buckley-Geer}, {Capozzi}, {Carrasco Kind}, {Carretero}, {D'Andrea}, {Desai}, {Doel}, {Drlica-Wagner}, {Evrard}, {Fosalba}, {Garc{\'{\i}}a-Bellido}, {Gerdes}, {Gruendl}, {Gutierrez}, {Honscheid}, {James}, {Jeltema}, {Kuehn}, {Kuhlmann}, {Kuropatkin}, {Lahav}, {Lima}, {Marshall}, {Melchior}, {Menanteau}, {Plazas}, {Sanchez}, {Scarpine}, {Schindler}, {Schubnell}, {Sevilla-Noarbe}, {Smith}, {Smith}, {Suchyta}, {Swanson}, {Tarle}, \& {Walker}}]{Fau18}
{Fausti Neto}, A., {da Costa}, L.~N., {Carnero}, A., {et~al.} 2018, Astronomy and Computing, 24, 52

\bibitem[{{Felten} {et~al.}(1966){Felten}, {Gould}, {Stein}, \& {Woolf}}]{Felten66}
{Felten}, J.~E., {Gould}, R.~J., {Stein}, W.~A., \& {Woolf}, N.~J. 1966, \apj, 146, 955

\bibitem[{{Flaugher} {et~al.}(2015){Flaugher}, {Diehl}, {Honscheid}, {et~al.}}]{Fla15}
{Flaugher}, B., {Diehl}, H.~T., {Honscheid}, K., {et~al.} 2015, \aj, 150, 150

\bibitem[{{Foreman-Mackey} {et~al.}(2013){Foreman-Mackey}, {Hogg}, {Lang}, \& {Goodman}}]{emcee}
{Foreman-Mackey}, D., {Hogg}, D.~W., {Lang}, D., \& {Goodman}, J. 2013, \pasp, 125, 306

\bibitem[{{Fumagalli} {et~al.}(2024){Fumagalli}, {Costanzi}, {Saro}, {Castro}, \& {Borgani}}]{Fumagalli23}
{Fumagalli}, A., {Costanzi}, M., {Saro}, A., {Castro}, T., \& {Borgani}, S. 2024, \aap, 682, A148

\bibitem[{{Ghirardini} {et~al.}(2024){Ghirardini}, {Bulbul}, {Artis}, {Clerc}, {Garrel}, {Grandis}, {Kluge}, {Liu}, {Bahar}, {Balzer}, {Chiu}, {Comparat}, {Gruen}, {Kleinebreil}, {Krippendorf}, {Merloni}, {Nandra}, {Okabe}, {Pacaud}, {Predehl}, {Ramos-Ceja}, {Reiprich}, {Sanders}, {Schrabback}, {Seppi}, {Zelmer}, {Zhang}, {Bornemann}, {Brunner}, {Burwitz}, {Coutinho}, {Dennerl}, {Freyberg}, {Friedrich}, {Gaida}, {Gueguen}, {Haberl}, {Kink}, {Lamer}, {Li}, {Liu}, {Maitra}, {Meidinger}, {Mueller}, {Miyatake}, {Miyazaki}, {Robrade}, {Schwope}, \& {Stewart}}]{2024A&A...689A.298G}
{Ghirardini}, V., {Bulbul}, E., {Artis}, E., {et~al.} 2024, \aap, 689, A298

\bibitem[{{G{\'o}rski} {et~al.}(2005){G{\'o}rski}, {Hivon}, {Banday}, {Wandelt}, {Hansen}, {Reinecke}, \& {Bartelmann}}]{healpix}
{G{\'o}rski}, K.~M., {Hivon}, E., {Banday}, A.~J., {et~al.} 2005, \apj, 622, 759

\bibitem[{{Goto} {et~al.}(2002){Goto}, {Sekiguchi}, {Nichol}, {Bahcall}, {Kim}, {Annis}, {Ivezi{\'c}}, {Brinkmann}, {Hennessy}, {Szokoly}, \& {Tucker}}]{goto2002}
{Goto}, T., {Sekiguchi}, M., {Nichol}, R.~C., {et~al.} 2002, \aj, 123, 1807

\bibitem[{{Grandis} {et~al.}(2021){Grandis}, {Mohr}, {Costanzi}, {Saro}, {Bocquet}, {Klein}, {Aguena}, {Allam}, {Annis}, {Ansarinejad}, {Bacon}, {Bertin}, {Bleem}, {Brooks}, {Burke}, {Carnero Rosel}, {Carrasco Kind}, {Carretero}, {Castander}, {Choi}, {da Costa}, {De Vincente}, {Desai}, {Diehl}, {Dietrich}, {Doel}, {Eifler}, {Everett}, {Ferrero}, {Floyd}, {Fosalba}, {Frieman}, {Garc{\'\i}a-Bellido}, {Gaztanaga}, {Gruen}, {Gruendl}, {Gschwend}, {Gupta}, {Gutierrez}, {Hinton}, {Hollowood}, {Honscheid}, {James}, {Jeltema}, {Kuehn}, {Lahav}, {Lidman}, {Lima}, {Maia}, {March}, {Marshall}, {Melchior}, {Menanteau}, {Miquel}, {Morgan}, {Myles}, {Ogando}, {Palmese}, {Paz-Chinch{\'o}n}, {Plazas}, {Reichardt}, {Romer}, {Sanchez}, {Scarpine}, {Serrano}, {Sevilla-Noarbe}, {Singh}, {Smith}, {Suchyta}, {Swanson}, {Tarle}, {Thomas}, {To}, {Weller}, {Wilkinson}, \& {Wu}}]{Grandis21}
{Grandis}, S., {Mohr}, J.~J., {Costanzi}, M., {et~al.} 2021, \mnras, 504, 1253

\bibitem[{{Gschwend} {et~al.}(2018){Gschwend}, {Rossel}, {Ogando}, {Neto}, {Maia}, {da Costa}, {Lima}, {Pellegrini}, {Campisano}, {Singulani}, {Adean}, {Benoist}, {Aguena}, {Carrasco Kind}, {Davis}, {de Vicente}, {Hartley}, {Hoyle}, {Palmese}, {Sadeh}, {Abbott}, {Abdalla}, {Allam}, {Annis}, {Asorey}, {Brooks}, {Calcino}, {Carollo}, {Castander}, {D'Andrea}, {Desai}, {Evrard}, {Fosalba}, {Frieman}, {Garc{\'{\i}}a-Bellido}, {Glazebrook}, {Gerdes}, {Gruendl}, {Gutierrez}, {Hinton}, {Hollowood}, {Honscheid}, {Hoormann}, {James}, {Kuehn}, {Kuropatkin}, {Lahav}, {Lewis}, {Lidman}, {Lin}, {Macaulay}, {Marshall}, {Melchior}, {Miquel}, {M{\"o}ller}, {Plazas}, {Sanchez}, {Santiago}, {Scarpine}, {Schindler}, {Sevilla-Noarbe}, {Smith}, {Sobreira}, {Sommer}, {Suchyta}, {Swanson}, {Tarle}, {Tucker}, {Tucker}, {Uddin}, \& {Walker}}]{GSC18}
{Gschwend}, J., {Rossel}, A.~C., {Ogando}, R.~L.~C., {et~al.} 2018, Astronomy and Computing, 25, 58

\bibitem[{{Hartley} {et~al.}(2022){Hartley}, {Choi}, {Amon}, {Gruendl}, {Sheldon}, {Harrison}, {Bernstein}, {Sevilla-Noarbe}, {Yanny}, {Eckert}, {Diehl}, {Alarcon}, {Banerji}, {Bechtol}, {Buchs}, {Cantu}, {Conselice}, {Cordero}, {Davis}, {Davis}, {Dodelson}, {Drlica-Wagner}, {Everett}, {Fert{\'e}}, {Gruen}, {Honscheid}, {Jarvis}, {Johnson}, {Kokron}, {MacCrann}, {Myles}, {Pace}, {Palmese}, {Paz-Chinch{\'o}n}, {Pereira}, {Plazas}, {Prat}, {Rodriguez-Monroy}, {Rykoff}, {Samuroff}, {S{\'a}nchez}, {Secco}, {Tarsitano}, {Tong}, {Troxel}, {Vasquez}, {Wang}, {Zhou}, {Abbott}, {Aguena}, {Allam}, {Annis}, {Bacon}, {Bertin}, {Bhargava}, {Brooks}, {Burke}, {Carnero Rosell}, {Carrasco Kind}, {Carretero}, {Castander}, {Costanzi}, {Crocce}, {da Costa}, {De Vicente}, {DeRose}, {Desai}, {Dietrich}, {Eifler}, {Elvin-Poole}, {Ferrero}, {Flaugher}, {Fosalba}, {Garc{\'\i}a-Bellido}, {Gaztanaga}, {Gerdes}, {Gschwend}, {Gutierrez}, {Hinton}, {Hollowood}, {Huterer}, {James}, {Kent}, {Krause}, {Kuehn}, {Kuropatkin}, {Lahav}, {Lin},
  {Maia}, {March}, {Marshall}, {Martini}, {Melchior}, {Menanteau}, {Miquel}, {Mohr}, {Morgan}, {Neilsen}, {Ogando}, {Pandey}, {Romer}, {Roodman}, {Sako}, {Sanchez}, {Scarpine}, {Serrano}, {Smith}, {Soares-Santos}, {Suchyta}, {Swanson}, {Tarle}, {Thomas}, {To}, {Varga}, {Walker}, {Wester}, {Wilkinson}, {Zuntz}, {Zuntz}, \& {DES Collaboration}}]{2022MNRAS.509.3547H}
{Hartley}, W.~G., {Choi}, A., {Amon}, A., {et~al.} 2022, \mnras, 509, 3547

\bibitem[{{Hilton} {et~al.}(2021){Hilton}, {Sif{\'o}n}, {Naess}, {Madhavacheril}, {Oguri}, {Rozo}, {Rykoff}, {Abbott}, {Adhikari}, {Aguena}, {Aiola}, {Allam}, {Amodeo}, {Amon}, {Annis}, {Ansarinejad}, {Aros-Bunster}, {Austermann}, {Avila}, {Bacon}, {Battaglia}, {Beall}, {Becker}, {Bernstein}, {Bertin}, {Bhandarkar}, {Bhargava}, {Bond}, {Brooks}, {Burke}, {Calabrese}, {Carrasco Kind}, {Carretero}, {Choi}, {Choi}, {Conselice}, {da Costa}, {Costanzi}, {Crichton}, {Crowley}, {D{\"u}nner}, {Denison}, {Devlin}, {Dicker}, {Diehl}, {Dietrich}, {Doel}, {Duff}, {Duivenvoorden}, {Dunkley}, {Everett}, {Ferraro}, {Ferrero}, {Fert{\'e}}, {Flaugher}, {Frieman}, {Gallardo}, {Garc{\'\i}a-Bellido}, {Gaztanaga}, {Gerdes}, {Giles}, {Golec}, {Gralla}, {Grandis}, {Gruen}, {Gruendl}, {Gschwend}, {Gutierrez}, {Han}, {Hartley}, {Hasselfield}, {Hill}, {Hilton}, {Hincks}, {Hinton}, {Ho}, {Honscheid}, {Hoyle}, {Hubmayr}, {Huffenberger}, {Hughes}, {Jaelani}, {Jain}, {James}, {Jeltema}, {Kent}, {Knowles}, {Koopman}, {Kuehn}, {Lahav},
  {Lima}, {Lin}, {Lokken}, {Loubser}, {MacCrann}, {Maia}, {Marriage}, {Martin}, {McMahon}, {Melchior}, {Menanteau}, {Miquel}, {Miyatake}, {Moodley}, {Morgan}, {Mroczkowski}, {Nati}, {Newburgh}, {Niemack}, {Nishizawa}, {Ogando}, {Orlowski-Scherer}, {Page}, {Palmese}, {Partridge}, {Paz-Chinch{\'o}n}, {Phakathi}, {Plazas}, {Robertson}, {Romer}, {Carnero Rosell}, {Salatino}, {Sanchez}, {Schaan}, {Schillaci}, {Sehgal}, {Serrano}, {Shin}, {Simon}, {Smith}, {Soares-Santos}, {Spergel}, {Staggs}, {Storer}, {Suchyta}, {Swanson}, {Tarle}, {Thomas}, {To}, {Trac}, {Ullom}, {Vale}, {Van Lanen}, {Vavagiakis}, {De Vicente}, {Wilkinson}, {Wollack}, {Xu}, \& {Zhang}}]{Hil21}
{Hilton}, M., {Sif{\'o}n}, C., {Naess}, S., {et~al.} 2021, \apjs, 253, 3

\bibitem[{{Huang} {et~al.}(2020){Huang}, {Bleem}, {Stalder}, {Ade}, {Allen}, {Anderson}, {Austermann}, {Avva}, {Beall}, {Bender}, {Benson}, {Bianchini}, {Bocquet}, {Brodwin}, {Carlstrom}, {Chang}, {Chiang}, {Citron}, {Moran}, {Crawford}, {Crites}, {Haan}, {Dobbs}, {Everett}, {Floyd}, {Gallicchio}, {George}, {Gilbert}, {Gladders}, {Guns}, {Gupta}, {Halverson}, {Harrington}, {Henning}, {Hilton}, {Holder}, {Holzapfel}, {Hrubes}, {Hubmayr}, {Irwin}, {Khullar}, {Knox}, {Lee}, {Li}, {Lowitz}, {McDonald}, {McMahon}, {Meyer}, {Mocanu}, {Montgomery}, {Nadolski}, {Natoli}, {Nibarger}, {Noble}, {Novosad}, {Padin}, {Patil}, {Pryke}, {Reichardt}, {Ruhl}, {Saliwanchik}, {Saro}, {Sayre}, {Schaffer}, {Sharon}, {Sievers}, {Smecher}, {Stark}, {Story}, {Tucker}, {Vanderlinde}, {Veach}, {Vieira}, {Wang}, {Whitehorn}, {Wu}, \& {Yefremenko}}]{Hua20}
{Huang}, N., {Bleem}, L.~E., {Stalder}, B., {et~al.} 2020, \aj, 159, 110

\bibitem[{{Kim} {et~al.}(2002){Kim}, {Kepner}, {Postman}, {Strauss}, {Bahcall}, {Gunn}, {Lupton}, {Annis}, {Nichol}, {Castander}, {Brinkmann}, {Brunner}, {Connolly}, {Csabai}, {Hindsley}, {Ivezi{\'c}}, {Vogeley}, \& {York}}]{kim2002}
{Kim}, R.~S.~J., {Kepner}, J.~V., {Postman}, M., {et~al.} 2002, \aj, 123, 20

\bibitem[{{Lesci} {et~al.}(2022){Lesci}, {Marulli}, {Moscardini}, {Sereno}, {Veropalumbo}, {Maturi}, {Giocoli}, {Radovich}, {Bellagamba}, {Roncarelli}, {Bardelli}, {Contarini}, {Covone}, {Ingoglia}, {Nanni}, \& {Puddu}}]{Lesci22}
{Lesci}, G.~F., {Marulli}, F., {Moscardini}, L., {et~al.} 2022, \aap, 659, A88

\bibitem[{{LSST Science Collaboration} {et~al.}(2009){LSST Science Collaboration}, {Abell}, {Allison}, {Anderson}, {Andrew}, {Angel}, {Armus}, {Arnett}, {Asztalos}, \& {Axelrod}}]{LSST2009}
{LSST Science Collaboration}, {Abell}, P.~A., {Allison}, J., {et~al.} 2009, arXiv e-prints [\eprint[arXiv]{0912.0201}]

\bibitem[{{Mantz} {et~al.}(2015){Mantz}, {von der Linden}, {Allen}, {Applegate}, {Kelly}, {Morris}, {Rapetti}, {Schmidt}, {Adhikari}, {Allen}, {Burchat}, {Burke}, {Cataneo}, {Donovan}, {Ebeling}, {Shandera}, \& {Wright}}]{Mantz15}
{Mantz}, A.~B., {von der Linden}, A., {Allen}, S.~W., {et~al.} 2015, \mnras, 446, 2205

\bibitem[{{{McClintock}, T. \& {Varga}, T.~N.} {et~al.}(2019){{McClintock}, T. \& {Varga}, T.~N.}, {Gruen}, {Rozo}, {Rykoff}, {Shin}, {Melchior}, {DeRose}, {Seitz}, {Dietrich}, {Sheldon}, {Zhang}, {von der Linden}, {Jeltema}, {Mantz}, {Romer}, {Allen}, {Becker}, {Bermeo}, {Bhargava}, {Costanzi}, {Everett}, {Farahi}, {Hamaus}, {Hartley}, {Hollowood}, {Hoyle}, {Israel}, {Li}, {MacCrann}, {Morris}, {Palmese}, {Plazas}, {Pollina}, {Rau}, {Simet}, {Soares-Santos}, {Troxel}, {Vergara Cervantes}, {Wechsler}, {Zuntz}, {Abbott}, {Abdalla}, {Allam}, {Annis}, {Avila}, {Bridle}, {Brooks}, {Burke}, {Carnero Rosell}, {Carrasco Kind}, {Carretero}, {Castander}, {Crocce}, {Cunha}, {D'Andrea}, {da Costa}, {Davis}, {De Vicente}, {Diehl}, {Doel}, {Drlica-Wagner}, {Evrard}, {Flaugher}, {Fosalba}, {Frieman}, {Garc{\'\i}a-Bellido}, {Gaztanaga}, {Gerdes}, {Giannantonio}, {Gruendl}, {Gutierrez}, {Honscheid}, {James}, {Kirk}, {Krause}, {Kuehn}, {Lahav}, {Li}, {Lima}, {March}, {Marshall}, {Menanteau}, {Miquel}, {Mohr}, {Nord},
  {Ogando}, {Roodman}, {Sanchez}, {Scarpine}, {Schindler}, {Sevilla-Noarbe}, {Smith}, {Smith}, {Sobreira}, {Suchyta}, {Swanson}, {Tarle}, {Tucker}, {Vikram}, {Walker}, {Weller}, \& {DES Collaboration}}]{McC19}
{{McClintock}, T. \& {Varga}, T.~N.}, {Gruen}, D., {Rozo}, E., {et~al.} 2019, \mnras, 482, 1352

\bibitem[{{Merloni} {et~al.}(2012){Merloni}, {Predehl}, {Becker}, {B{\"o}hringer}, {Boller}, {Brunner}, {Brusa}, {Dennerl}, {Freyberg}, {Friedrich}, {Georgakakis}, {Haberl}, {Hasinger}, {Meidinger}, {Mohr}, {Nandra}, {Rau}, {Reiprich}, {Robrade}, {Salvato}, {Santangelo}, {Sasaki}, {Schwope}, {Wilms}, \& {German eROSITA Consortium}}]{Merloni12}
{Merloni}, A., {Predehl}, P., {Becker}, W., {et~al.} 2012, arXiv e-prints, arXiv:1209.3114

\bibitem[{{Planck Collaboration} {et~al.}(2020){Planck Collaboration}, {Aghanim}, {Akrami}, {Ashdown}, {Aumont}, {Baccigalupi}, {Ballardini}, {Banday}, {Barreiro}, {Bartolo}, {Basak}, {Benabed}, {Bernard}, {Bersanelli}, {Bielewicz}, {Bond}, {Borrill}, {Bouchet}, {Boulanger}, {Bucher}, {Burigana}, {Calabrese}, {Cardoso}, {Carron}, {Challinor}, {Chiang}, {Colombo}, {Combet}, {Couchot}, {Crill}, {Cuttaia}, {de Bernardis}, {de Rosa}, {de Zotti}, {Delabrouille}, {Delouis}, {Di Valentino}, {Diego}, {Dor{\'e}}, {Douspis}, {Ducout}, {Dupac}, {Efstathiou}, {Elsner}, {En{\ss}lin}, {Eriksen}, {Falgarone}, {Fantaye}, {Finelli}, {Frailis}, {Fraisse}, {Franceschi}, {Frolov}, {Galeotta}, {Galli}, {Ganga}, {G{\'e}nova-Santos}, {Gerbino}, {Ghosh}, {Gonz{\'a}lez-Nuevo}, {G{\'o}rski}, {Gratton}, {Gruppuso}, {Gudmundsson}, {Handley}, {Hansen}, {Henrot-Versill{\'e}}, {Herranz}, {Hivon}, {Huang}, {Jaffe}, {Jones}, {Karakci}, {Keih{\"a}nen}, {Keskitalo}, {Kiiveri}, {Kim}, {Kisner}, {Krachmalnicoff}, {Kunz}, {Kurki-Suonio},
  {Lagache}, {Lamarre}, {Lasenby}, {Lattanzi}, {Lawrence}, {Levrier}, {Liguori}, {Lilje}, {Lindholm}, {L{\'o}pez-Caniego}, {Ma}, {Mac{\'\i}as-P{\'e}rez}, {Maggio}, {Maino}, {Mandolesi}, {Mangilli}, {Martin}, {Mart{\'\i}nez-Gonz{\'a}lez}, {Matarrese}, {Mauri}, {McEwen}, {Melchiorri}, {Mennella}, {Migliaccio}, {Miville-Desch{\^e}nes}, {Molinari}, {Moneti}, {Montier}, {Morgante}, {Moss}, {Mottet}, {Natoli}, {Pagano}, {Paoletti}, {Partridge}, {Patanchon}, {Patrizii}, {Perdereau}, {Perrotta}, {Pettorino}, {Piacentini}, {Puget}, {Rachen}, {Reinecke}, {Remazeilles}, {Renzi}, {Rocha}, {Roudier}, {Salvati}, {Sandri}, {Savelainen}, {Scott}, {Sirignano}, {Sirri}, {Spencer}, {Sunyaev}, {Suur-Uski}, {Tauber}, {Tavagnacco}, {Tenti}, {Toffolatti}, {Tomasi}, {Tristram}, {Trombetti}, {Valiviita}, {Vansyngel}, {Van Tent}, {Vibert}, {Vielva}, {Villa}, {Vittorio}, {Wandelt}, {Wehus}, \& {Zonca}}]{planck_HFI}
{Planck Collaboration}, {Aghanim}, N., {Akrami}, Y., {et~al.} 2020, \aap, 641, A3

\bibitem[{{Rykoff} {et~al.}(2012){Rykoff}, {Koester}, {Rozo}, {Annis}, {Evrard}, {Hansen}, {Hao}, {Johnston}, {McKay}, \& {Wechsler}}]{Ryk12}
{Rykoff}, E.~S., {Koester}, B.~P., {Rozo}, E., {et~al.} 2012, \apj, 746, 178

\bibitem[{{Rykoff} {et~al.}(2014){Rykoff}, {Rozo}, {Busha}, {Cunha}, {Finoguenov}, {Evrard}, {Hao}, {Koester}, {Leauthaud}, {Nord}, {Pierre}, {Reddick}, {Sadibekova}, {Sheldon}, \& {Wechsler}}]{Ryk14}
{Rykoff}, E.~S., {Rozo}, E., {Busha}, M.~T., {et~al.} 2014, \apj, 785, 104

\bibitem[{{S{\'a}nchez} {et~al.}(2014){S{\'a}nchez}, {Carrasco Kind}, {Lin}, {Miquel}, {Abdalla}, {Amara}, {Banerji}, {Bonnett}, {Brunner}, {Capozzi}, {Carnero}, {Castander}, {da Costa}, {Cunha}, {Fausti}, {Gerdes}, {Greisel}, {Gschwend}, {Hartley}, {Jouvel}, {Lahav}, {Lima}, {Maia}, {Mart{\'{\i}}}, {Ogando}, {Ostrovski}, {Pellegrini}, {Rau}, {Sadeh}, {Seitz}, {Sevilla-Noarbe}, {Sypniewski}, {de Vicente}, {Abbot}, {Allam}, {Atlee}, {Bernstein}, {Bernstein}, {Buckley-Geer}, {Burke}, {Childress}, {Davis}, {DePoy}, {Dey}, {Desai}, {Diehl}, {Doel}, {Estrada}, {Evrard}, {Fern{\'a}ndez}, {Finley}, {Flaugher}, {Frieman}, {Gaztanaga}, {Glazebrook}, {Honscheid}, {Kim}, {Kuehn}, {Kuropatkin}, {Lidman}, {Makler}, {Marshall}, {Nichol}, {Roodman}, {S{\'a}nchez}, {Santiago}, {Sako}, {Scalzo}, {Smith}, {Swanson}, {Tarle}, {Thomas}, {Tucker}, {Uddin}, {Vald{\'e}s}, {Walker}, {Yuan}, \& {Zuntz}}]{San14}
{S{\'a}nchez}, C., {Carrasco Kind}, M., {Lin}, H., {et~al.} 2014, \mnras, 445, 1482

\bibitem[{{Saro} {et~al.}(2015){Saro}, {Bocquet}, {Rozo}, {Benson}, {Mohr}, {Rykoff}, {Soares-Santos}, {Bleem}, {Dodelson}, {Melchior}, {Sobreira}, {Upadhyay}, {Weller}, {Abbott}, {Abdalla}, {Allam}, {Armstrong}, {Banerji}, {Bauer}, {Bayliss}, {Benoit-L{\'e}vy}, {Bernstein}, {Bertin}, {Brodwin}, {Brooks}, {Buckley-Geer}, {Burke}, {Carlstrom}, {Capasso}, {Capozzi}, {Carnero Rosell}, {Carrasco Kind}, {Chiu}, {Covarrubias}, {Crawford}, {Crocce}, {D'Andrea}, {da Costa}, {DePoy}, {Desai}, {de Haan}, {Diehl}, {Dietrich}, {Doel}, {Cunha}, {Eifler}, {Evrard}, {Fausti Neto}, {Fernandez}, {Flaugher}, {Fosalba}, {Frieman}, {Gangkofner}, {Gaztanaga}, {Gerdes}, {Gruen}, {Gruendl}, {Gupta}, {Hennig}, {Holzapfel}, {Honscheid}, {Jain}, {James}, {Kuehn}, {Kuropatkin}, {Lahav}, {Li}, {Lin}, {Maia}, {March}, {Marshall}, {Martini}, {McDonald}, {Miller}, {Miquel}, {Nord}, {Ogando}, {Plazas}, {Reichardt}, {Romer}, {Roodman}, {Sako}, {Sanchez}, {Schubnell}, {Sevilla}, {Smith}, {Stalder}, {Stark}, {Strazzullo}, {Suchyta}, {Swanson},
  {Tarle}, {Thaler}, {Thomas}, {Tucker}, {Vikram}, {von der Linden}, {Walker}, {Wechsler}, {Wester}, {Zenteno}, \& {Ziegler}}]{Sar15}
{Saro}, A., {Bocquet}, S., {Rozo}, E., {et~al.} 2015, \mnras, 454, 2305

\bibitem[{{Seppi} {et~al.}(2021){Seppi}, {Comparat}, {Nandra}, {Bulbul}, {Prada}, {Klypin}, {Merloni}, {Predehl}, \& {Ider Chitham}}]{Seppi2021}
{Seppi}, R., {Comparat}, J., {Nandra}, K., {et~al.} 2021, \aap, 652, A155

\bibitem[{{Song} {et~al.}(2012){Song}, {Zenteno}, {Stalder}, {Desai}, {Bleem}, {Aird}, {Armstrong}, {Ashby}, {Bayliss}, {Bazin}, {Benson}, {Bertin}, {Brodwin}, {Carlstrom}, {Chang}, {Cho}, {Clocchiatti}, {Crawford}, {Crites}, {de Haan}, {Dobbs}, {Dudley}, {Foley}, {George}, {Gettings}, {Gladders}, {Gonzalez}, {Halverson}, {Harrington}, {High}, {Holder}, {Holzapfel}, {Hoover}, {Hrubes}, {Joy}, {Keisler}, {Knox}, {Lee}, {Leitch}, {Liu}, {Lueker}, {Luong-Van}, {Marrone}, {McDonald}, {McMahon}, {Mehl}, {Meyer}, {Mocanu}, {Mohr}, {Montroy}, {Natoli}, {Nurgaliev}, {Padin}, {Plagge}, {Pryke}, {Reichardt}, {Rest}, {Ruel}, {Ruhl}, {Saliwanchik}, {Saro}, {Sayre}, {Schaffer}, {Shaw}, {Shirokoff}, {{\v{S}}uhada}, {Spieler}, {Stanford}, {Staniszewski}, {Stark}, {Story}, {Stubbs}, {van Engelen}, {Vanderlinde}, {Vieira}, {Williamson}, \& {Zahn}}]{song12}
{Song}, J., {Zenteno}, A., {Stalder}, B., {et~al.} 2012, \apj, 761, 22

\bibitem[{{Sunayama} {et~al.}(2024){Sunayama}, {Miyatake}, {Sugiyama}, {More}, {Li}, {Dalal}, {Rau}, {Shi}, {Chiu}, {Shirasaki}, {Zhang}, \& {Nishizawa}}]{Sunayama23}
{Sunayama}, T., {Miyatake}, H., {Sugiyama}, S., {et~al.} 2024, \prd, 110, 083511

\bibitem[{{Sunyaev} \& {Zeldovich}(1972)}]{SZE72}
{Sunyaev}, R.~A. \& {Zeldovich}, Y.~B. 1972, Comments on Astrophysics and Space Physics, 4, 173

\bibitem[{{The Dark Energy Survey Collaboration}(2005)}]{DES05}
{The Dark Energy Survey Collaboration}. 2005, arXiv e-prints, astro

\bibitem[{{To} {et~al.}(2021){To}, {Krause}, {Rozo}, {Wu}, {Gruen}, {Wechsler}, {Eifler}, {Rykoff}, {Costanzi}, {Becker}, {Bernstein}, {Blazek}, {Bocquet}, {Bridle}, {Cawthon}, {Choi}, {Crocce}, {Davis}, {DeRose}, {Drlica-Wagner}, {Elvin-Poole}, {Fang}, {Farahi}, {Friedrich}, {Gatti}, {Gaztanaga}, {Giannantonio}, {Hartley}, {Hoyle}, {Jarvis}, {MacCrann}, {McClintock}, {Miranda}, {Pereira}, {Park}, {Porredon}, {Prat}, {Rau}, {Ross}, {Samuroff}, {S{\'a}nchez}, {Sevilla-Noarbe}, {Sheldon}, {Troxel}, {Varga}, {Vielzeuf}, {Zhang}, {Zuntz}, {Abbott}, {Aguena}, {Amon}, {Annis}, {Avila}, {Bertin}, {Bhargava}, {Brooks}, {Burke}, {Carnero Rosell}, {Carrasco Kind}, {Carretero}, {Chang}, {Conselice}, {da Costa}, {Davis}, {Desai}, {Diehl}, {Dietrich}, {Everett}, {Evrard}, {Ferrero}, {Flaugher}, {Fosalba}, {Frieman}, {Garc{\'\i}a-Bellido}, {Gruendl}, {Gutierrez}, {Hinton}, {Hollowood}, {Honscheid}, {Huterer}, {James}, {Jeltema}, {Kron}, {Kuehn}, {Kuropatkin}, {Lima}, {Maia}, {Marshall}, {Menanteau}, {Miquel}, {Morgan},
  {Muir}, {Myles}, {Palmese}, {Paz-Chinch{\'o}n}, {Plazas}, {Romer}, {Roodman}, {Sanchez}, {Santiago}, {Scarpine}, {Serrano}, {Smith}, {Suchyta}, {Swanson}, {Tarle}, {Thomas}, {Tucker}, {Weller}, {Wester}, {Wilkinson}, \& {DES Collaboration}}]{To21}
{To}, C., {Krause}, E., {Rozo}, E., {et~al.} 2021, \prl, 126, 141301

\bibitem[{{To} {et~al.}(2024){To}, {DeRose}, {Wechsler}, {Rykoff}, {Wu}, {Adhikari}, {Krause}, {Rozo}, \& {Weinberg}}]{To24}
{To}, C.-H., {DeRose}, J., {Wechsler}, R.~H., {et~al.} 2024, \apj, 961, 59

\bibitem[{{Upsdell} {et~al.}(2023){Upsdell}, {Giles}, {Romer}, {Wilkinson}, {Turner}, {Hilton}, {Rykoff}, {Farahi}, {Bhargava}, {Jeltema}, {Klein}, {Bermeo}, {Collins}, {Ebrahimpour}, {Hollowood}, {Mann}, {Manolopoulou}, {Miller}, {Rooney}, {Sahl{\'e}n}, {Stott}, {Viana}, {Allam}, {Alves}, {Bacon}, {Bertin}, {Bocquet}, {Brooks}, {Burke}, {Carrasco Kind}, {Carretero}, {Costanzi}, {da Costa}, {Pereira}, {De Vicente}, {Desai}, {Diehl}, {Dietrich}, {Everett}, {Ferrero}, {Frieman}, {Garc{\'\i}a-Bellido}, {Gerdes}, {Gutierrez}, {Hinton}, {Honscheid}, {James}, {Kuehn}, {Kuropatkin}, {Lima}, {Marshall}, {Mena-Fern{\'a}ndez}, {Menanteau}, {Miquel}, {Mohr}, {Ogando}, {Pieres}, {Raveri}, {Rodriguez-Monroy}, {Sanchez}, {Scarpine}, {Sevilla-Noarbe}, {Smith}, {Suchyta}, {Swanson}, {Tarle}, {To}, {Weaverdyck}, {Weller}, \& {Wiseman}}]{Upsdell23}
{Upsdell}, E.~W., {Giles}, P.~A., {Romer}, A.~K., {et~al.} 2023, \mnras, 522, 5267

\bibitem[{{Weinberg} {et~al.}(2013){Weinberg}, {Mortonson}, {Eisenstein}, {Hirata}, {Riess}, \& {Rozo}}]{Wei13}
{Weinberg}, D.~H., {Mortonson}, M.~J., {Eisenstein}, D.~J., {et~al.} 2013, \physrep, 530, 87

\bibitem[{{Wen} \& {Han}(2024)}]{2024ApJS..272...39W}
{Wen}, Z.~L. \& {Han}, J.~L. 2024, \apjs, 272, 39

\bibitem[{{Yantovski-Barth} {et~al.}(2024){Yantovski-Barth}, {Newman}, {Dey}, {Andrews}, {Eracleous}, {Golden-Marx}, \& {Zhou}}]{2024MNRAS.531.2285Y}
{Yantovski-Barth}, M.~J., {Newman}, J.~A., {Dey}, B., {et~al.} 2024, \mnras, 531, 2285

\bibitem[{{Zenteno} {et~al.}(2020){Zenteno}, {Hern{\'a}ndez-Lang}, {Klein}, {Vergara Cervantes}, {Hollowood}, {Bhargava}, {Palmese}, {Strazzullo}, {Romer}, {Mohr}, {Jeltema}, {Saro}, {Lidman}, {Gruen}, {Ojeda}, {Katzenberger}, {Aguena}, {Allam}, {Avila}, {Bayliss}, {Bertin}, {Brooks}, {Buckley-Geer}, {Burke}, {Capasso}, {Carnero Rosell}, {Carrasco Kind}, {Carretero}, {Castander}, {Costanzi}, {da Costa}, {De Vicente}, {Desai}, {Diehl}, {Doel}, {Eifler}, {Evrard}, {Flaugher}, {Floyd}, {Fosalba}, {Frieman}, {Garc{\'\i}a-Bellido}, {Gerdes}, {Gonzalez}, {Gruendl}, {Gschwend}, {Gutierrez}, {Hartley}, {Hinton}, {Honscheid}, {James}, {Kuehn}, {Lahav}, {Lima}, {McDonald}, {Maia}, {March}, {Melchior}, {Menanteau}, {Miquel}, {Ogando}, {Paz-Chinch{\'o}n}, {Plazas}, {Roodman}, {Rykoff}, {Sanchez}, {Scarpine}, {Schubnell}, {Serrano}, {Sevilla-Noarbe}, {Smith}, {Soares-Santos}, {Suchyta}, {Swanson}, {Tarle}, {Thomas}, {Varga}, {Walker}, {Wilkinson}, \& {DES Collaboration}}]{Zen20}
{Zenteno}, A., {Hern{\'a}ndez-Lang}, D., {Klein}, M., {et~al.} 2020, \mnras, 495, 705

\end{thebibliography}
%%%%%%%%%%%%%%%%%%%%%
\section*{Affiliations}{
{\tiny

  1. Université C\^{o}te d'Azur, OCA, CNRS, Lagrange, UMR 7293, CS 34229, 06304, Nice Cedex 4, France
 \\  2. Laborat\'orio Interinstitucional de e-Astronomia - LIneA, Av. Pastor Martin Luther King Jr, 126 Del Castilho, Nova Am\'erica Offices, Torre 3000/sala 817 CEP: 20765-000, Brazil
 \\  3. INAF-Osservatorio Astronomico di Trieste, via G. B. Tiepolo 11, I-34143 Trieste, Italy
 \\  4. Fermi National Accelerator Laboratory, P. O. Box 500, Batavia, IL 60510, USA
 \\  5. Department of Physics, University of Michigan, Ann Arbor, MI 48109, USA
 \\  6. Physik-Institut, University of Zürich, Winterthurerstrasse 190, CH-8057 Zürich, Switzerland
 \\  7. Institute of Cosmology and Gravitation, University of Portsmouth, Portsmouth, PO1 3FX, UK
 \\  8. Argonne National Laboratory, 9700 South Cass Avenue, Lemont, IL 60439, USA
 \\  9. Department of Physics \& Astronomy, University College London, Gower Street, London, WC1E 6BT, UK
 \\ 10. Instituto de Astrofisica de Canarias, E-38205 La Laguna, Tenerife, Spain
 \\ 11. Universidad de La Laguna, Dpto. Astrofísica, E-38206 La Laguna, Tenerife, Spain
 \\ 12. Institut de F\'{\i}sica d'Altes Energies (IFAE), The Barcelona Institute of Science and Technology, Campus UAB, 08193 Bellaterra (Barcelona) Spain
 \\ 13. Institut d'Estudis Espacials de Catalunya (IEEC), 08034 Barcelona, Spain
 \\ 14. Institute of Space Sciences (ICE, CSIC),  Campus UAB, Carrer de Can Magrans, s/n,  08193 Barcelona, Spain
 \\ 15. Astronomy Unit, Department of Physics, University of Trieste, via Tiepolo 11, I-34131 Trieste, Italy
 \\ 16. Institute for Fundamental Physics of the Universe, Via Beirut 2, 34014 Trieste, Italy
 \\ 17. Centro de Investigaciones Energ\'eticas, Medioambientales y Tecnol\'ogicas (CIEMAT), Madrid, Spain
 \\ 18. Department of Physics, IIT Hyderabad, Kandi, Telangana 502285, India
 \\ 19. Department of Astronomy and Astrophysics, University of Chicago, Chicago, IL 60637, USA
 \\ 20. Kavli Institute for Cosmological Physics, University of Chicago, Chicago, IL 60637, USA
 \\ 21. California Institute of Technology, 1200 East California Blvd, MC 249-17, Pasadena, CA 91125, USA
 \\ 22. Instituto de Fisica Teorica UAM/CSIC, Universidad Autonoma de Madrid, 28049 Madrid, Spain
 \\ 23. Department of Physics and Astronomy, Pevensey Building, University of Sussex, Brighton, BN1 9QH, UK
 \\ 24. Center for Astrophysical Surveys, National Center for Supercomputing Applications, 1205 West Clark St., Urbana, IL 61801, USA
 \\ 25. Department of Astronomy, University of Illinois at Urbana-Champaign, 1002 W. Green Street, Urbana, IL 61801, USA
 \\ 26. School of Mathematics and Physics, University of Queensland,  Brisbane, QLD 4072, Australia
 \\ 27. Santa Cruz Institute for Particle Physics, Santa Cruz, CA 95064, USA
 \\ 28. Center for Cosmology and Astro-Particle Physics, The Ohio State University, Columbus, OH 43210, USA
 \\ 29. Department of Physics, The Ohio State University, Columbus, OH 43210, USA
 \\ 30. Center for Astrophysics $\vert$ Harvard \& Smithsonian, 60 Garden Street, Cambridge, MA 02138, USA
 \\ 31. Australian Astronomical Optics, Macquarie University, North Ryde, NSW 2113, Australia
 \\ 32. Lowell Observatory, 1400 Mars Hill Rd, Flagstaff, AZ 86001, USA
 \\ 33. Jet Propulsion Laboratory, California Institute of Technology, 4800 Oak Grove Dr., Pasadena, CA 91109, USA
 \\ 34. George P. and Cynthia Woods Mitchell Institute for Fundamental Physics and Astronomy, and Department of Physics and Astronomy, Texas A\&M University, College Station, TX 77843,  USA
 \\ 35. Universit\'e Grenoble Alpes, CNRS, LPSC-IN2P3, 38000 Grenoble, France
 \\ 36. Instituci\'o Catalana de Recerca i Estudis Avan\c{c}ats, E-08010 Barcelona, Spain
 \\ 37. Kavli Institute for Particle Astrophysics \& Cosmology, P. O. Box 2450, Stanford University, Stanford, CA 94305, USA
 \\ 38. SLAC National Accelerator Laboratory, Menlo Park, CA 94025, USA
 \\ 39. Instituto de F\'\i sica, UFRGS, Caixa Postal 15051, Porto Alegre, RS - 91501-970, Brazil
 \\ 40. Physics Department, Lancaster University, Lancaster, LA1 4YB, UK
 \\ 41. Computer Science and Mathematics Division, Oak Ridge National Laboratory, Oak Ridge, TN 37831
 \\ 42. Department of Astronomy, University of California, Berkeley,  501 Campbell Hall, Berkeley, CA 94720, USA
 \\ 43. Lawrence Berkeley National Laboratory, 1 Cyclotron Road, Berkeley, CA 94720, USA
 \\ 44. Max Planck Institute for Extraterrestrial Physics, Giessenbachstrasse, 85748 Garching, Germany
 \\ 45. Universit\"ats-Sternwarte, Fakult\"at f\"ur Physik, Ludwig-Maximilians Universit\"at M\"unchen, Scheinerstr. 1, 81679 M\"unchen, Germany
 \\ 46. Hamburger Sternwarte, Universit\"{a}t Hamburg, Gojenbergsweg 112, 21029 Hamburg, Germany
}}

\begin{appendix}

\section{Photometric redshifts quality and dust absorption}
\label{app:dust}

\begin{figure}[h]
\includegraphics[scale=0.5]{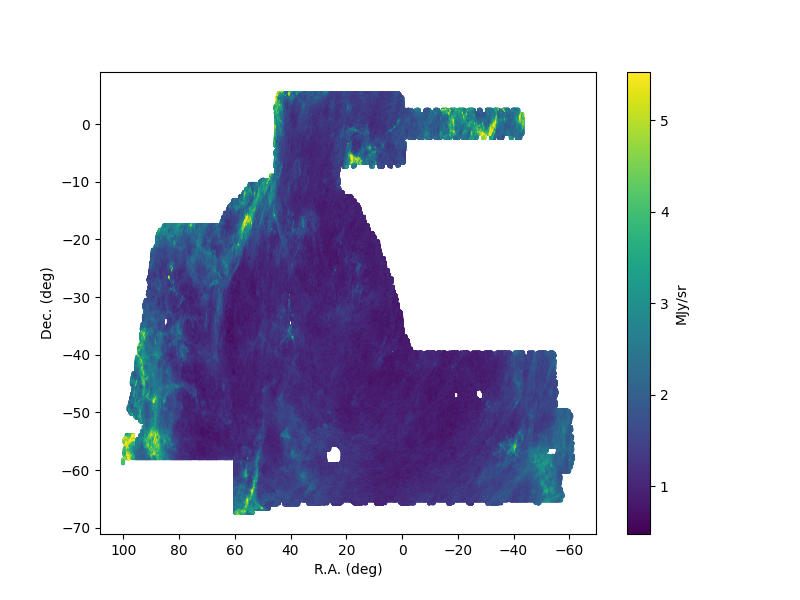}
\caption{The Planck All Sky Map at 857 GHz restricted to the DES footprint.}
\label{fig:dust}
\end{figure}

In Sec.~\ref{sec:homog}, we evaluate the spatial homogeneity of the photometric redshift errors used in this work (Fig.~\ref{fig:pdz_err_radec}). In Fig.~\ref{fig:dust}, the comparison with the Planck All Sky Map at 857 GHz \citep{planck_HFI} shows that the few regions with larger errors correspond to larger galactic absorption. 

\section{Examples of \wazp\ detections in DES-Y6 }
\label{app:wazp_ex}

In section~\ref{sec:sz_to_opt}, we present the cross-match between \wazp\ optically selected clusters and \sz\ \act\ and \spt\ clusters. In Fig.~\ref{app:wazp_ex}, we show examples of such systems that span three redshift and three richness ranges. 

\begin{figure*}
\centering
    \includegraphics[scale=0.28]{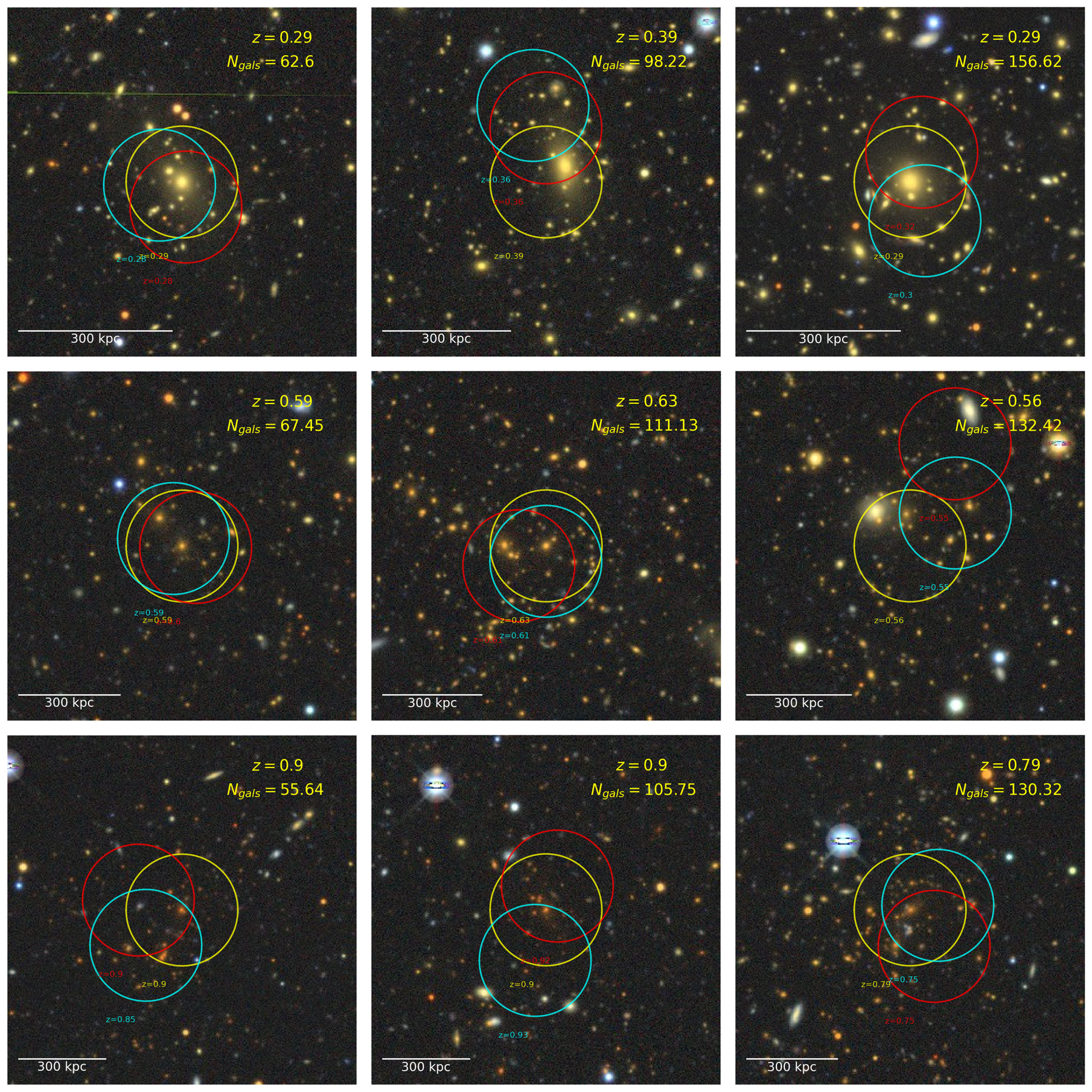}
\caption{Mosaic of 9 \wazp\ detections (yellow) with their \sz\ counterparts (\spt\ in blue and \act\ in red). Each (g, r, i) color postage stamp has a size of $2.5 \times 2.5$\,arcmin and is generated from the Legacy Survey DR10 cutout service. These detections are selected to sample three redshift ranges (vertically) and three richness ranges (horizontally). }
\label{fig:9examples}
\end{figure*}

\end{appendix}

\end{document}